%% file: ms.tex
\documentclass[iop,twocolumn]{emulateapj}
\usepackage{color}

\newcommand{\Msol}{M$_{\odot}$}

\newcommand{\mum}{$\mu$m\ }
\newcommand{\None}{NGC~2023/2024}
\newcommand{\Ntwo}{NGC~2068/2071}

\defcitealias{Gutermuth09}{G09}
\defcitealias{Kirk11}{KM11}

\begin{document}
\title{The JCMT Gould Belt Survey: A First Look at Dense Cores in Orion~B}
\author{
H. Kirk\altaffilmark{1}, 
J. Di Francesco\altaffilmark{1, 2}, 
D. Johnstone\altaffilmark{1, 2, 3}, 
A. Duarte-Cabral\altaffilmark{4}, 
S. Sadavoy\altaffilmark{5},
J. Hatchell\altaffilmark{4}, 
J.C. Mottram\altaffilmark{5, 6}, 
J. Buckle\altaffilmark{7, 8}, 
D.S. Berry\altaffilmark{3}, 
H. Broekhoven-Fiene\altaffilmark{2}, 
M.J. Currie\altaffilmark{3}, 
M. Fich\altaffilmark{9}, 
T. Jenness\altaffilmark{3, 10}, 
D. Nutter\altaffilmark{11}, 
K. Pattle\altaffilmark{12}, 
J.E. Pineda\altaffilmark{13, 14, 15}, 
C. Quinn\altaffilmark{11}, 
C. Salji\altaffilmark{7, 8}, 
S. Tisi\altaffilmark{9}, 
M.R. Hogerheijde\altaffilmark{6}, 
D. Ward-Thompson\altaffilmark{12}, 
P. Bastien\altaffilmark{16}, 
D. Bresnahan\altaffilmark{12}, 
H. Butner\altaffilmark{17}, 
M. Chen\altaffilmark{2}, 
A. Chrysostomou\altaffilmark{18}, 
S. Coude\altaffilmark{16}, 
C.J. Davis\altaffilmark{19}, 
E. Drabek-Maunder\altaffilmark{20}, 
J. Fiege\altaffilmark{21}, 
P. Friberg\altaffilmark{3}, 
R. Friesen\altaffilmark{22}, 
G.A. Fuller\altaffilmark{14}, 
S. Graves\altaffilmark{3}, 
J. Greaves\altaffilmark{23}, 
J. Gregson\altaffilmark{24, 25}, 
W. Holland\altaffilmark{26, 27}, 
G. Joncas\altaffilmark{28}, 
J.M. Kirk\altaffilmark{12}, 
L.B.G. Knee\altaffilmark{1}, 
S. Mairs\altaffilmark{2}, 
K. Marsh\altaffilmark{11}, 
B.C. Matthews\altaffilmark{1, 2}, 
G. Moriarty-Schieven\altaffilmark{1}, 
C. Mowat\altaffilmark{4}, 
J. Rawlings\altaffilmark{29}, 
J. Richer\altaffilmark{7, 8}, 
D. Robertson\altaffilmark{30}, 
E. Rosolowsky\altaffilmark{31}, 
D. Rumble\altaffilmark{4},
H. Thomas\altaffilmark{3},
N. Tothill\altaffilmark{32},
S. Viti\altaffilmark{29},
G.J. White\altaffilmark{24, 25},
J. Wouterloot\altaffilmark{3},
J. Yates\altaffilmark{29},
M. Zhu\altaffilmark{33}}

\altaffiltext{1}{NRC Herzberg Astronomy and Astrophysics, 5071 West Saanich Rd, Victoria, BC, V9E 2E7, Canada}
\altaffiltext{2}{Department of Physics and Astronomy, University of Victoria, Victoria, BC, V8P 1A1, Canada}
\altaffiltext{3}{Joint Astronomy Centre, 660 N. A`oh\={o}k\={u} Place, University Park, Hilo, Hawaii 96720, USA}
\altaffiltext{4}{Physics and Astronomy, University of Exeter, Stocker Road, Exeter EX4 4QL, UK}
\altaffiltext{5}{Max Planck Institute for Astronomy, K\"{o}nigstuhl 17, D-69117 Heidelberg, Germany}
\altaffiltext{6}{Leiden Observatory, Leiden University, PO Box 9513, 2300 RA Leiden, The Netherlands}
\altaffiltext{7}{Astrophysics Group, Cavendish Laboratory, J J Thomson Avenue, Cambridge, CB3 0HE, UK}
\altaffiltext{8}{Kavli Institute for Cosmology, Institute of Astronomy, University of Cambridge, Madingley Road, Cambridge, CB3 0HA, UK}
\altaffiltext{9}{Department of Physics and Astronomy, University of Waterloo, Waterloo, Ontario, N2L 3G1, Canada  }
\altaffiltext{10}{LSST Project Office, 933 N. Cherry Ave, Tucson, AZ 85719, USA}
\altaffiltext{11}{School of Physics and Astronomy, Cardiff University, The Parade, Cardiff, CF24 3AA, UK}
\altaffiltext{12}{Jeremiah Horrocks Institute, University of Central Lancashire, Preston, Lancashire, PR1 2HE, UK}
\altaffiltext{13}{European Southern Observatory (ESO), Garching, Germany}
\altaffiltext{14}{Jodrell Bank Centre for Astrophysics, Alan Turing Building, School of Physics and Astronomy, University of Manchester, Oxford Road, Manchester, M13 9PL, UK}
\altaffiltext{15}{Current address: Max Planck Institute for Extraterrestrial Physics, Giessenbachstrasse 1, 85748 Garching, Germany}
\altaffiltext{16}{Universit\'e de Montr\'eal, Centre de Recherche en Astrophysique du Qu\'ebec et d\'epartement de physique, C.P. 6128, succ. centre-ville, Montr\'eal, QC, H3C 3J7, Canada}
\altaffiltext{17}{James Madison University, Harrisonburg, Virginia 22807, USA}
\altaffiltext{18}{School of Physics, Astronomy \& Mathematics, University of Hertfordshire, College Lane, Hatfield, HERTS AL10 9AB, UK}
\altaffiltext{19}{Astrophysics Research Institute, Liverpool John Moores University, Egerton Warf, Birkenhead, CH41 1LD, UK}
\altaffiltext{20}{Imperial College London, Blackett Laboratory, Prince Consort Rd, London SW7 2BB, UK}
\altaffiltext{21}{Dept of Physics \& Astronomy, University of Manitoba, Winnipeg, Manitoba, R3T 2N2 Canada}
\altaffiltext{22}{Dunlap Institute for Astronomy \& Astrophysics, University of Toronto, 50 St. George St., Toronto ON M5S 3H4 Canada}
\altaffiltext{23}{Physics \& Astronomy, University of St Andrews, North Haugh, St Andrews, Fife KY16 9SS, UK}
\altaffiltext{24}{Dept. of Physical Sciences, The Open University, Milton Keynes MK7 6AA, UK}
\altaffiltext{25}{The Rutherford Appleton Laboratory, Chilton, Didcot, OX11 0NL, UK.}
\altaffiltext{26}{UK Astronomy Technology Centre, Royal Observatory, Blackford Hill, Edinburgh EH9 3HJ, UK}
\altaffiltext{27}{Institute for Astronomy, Royal Observatory, University of Edinburgh, Blackford Hill, Edinburgh EH9 3HJ, UK}
\altaffiltext{28}{Centre de recherche en astrophysique du Qu\'ebec et D\'epartement de physique, de g\'enie physique et d'optique, Universit\'e Laval, 1045 avenue de la m\'edecine, Qu\'ebec, G1V 0A6, Canada}
\altaffiltext{29}{Department of Physics and Astronomy, UCL, Gower St, London, WC1E 6BT, UK}
\altaffiltext{30}{Department of Physics and Astronomy, McMaster University, Hamilton, ON, L8S 4M1, Canada}
\altaffiltext{31}{Department of Physics, University of Alberta, Edmonton, AB T6G 2E1, Canada}
\altaffiltext{32}{University of Western Sydney, Locked Bag 1797, Penrith NSW 2751, Australia}
\altaffiltext{33}{National Astronomical Observatory of China, 20A Datun Road, Chaoyang District, Beijing 100012, China}

\slugcomment{\today}

\begin{abstract}
We present a first look at the SCUBA-2 observations of three sub-regions of the
Orion~B molecular cloud: 
LDN~1622, \None, and \Ntwo, from the JCMT Gould Belt Legacy Survey.  We identify 
29, 564, and 322 dense
cores in L1622, \None, and \Ntwo\ respectively,
using the SCUBA-2 850~\mum map, and present their basic properties,
including their peak fluxes, total fluxes, and sizes, and an estimate of the 
corresponding 450~\mum peak fluxes and total fluxes, using the FellWalker
source extraction algorithm. 
Assuming a constant temperature of 20~K, the starless dense cores
have a mass function similar to that found in previous dense core analyses,
with a Salpeter-like slope at the high-mass end.  The majority
of cores appear stable to gravitational collapse when considering only thermal pressure;
indeed,
most of the cores which have masses above the thermal Jeans mass are already associated
with at least one protostar.  At higher cloud column densities, 
above $1-2 \times 10^{23}$~cm$^{-2}$, most of the mass is 
found within dense cores, while at lower cloud column densities, 
below $1 \times 10^{23}$~cm$^{-2}$, this fraction drops to
10\% or lower.  Overall, the fraction of dense cores associated with a 
protostar is quite small ($<8\%$), 
but becomes larger for the densest and most centrally concentrated
cores.  \None\ and \Ntwo\ appear to be on the path to forming a significant number
of stars in the future, while L1622 has little additional mass in dense cores 
to form many new stars.
\end{abstract}

\section{INTRODUCTION}
The James Clerk Maxwell Telescope Gould Belt Survey has mapped nearly all of the
nearby ($\sim$500~pc)
significant star-forming regions visible from Hawaii with the SCUBA-2
instrument \citep{Holland13}, tracing thermal emission from dust grains at 850~\mum and 450~\mum 
\citep{WardThomp07}.  
A subset of these star-forming regions has also been mapped in 
3--2 line emission of CO isotopologues using HARP 
\citep{Buckle09}.
With a variety of nearby star-forming regions mapped in a uniform manner, one of the
goals of the GBS is to characterize the properties of dense cores and their surroundings,
and determine the influence of the larger environment on their formation and evolution.
In this paper, we present a first look at the SCUBA-2 observations of the Orion~B molecular
cloud using SCUBA-2, identifying dense cores and analyzing their basic properties.
\citet{Buckle10} earlier presented a first-look analysis of the $^{12}$CO, 
$^{13}$CO, and C$^{18}$O line observations in Orion~B.

The Orion~B molecular cloud is part of the larger Orion complex, a large 
\citep[$\sim$100~pc long;][]{Maddalena86}, nearby 
\citep[$\sim$415~pc, e.g.,][]{AntTwarog82,Menten07} 
set of associated molecular clouds forming both 
low- and high- mass stars \citep[e.g.,][]{Bally08}.  The best-studied part of the
Orion complex is the Orion A cloud, which includes the Integral Shaped Filament
\citep[e.g.,][]{Bally87}
and the Orion Nebula Cluster \citep[e.g.,][]{Muench08}.  The Orion B cloud lies northeast of the
Orion A cloud and has a similar total mass of about $10^5$~\Msol\ 
\citep[e.g.,][]{Maddalena86,Meyer08}
but a smaller fraction of dense gas.  This lower fraction of dense gas also translates 
into a lower overall
star formation rate \citep[two to seven times lower;][]{Meyer08}. 
\citet{Lombardi14} found that the surface density
of young protostars varies roughly with the square of the extinction (or total column density)
in Orion.  The bulk of star formation in Orion~B is
concentrated within three clusters, 
NGC~2024, NGC~2068, and NGC~2071, which are estimated to contain 60\% to 90\% of
the current YSOs in Orion~B, while a fourth cluster, NGC~2023, is forming a smaller number of
stars \citep[e.g.,][]{Lada91,Meyer08}.
The most active parts of these four regions have been analyzed using 
prior submillimetre observations, including dust continuum maps from 
SCUBA \citep[e.g.,][]{Motte01,Mitchell01,Johnstone01,Johnstone06,Nutter07} and
the polarimeter attached to SCUBA \citep{Matthews02a,Matthews02b}.
Our SCUBA-2 observations cover a larger area around these four regions than the
original SCUBA data -- 2.1 and 1.7 square degrees were mapped by SCUBA-2 in \None\ and \Ntwo\ 
respectively, compared to 0.5 and 0.3 square degrees with SCUBA.  Our SCUBA-2
observations also cover a fifth
region, LDN~1622 (0.6 square degrees mapped), 
which contains roughly 30 YSOs \citep{Reipurth08}.
L1622 is formally part of `Orion East' 
and has a different typical CO centroid velocity than the neighbouring Orion~B
\citep[e.g., $\sim 1$~km~s$^{-1}$ versus $\sim$10~km~s$^{-1}$;][]{Maddalena86}.  
\citet{Reipurth08}, however, cite other evidence that suggests L1622 is still 
part of the same Orion complex at a similar distance as Orion~A and B, though 
a few observations
suggest a distance of less than 200~pc \citep[see discussion in][]{Reipurth08}. 

Star-forming regions tend to display hierarchical structure, as recent {\it Herschel}
Gould Belt Survey results \citep[e.g.,][]{Andre10,Andre14} have beautifully illustrated. 
The larger-scale (column) density distribution of material is often traced with
CO observations \citep[e.g.,][]{Maddalena86}, estimates of the dust column density based on
stellar reddening \citep[e.g.,][]{Lombardi11}, or more recently, 
combining {\it Herschel} and {\it Planck} measurements
of dust emission \citep[e.g.,][]{Lombardi14}.
SCUBA-2 is insensitive to the largest scale of (lower) column density, like any ground-based 
submillimetre instrument, but provides a higher-resolution view of smaller-scale
dense objects than the former measurements can usually provide.  
For example, Ward-Thompson et al (2015, in prep)
show that in the Taurus molecular cloud,
SCUBA-2 is particularly sensitive to the denser, more compact
objects that will likely become (or already are) the birthsites of protostars,
even when the effects of ground-based filtering are accounted for.

In our first-look analysis, we examine the dense cores detected by SCUBA-2 in the
context of the larger-scale column density \citep[using data from][]{Lombardi14},
as well as already-formed young protostars (using data from Megeath et al. 2012 and
Stutz et al. 2013).
In this paper, we describe the SCUBA-2 observations (Section~2), identify the dense cores 
therein (Section~\ref{sec_core_id}), analyze the basic properties 
of the cores including
their masses, gravitational stability, and relationship with the material in
the larger cloud 
(Section~\ref{sec_core_props}), discuss our results (Section~\ref{sec_disc}), and
summarize our conclusions (Section~\ref{sec_conc}).

\section{OBSERVATIONS}
\label{sec_obs}
\begin{figure*}[htbp]
\includegraphics[height=7in]{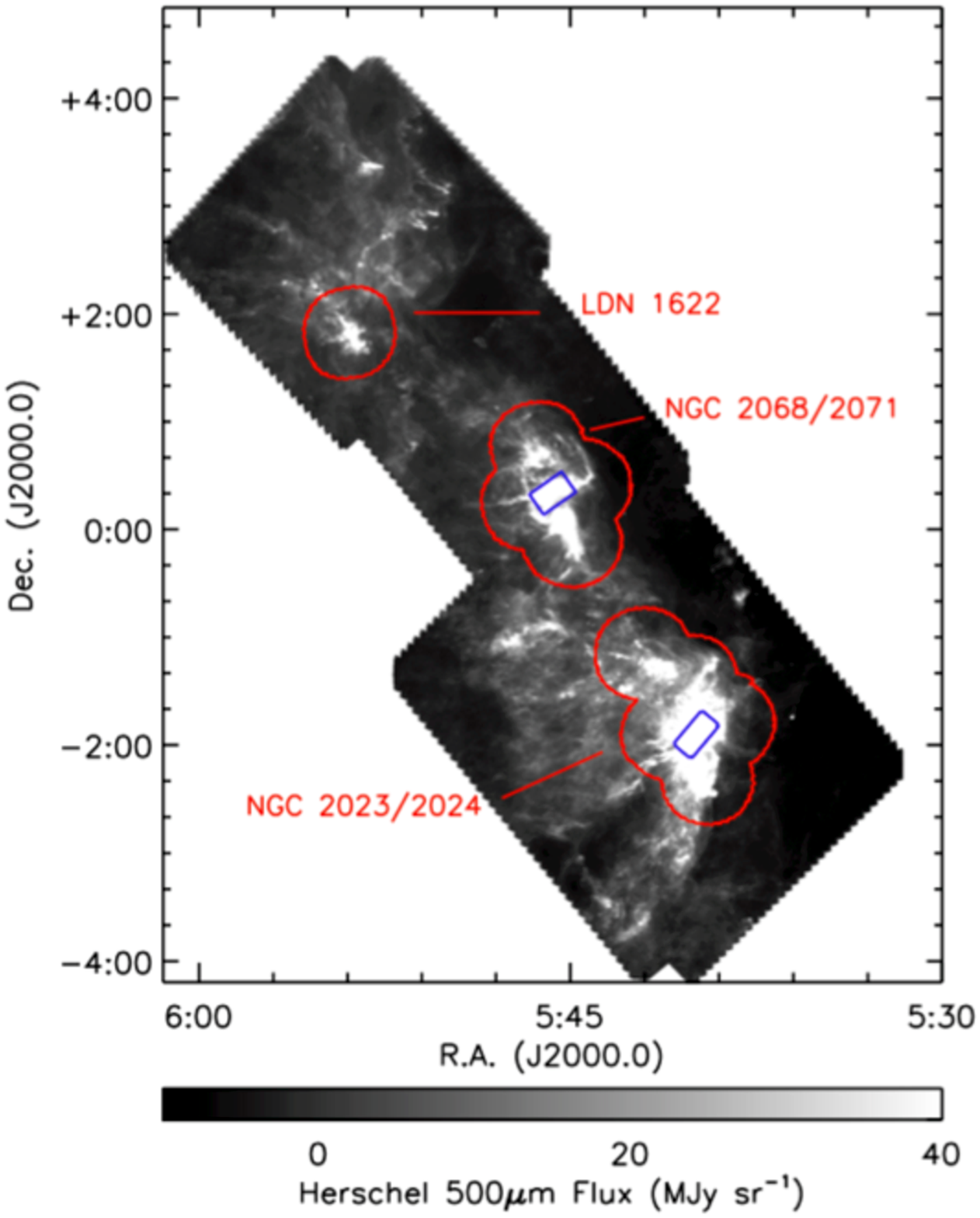}
\caption{The areas observed by SCUBA-2 in Orion~B.  The background image shows the 500~\mum
	flux measured by {\it Herschel}, 
	while the red contours show the areas observed with
	SCUBA-2, and the blue contours show the areas observed in CO(3--2) with HARP.}
\label{fig_overview}
\end{figure*}

Orion~B was observed with SCUBA-2 \citep{Holland13} at 850~\mum and 450~\mum as part of the 
JCMT Gould Belt Survey \citep{WardThomp07}.  Three separate regions were observed: the areas
around L1622, \None, and \Ntwo, as illustrated in 
Figure~\ref{fig_overview}.  Our SCUBA-2 observations
cover most of the high flux areas in the {\it Herschel} 500~\mum map from 
\citet{Schneider13}\footnote{We downloaded the {\it Herschel} 500~\mum map from 
{\tt http://www.herschel.fr/cea/gouldbelt/en/Phocea\\ /Vie\_des\_labos/Ast/ast\_visu.php?id\_ast=66}}.
The SCUBA-2 observations were obtained between February 2012 and November 2014 with some initial 
science
verification data taken in October 2011 and November 2011.  Most data 
were observed as fully sampled 30\arcmin\ diameter circular regions using the PONG 1800 mode 
\citep{Kackley10}.  Several science verification observations taken in the \None\ and 
\Ntwo\ regions were instead taken in PONG 900 mode, which fully samples a 15\arcmin\
diameter circular region \citep{Kackley10}.  
Each area of sky was observed between four 
to six times in PONG 1800 mode, with the number of repeats depending on weather 
conditions.  Neighbouring fields were set up to overlap slightly to create a more uniform
noise in the final mosaic.  
The PONG 900 observations are not included in the final mosaic that we analyze here, 
to maintain an approximately uniform noise
level and sensitivity to larger-scale structures across the areas observed.  

The data reduction used for the maps presented here follow the GBS Legacy Release 1
methodology, which is discussed in \citet{Mairs15}.  
The data presented here were reduced using an iterative map-making technique \citep[makemap 
in {\sc smurf}\footnote{{\sc smurf} is a software package used for reducing JCMT observations,
and is described in more detail in \citet{smurf}.};]
[]{Chapin13}, and 
gridded to 3\arcsec\ pixels at 
850~$\mu$m and 2\arcsec\ pixels at 450~$\mu$m.  The iterations were halted 
when the map pixels, on average, changed by $<$0.1\% of the estimated 
map rms. 
The initial reductions of each individual scan were coadded to 
form a mosaic from which a signal-to-noise mask was produced for each 
region.  The final mosaic was produced from a second reduction using 
this mask to define areas of emission.  In
Orion~B, the mask included all pixels with signal-to-noise ratio 
of 2 or higher at 850~$\mu$m.
Testing by our data reduction team showed
similar final maps using either an 850~\micron-based or a 450~\micron-based mask for the 
450~\mum
reduction, when using the SNR-based masking scheme described here.  Using identical masks
at both wavelengths for the reduction ensures that the same large-scale filtering 
is applied to
the observations at both wavelengths (e.g., maps of the ratio of fluxes at
both wavelengths are less susceptible to differing large-scale flux recovery).
Detection of emission structure and calibration accuracy are robust within the masked regions, 
but are less certain outside of the masked region \citep{Mairs15}.

Larger-scale structures are the most
poorly recovered outside of the masked areas, while point sources are better recovered.
A spatial filter of 600\arcsec\ is used during both the automask and external mask reductions,
and an additional filter of 200\arcsec\ is applied during 
the final iteration of both reductions
to the areas outside of the mask.
Further testing by our data reduction team
found that for 600\arcsec\ filtering, flux recovery is robust for sources with a Gaussian 
FWHM less than 2.5\arcmin, provided the mask is sufficiently large.  
Sources between 2.5\arcmin\ and 7.5\arcmin\ in diameter were 
detected, but both the flux and the size were underestimated because 
Fourier components representing scales greater than 5\arcmin\ were removed by the 
filtering process.  
Detection of sources larger than 7.5\arcmin\ is 
dependent on the mask used for reduction.  At a distance of 415~pc, 7.5\arcmin\ 
corresponds to 0.9~pc.

The data are calibrated in mJy per square arcsec using aperture 
flux conversion factors (FCFs) of 
2.34~Jy/pW/arcsec$^{2}$ and 
4.71~Jy/pW/arcsec$^{2}$ at 850~$\mu$m and 450~$\mu$m, respectively, 
as derived from average values of JCMT calibrators (Dempsey et al. 2013). 
The PONG scan pattern leads to lower 
noise in the map centre and mosaic overlap regions, while data reduction and 
emission artifacts can lead to small variations in the noise over the 
whole map.
The pointing accuracy of the JCMT is smaller than the pixel sizes we adopt,
with current rms pointing errors of 1.2\arcsec\ in azimuth and 1.6\arcsec\ in elevation
(see {\tt http://www.eaobservatory.org/JCMT/\\
telescope/pointing/pointing.html}); JCMT pointing
accuracy in the era of SCUBA is discussed in \citet{DiFrancesco08}. 

The observations for Orion~B were taken in both grade one
($\tau_{225GHz} < 0.05$) and grade two ($ 0.05 < \tau_{225GHz} < 0.08$) weather, corresponding
to $\tau_{850~\mu m} < 0.21$ and $0.21 < \tau_{850~\mu m} < 0.34$ respectively \citep{Dempsey13}, 
with a mean value of $\tau_{225GHz}$ of 0.06 $\pm$ 0.01.
At 850~$\mu$m, the final noise level in the mosaic is typically 
0.05~mJy~arcsec$^{-2}$ per 3\arcsec\ pixel, corresponding to 3.7~mJy per 14.6\arcsec\ beam.
At 450~$\mu$m, the final noise level is 1.2~mJy~arcsec$^{-2}$ per 2\arcsec\
pixel, corresponding to 59~mJy per 9.8\arcsec\ beam.  \citep[Note the beamsizes 
quoted here are the effective
beams determined by][and account for fact that the beam shape is well-represented by the
sum of a Gaussian primary beam shape and a fainter, larger Gaussian secondary beam]{Dempsey13}.
The noise levels for each PONG observing area in the final mosaic is given in
Table~\ref{tab_pong_noise} in terms of the typical rms in a pixel.

\input{tab1}

\begin{figure*}[p!]
\begin{tabular}{ccc}
\includegraphics[width=1.8in]{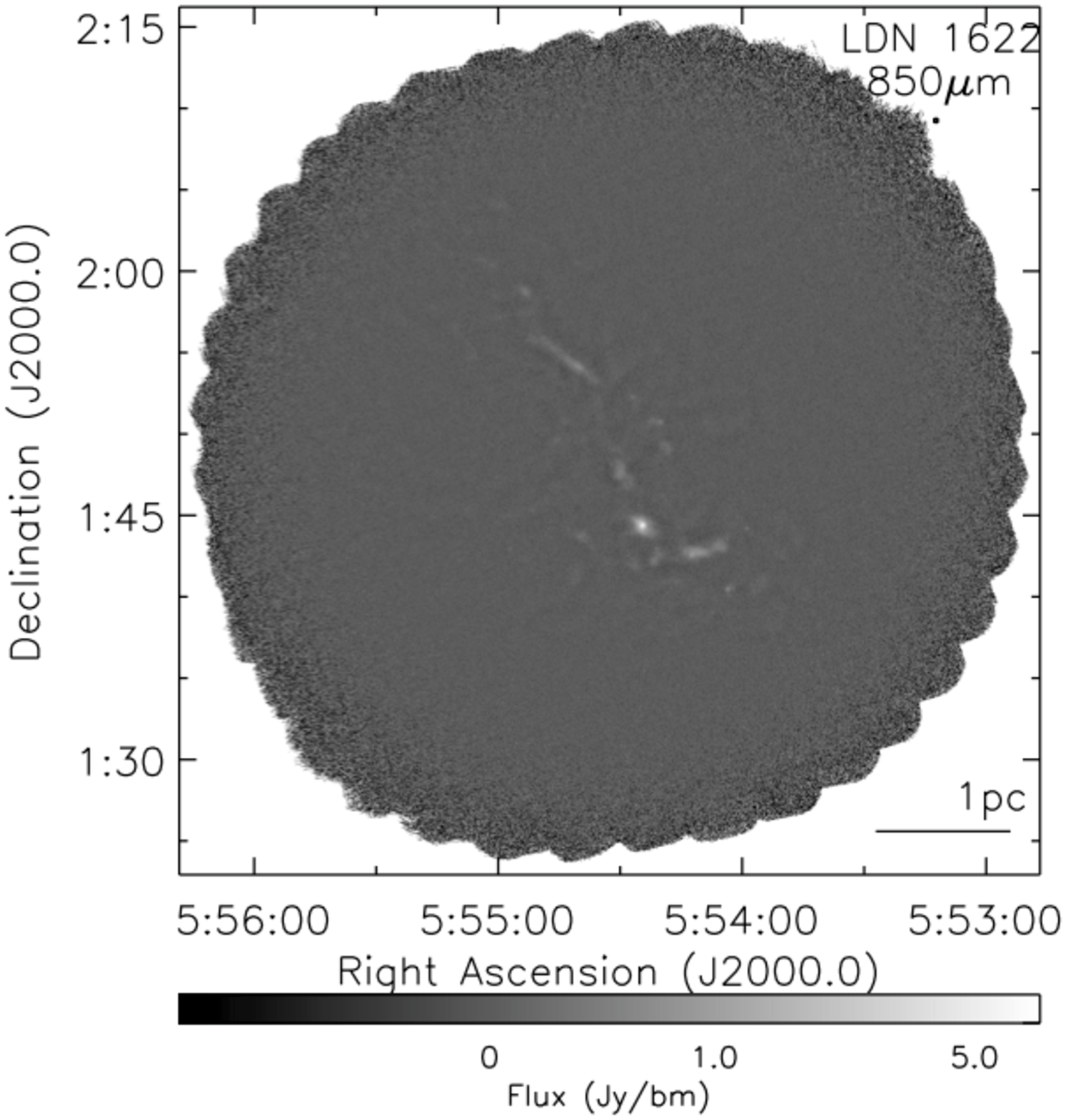} &
\includegraphics[width=1.8in]{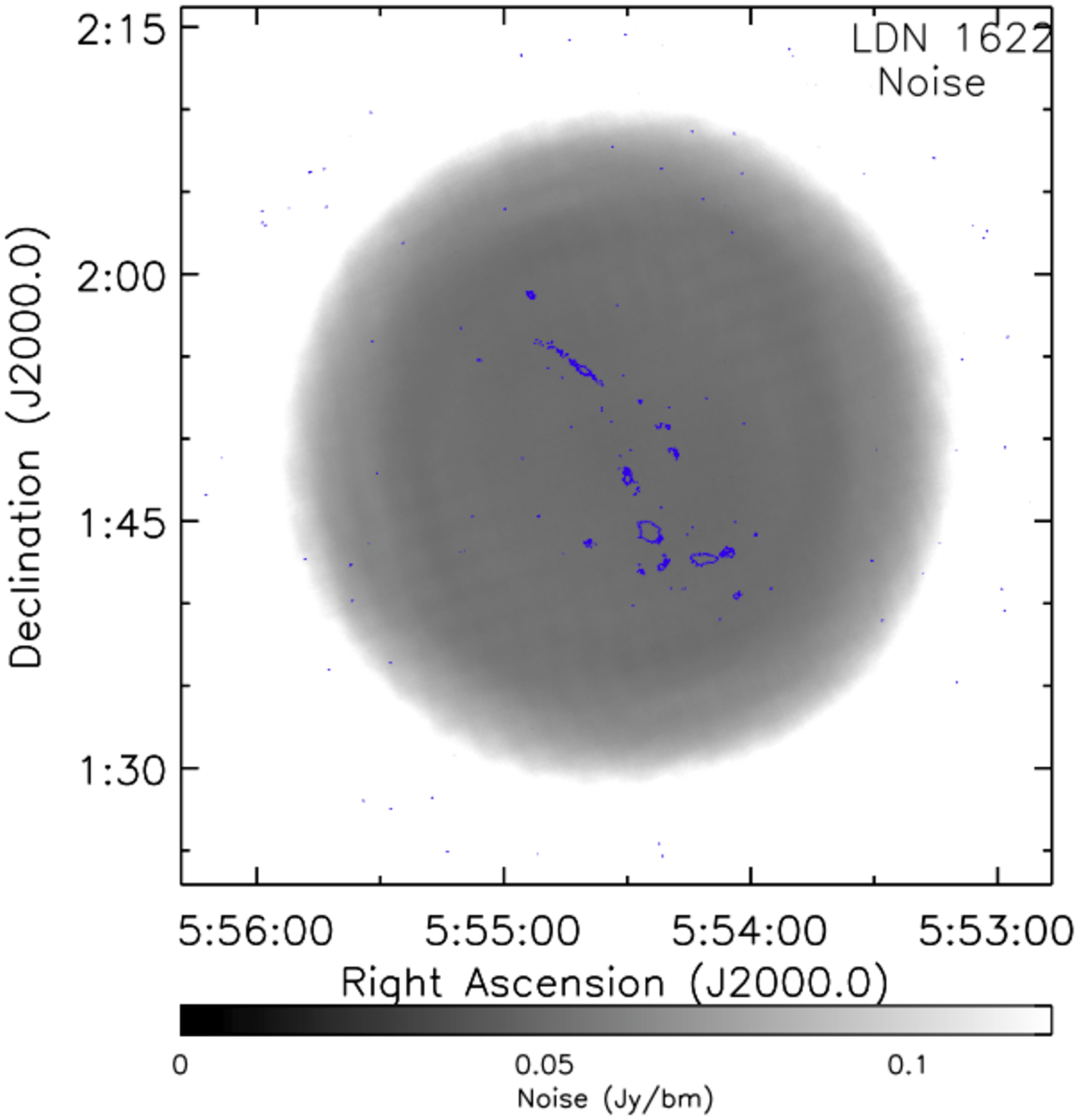} & 
\includegraphics[width=1.8in]{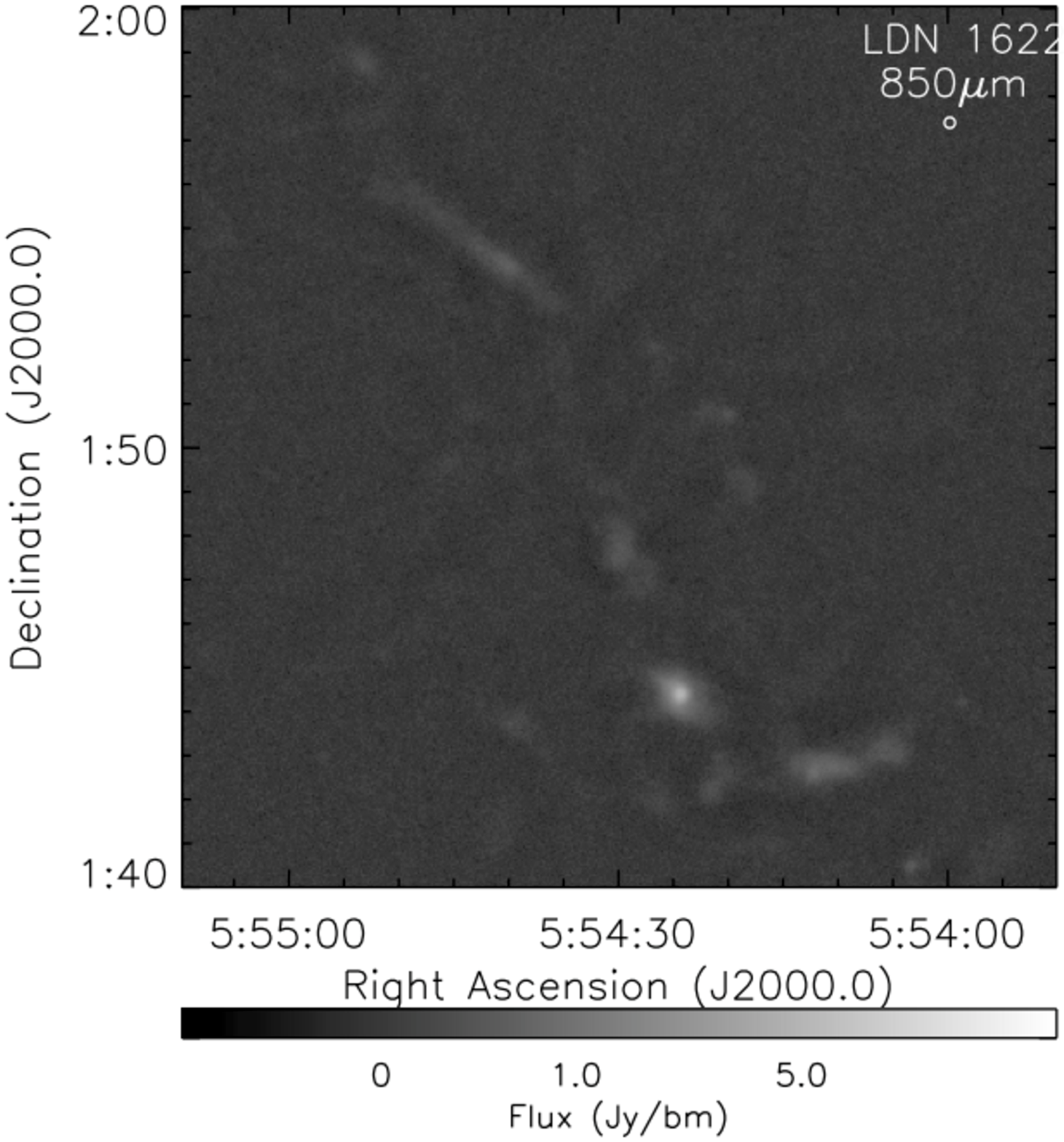} \\
\includegraphics[width=1.8in]{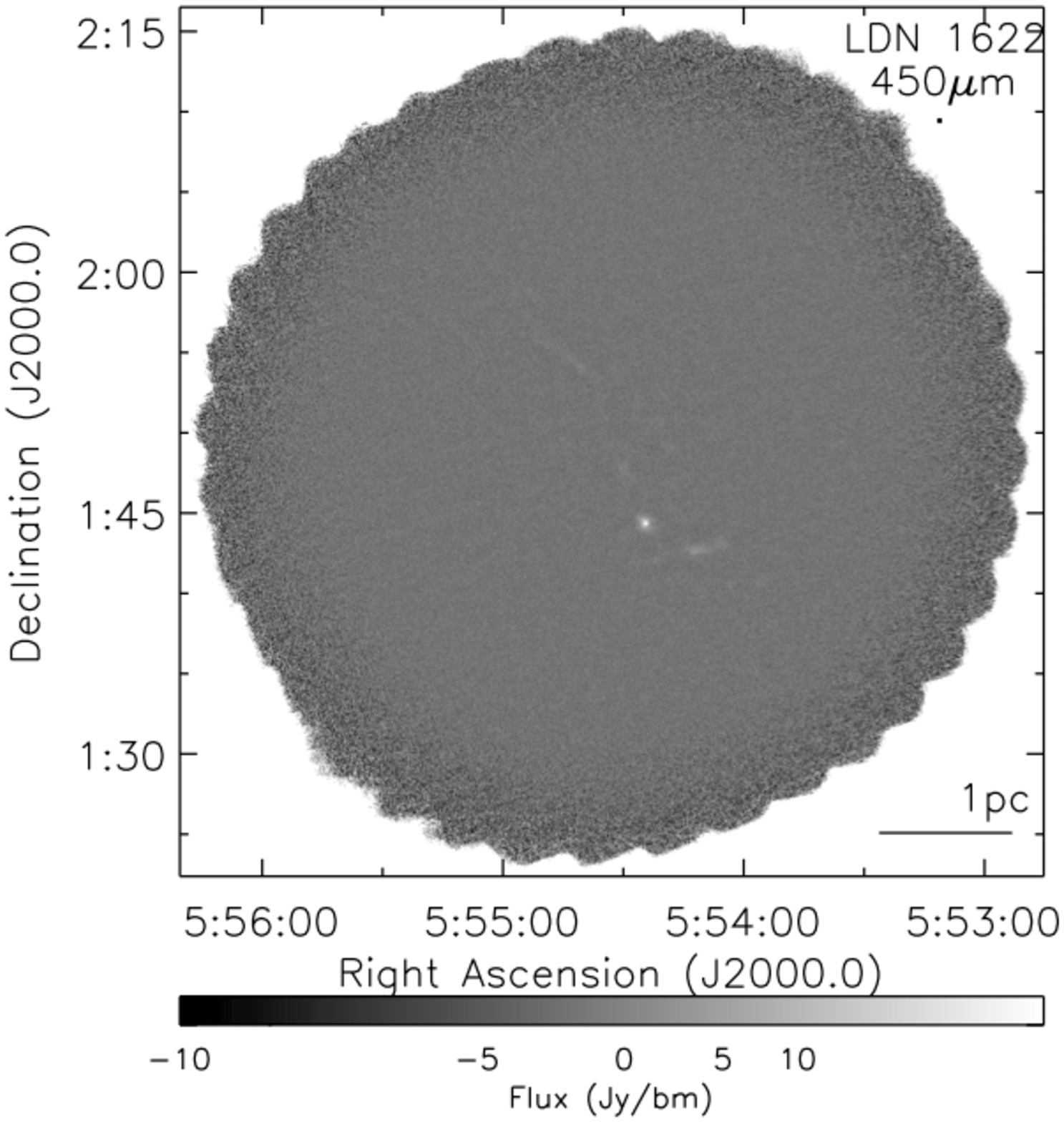} &
\includegraphics[width=1.8in]{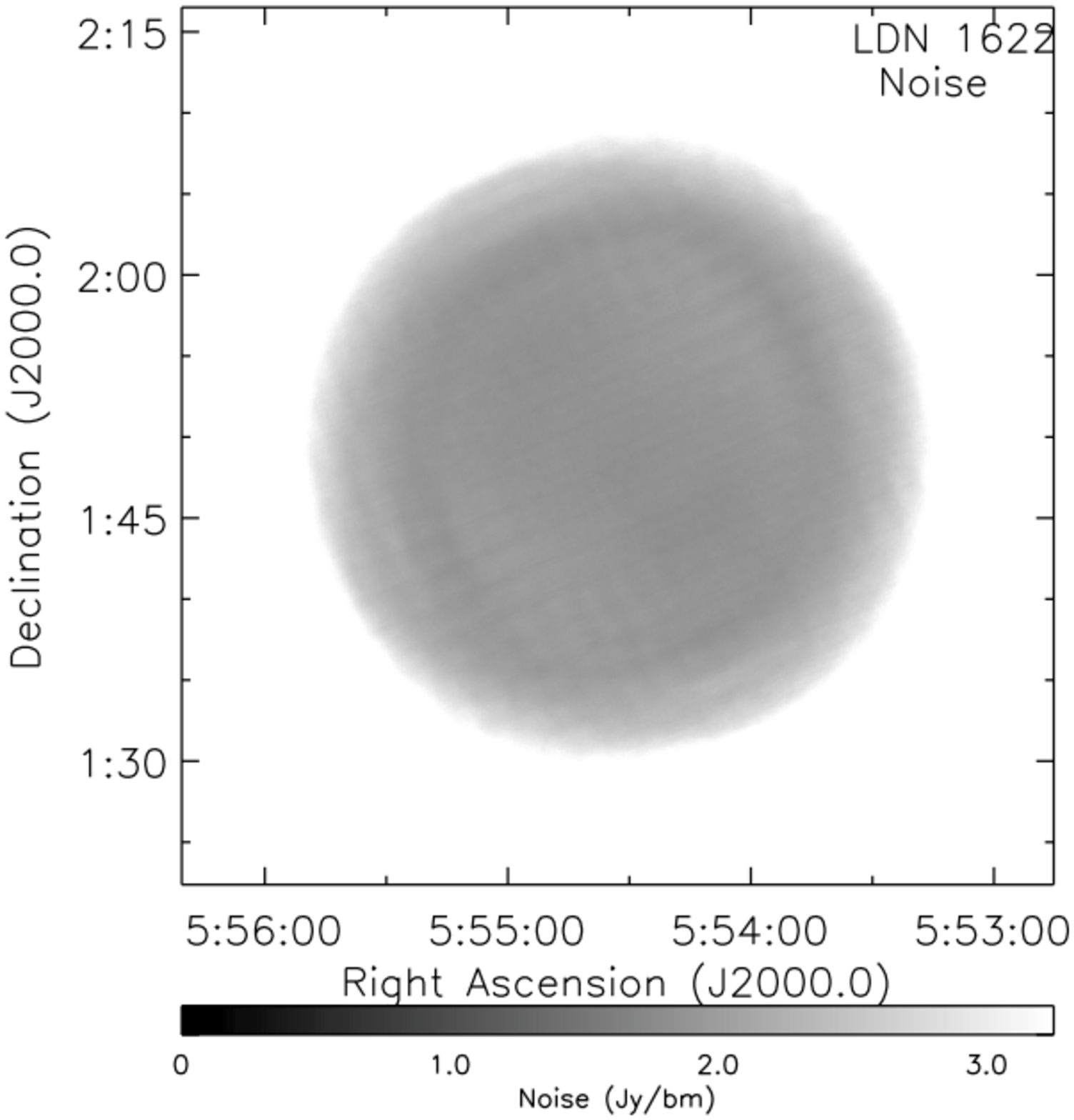} & 
\includegraphics[width=1.8in]{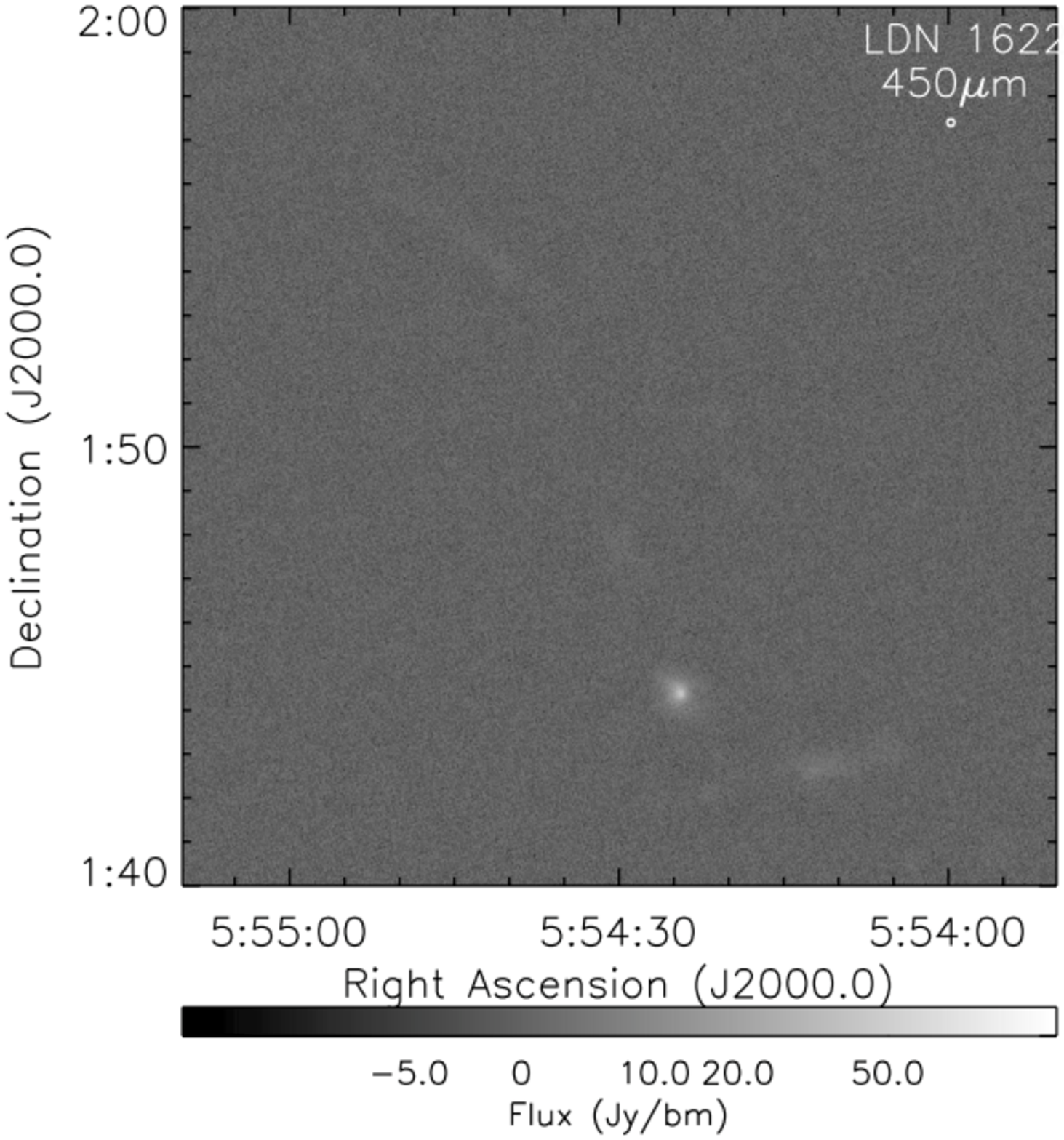} \\
\end{tabular}
\caption{The SCUBA-2 850~\mum (top) and 450~\mum (bottom) observations of L1622 in Orion B.
	The left panel on each row shows the entire map, while the middle panel shows the 
	noise, and the right panel shows a zoom on a zone of stronger emission.  
	In the left and right panels, 
	the scaling is approximately logarithmic, while the middle panel is shown
	with a linear scale.
	The black circle in the upper right corner shows the effective beamsize
	at each wavelength, while the scale bar at the bottom indicates 
	the angular distance corresponding to 1~pc
	at the assumed cloud distance of 415~pc.
	The external mask used in the reduction is indicated by the blue contour on the
	850~\mum noise map (signal-to-noise ratio $\ge 2$ at 850~\mum in the initial reduction).
	An identical mask was used at 450~$\mu$m.
	}
\label{fig_L1622}
\end{figure*}

\begin{figure*}[p!]
\begin{tabular}{ccc}
\includegraphics[width=1.8in]{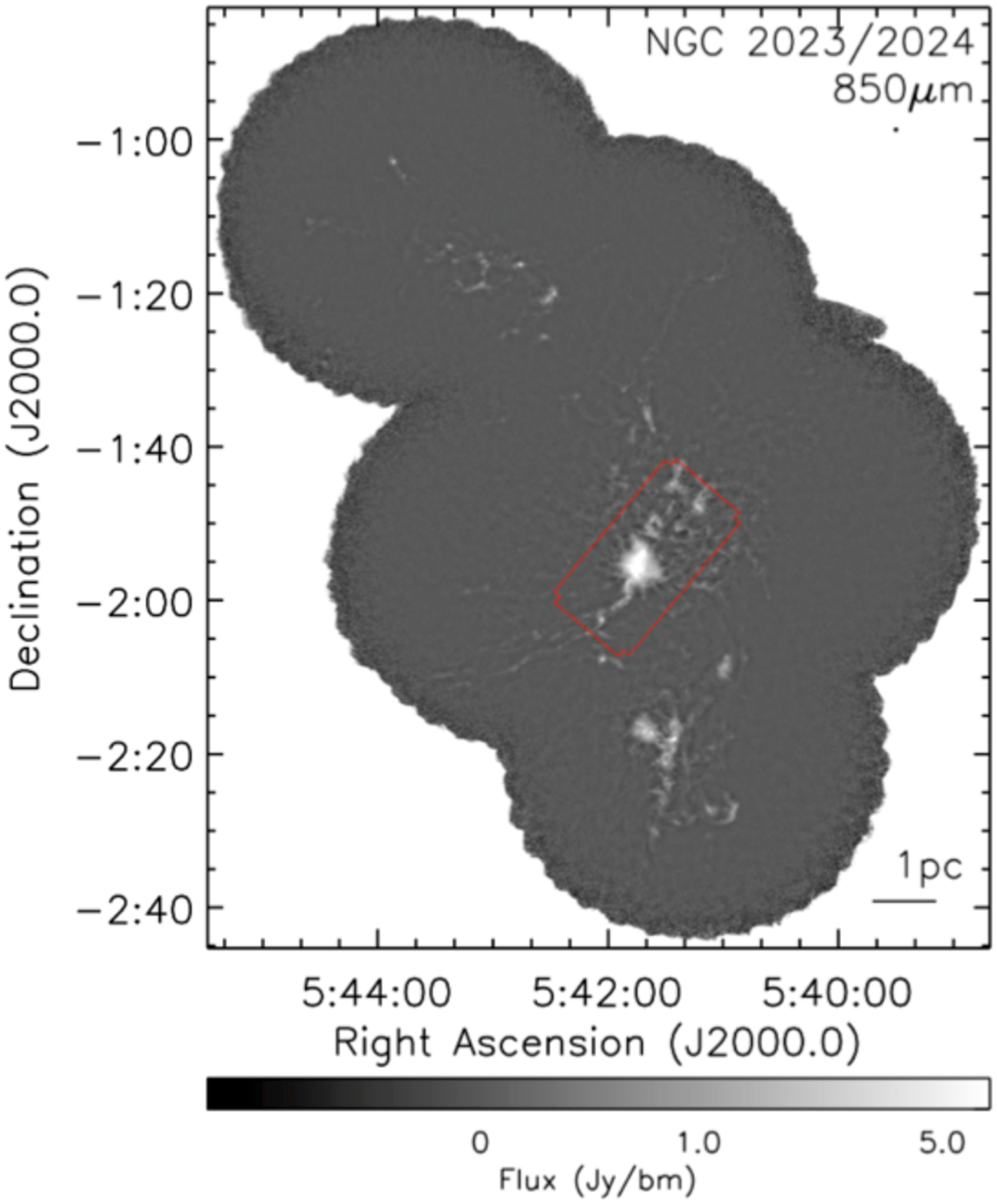} &
\includegraphics[width=1.8in]{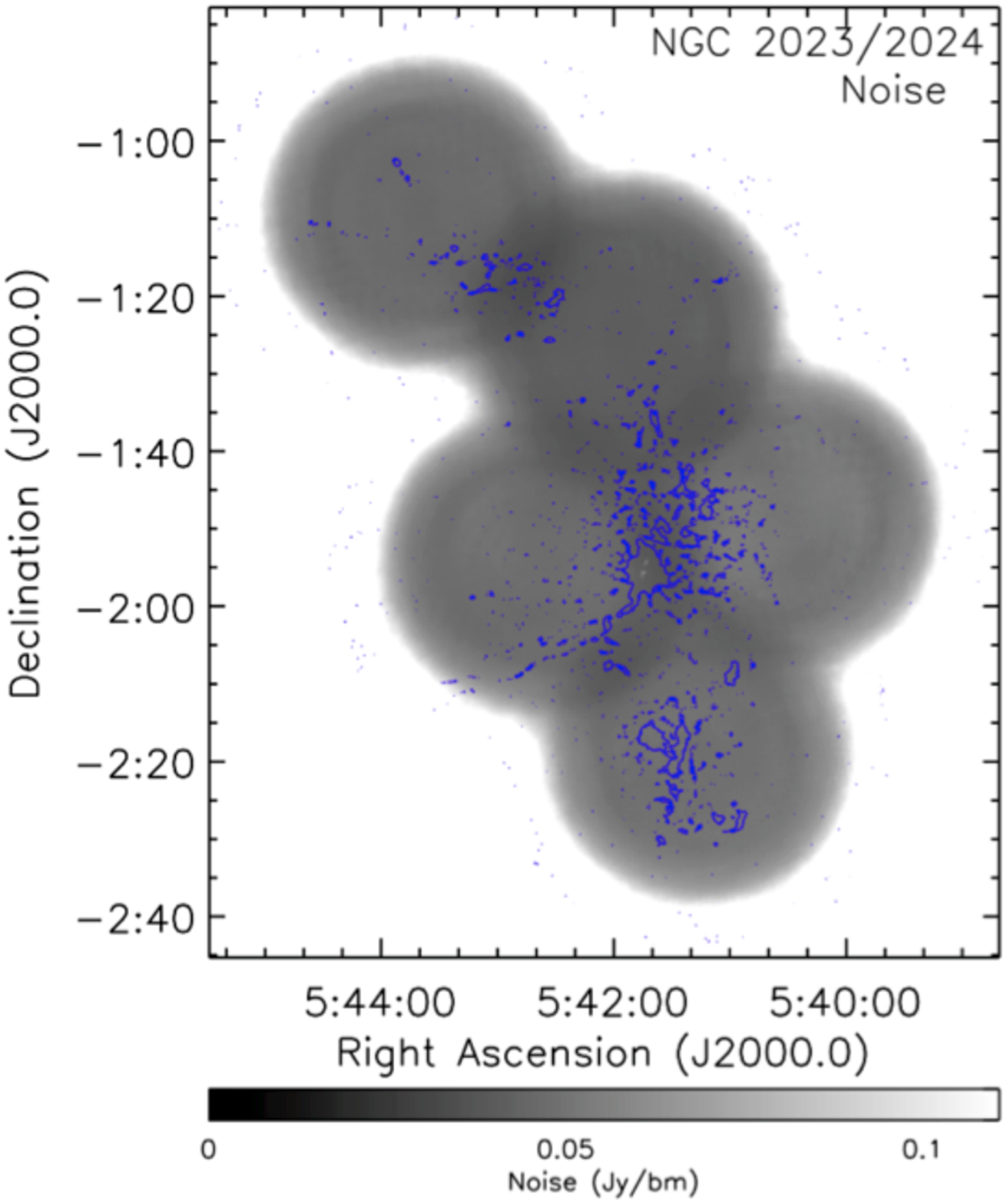} & 
\includegraphics[width=1.8in]{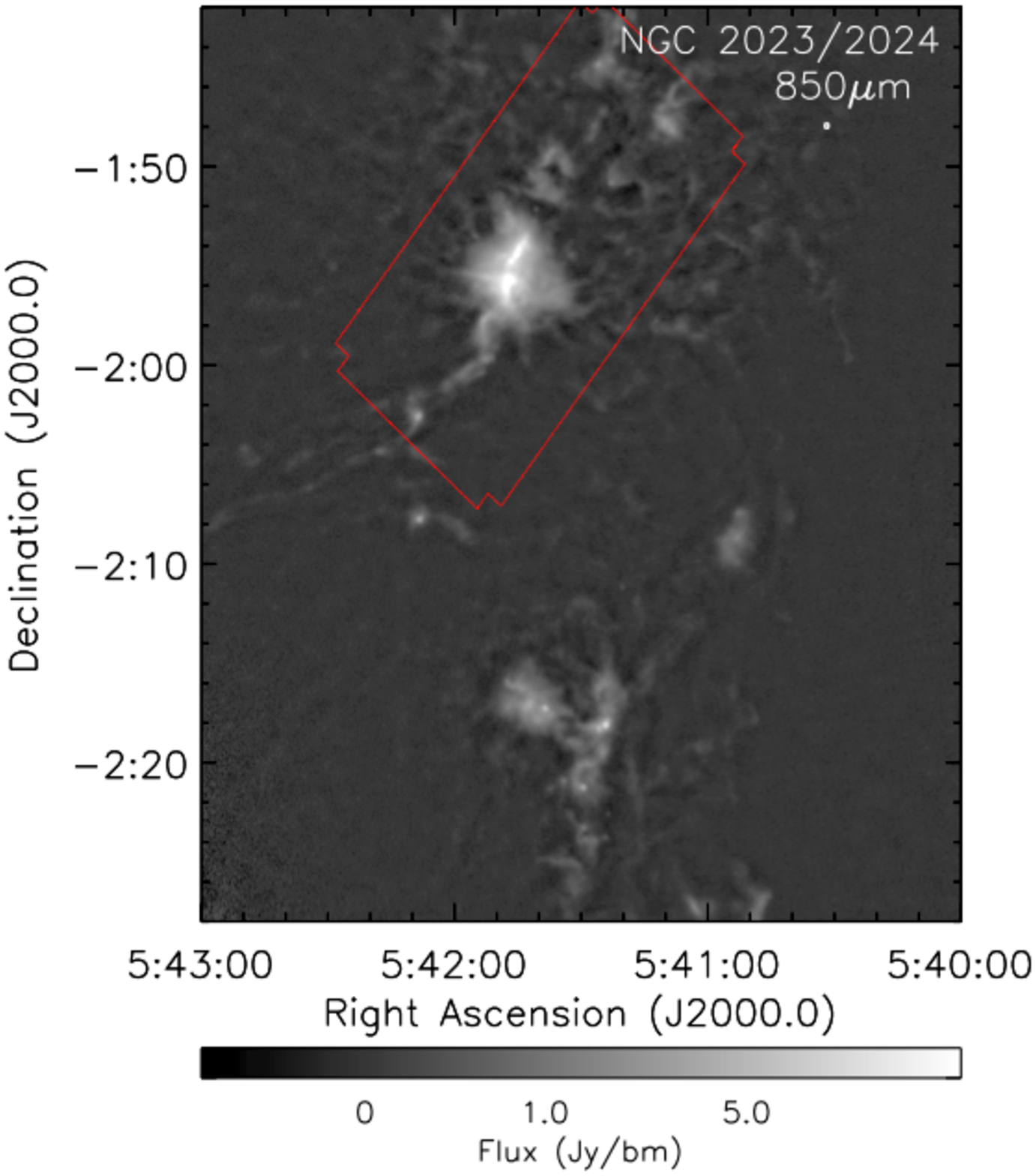} \\
\includegraphics[width=1.8in]{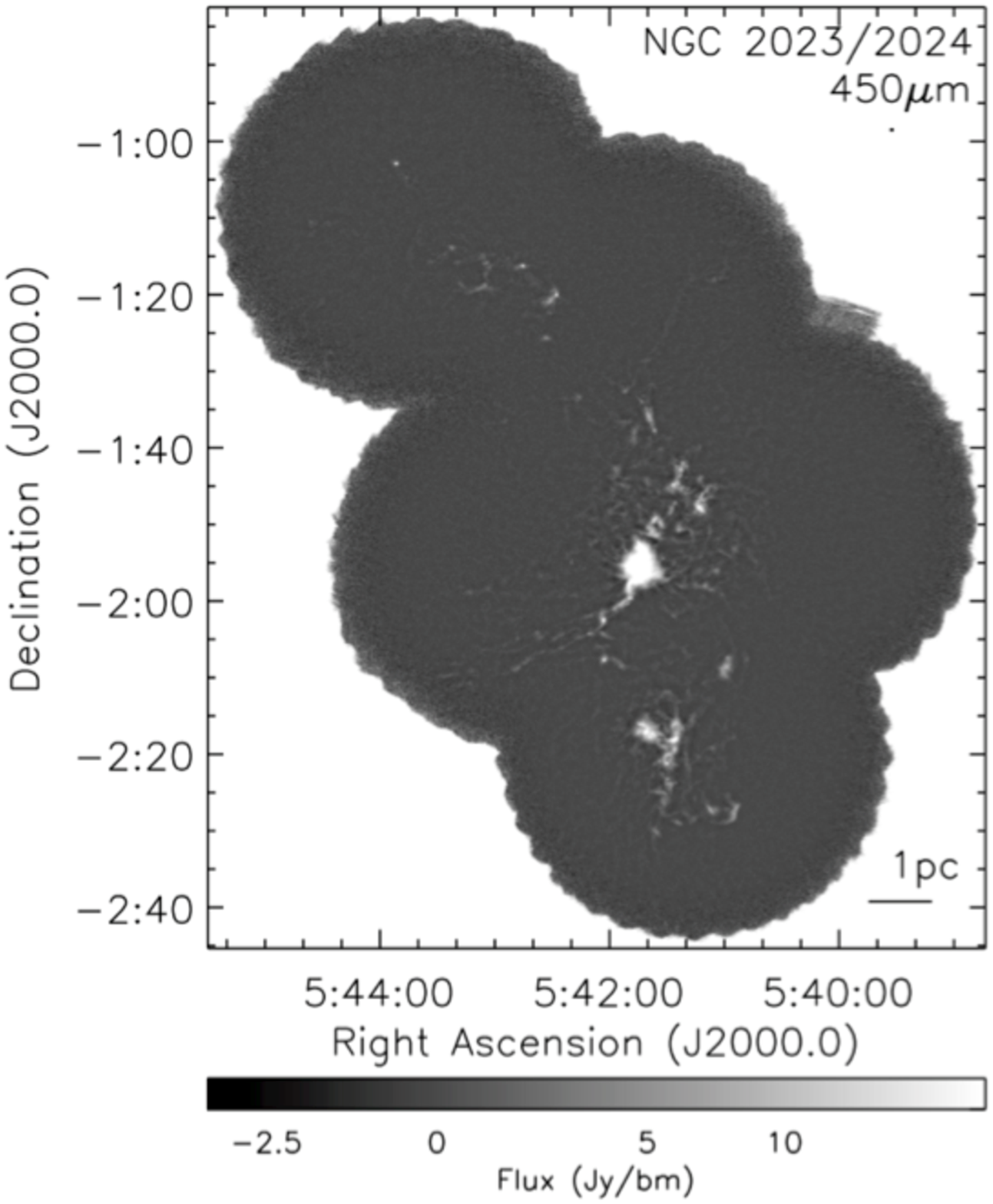} &
\includegraphics[width=1.8in]{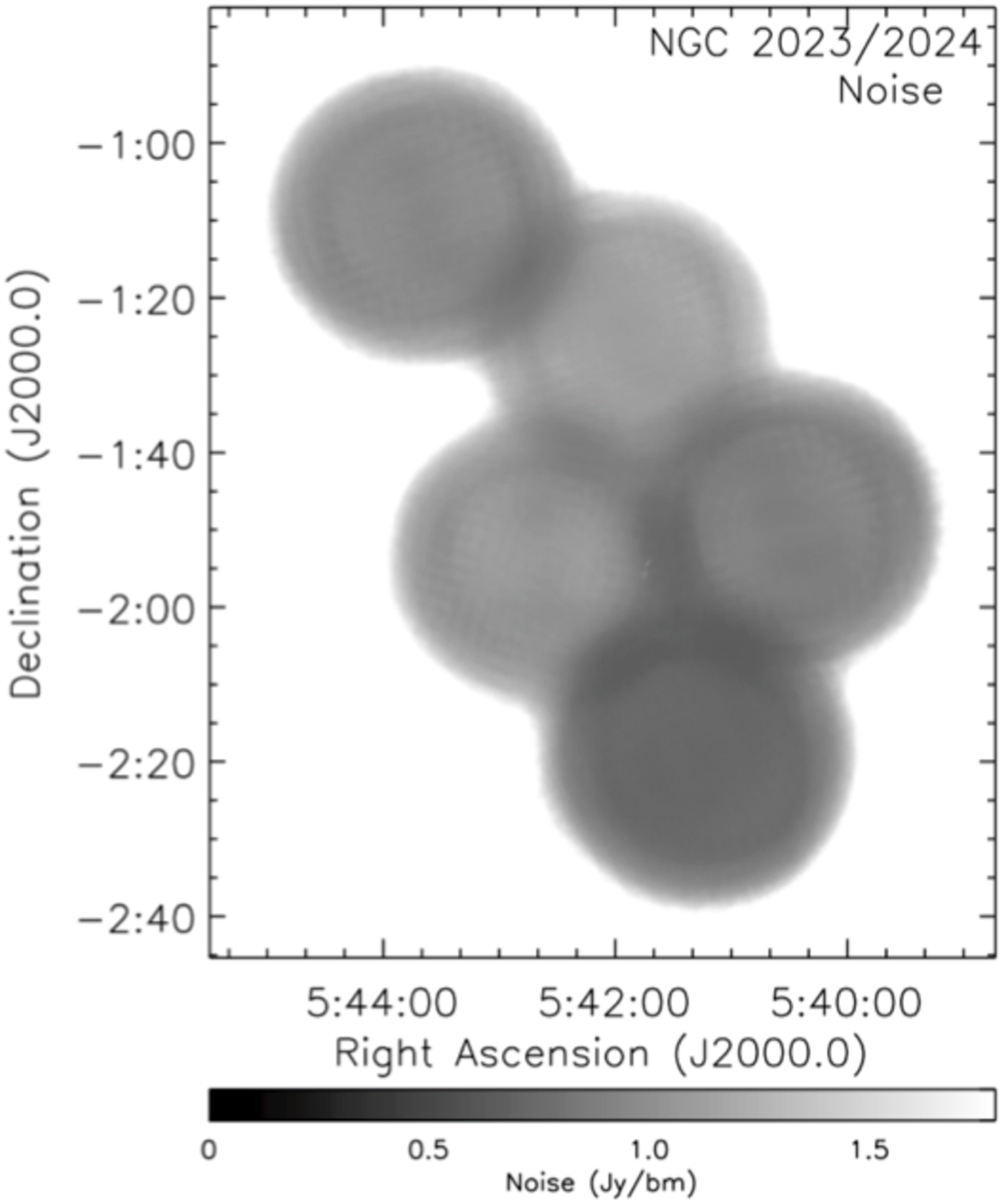} &
\includegraphics[width=1.8in]{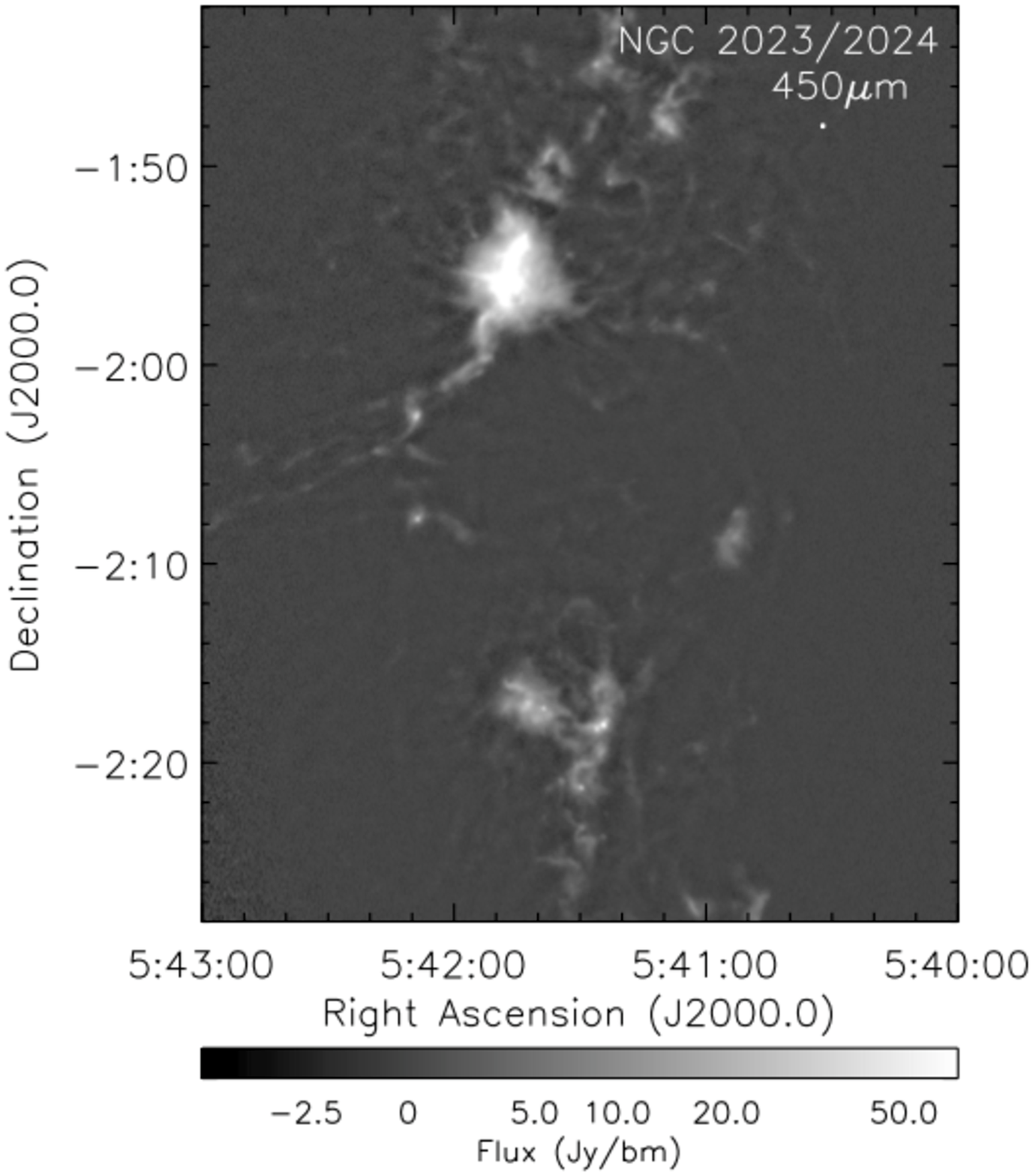} \\
\end{tabular}
\caption{The SCUBA-2 850~\mum (top) and 450~\mum (bottom) observations of \None\ in Orion B.
	See Figure~\ref{fig_L1622} for the plotting conventions used.  
	The red contours
	on the left and right panels indicate regions with GBS HARP CO observations.}
\label{fig_N2023}
\end{figure*}

\begin{figure*}[p!]
\begin{tabular}{ccc}
\includegraphics[width=1.8in]{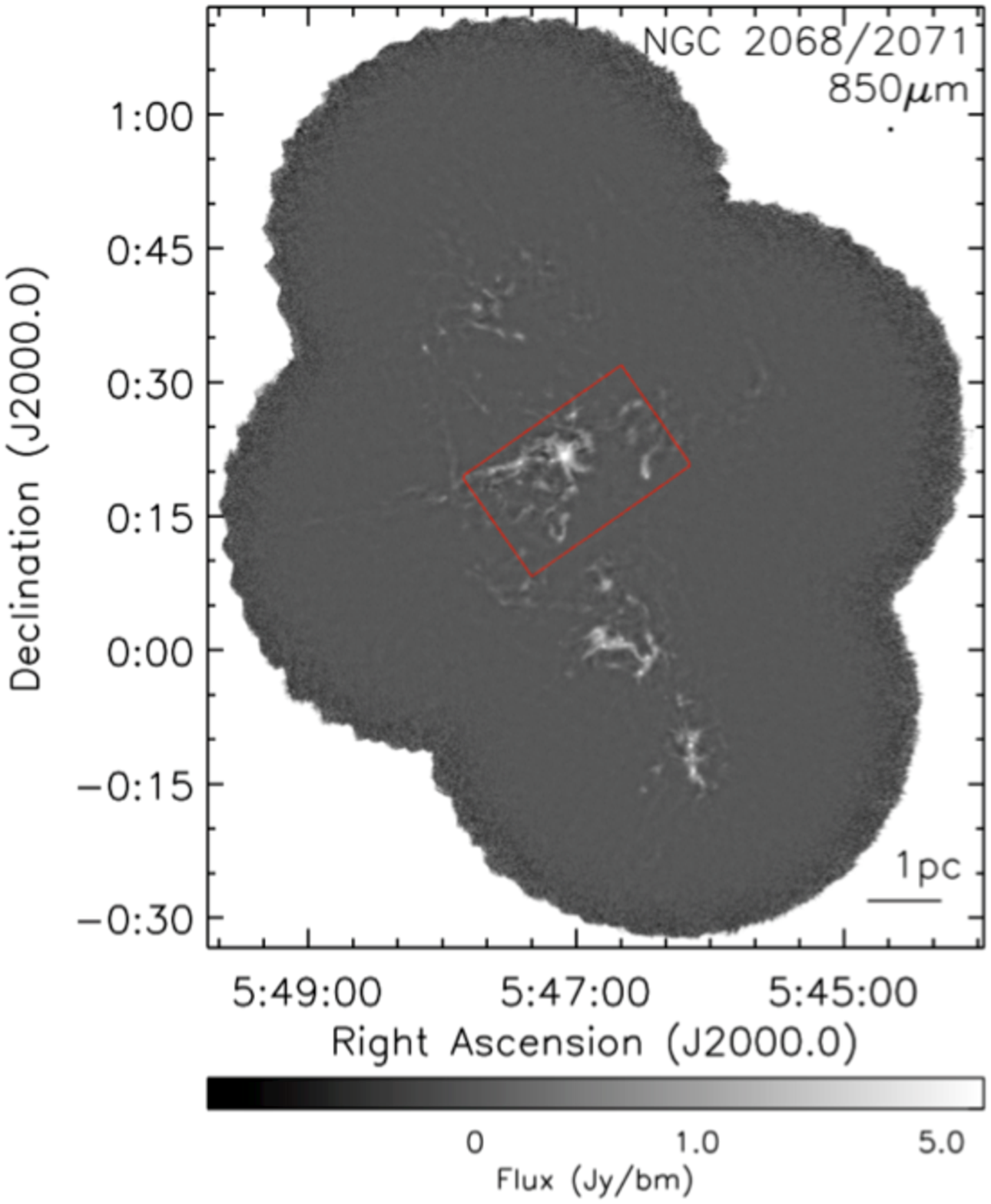} &
\includegraphics[width=1.8in]{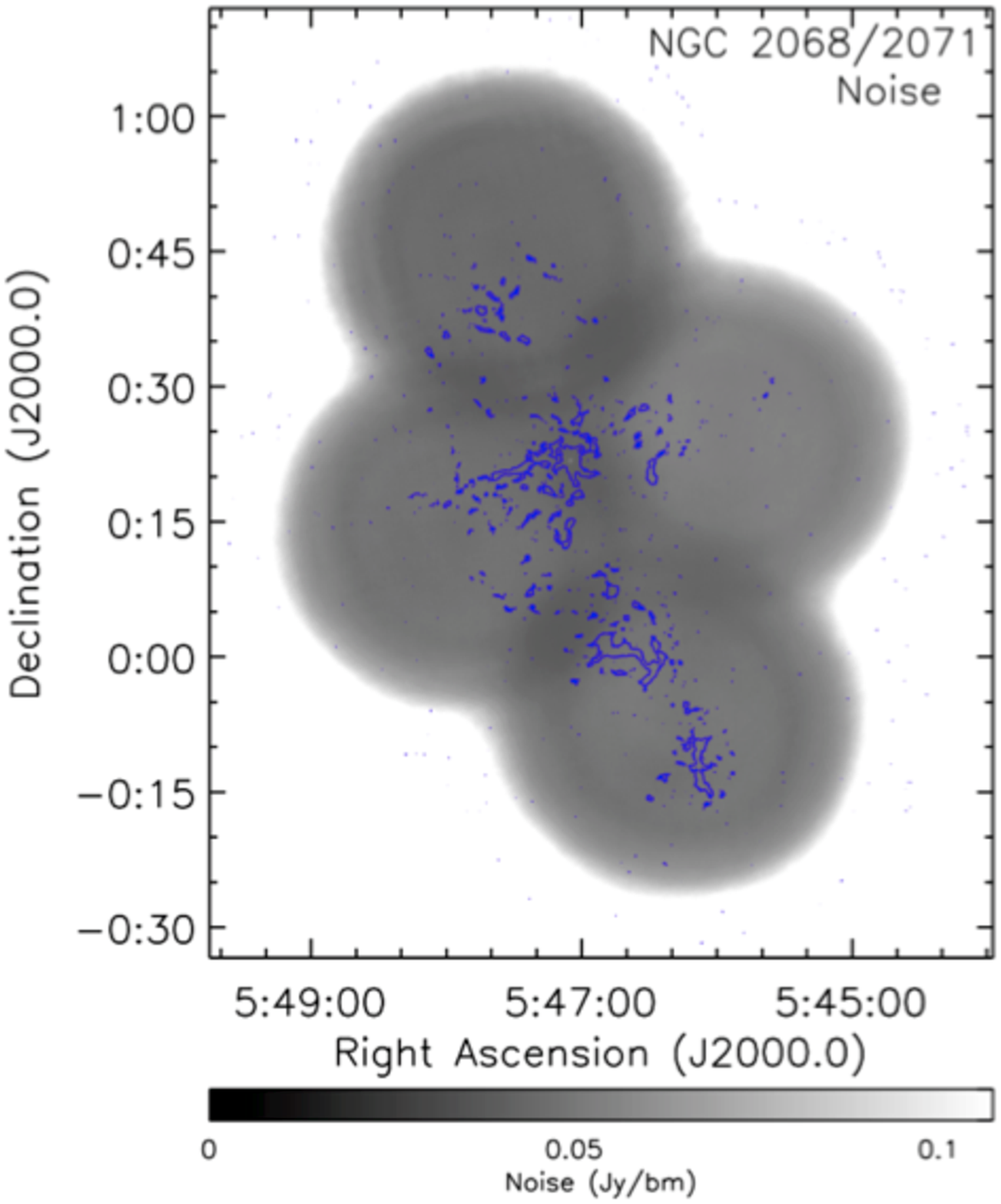} &
\includegraphics[width=1.8in]{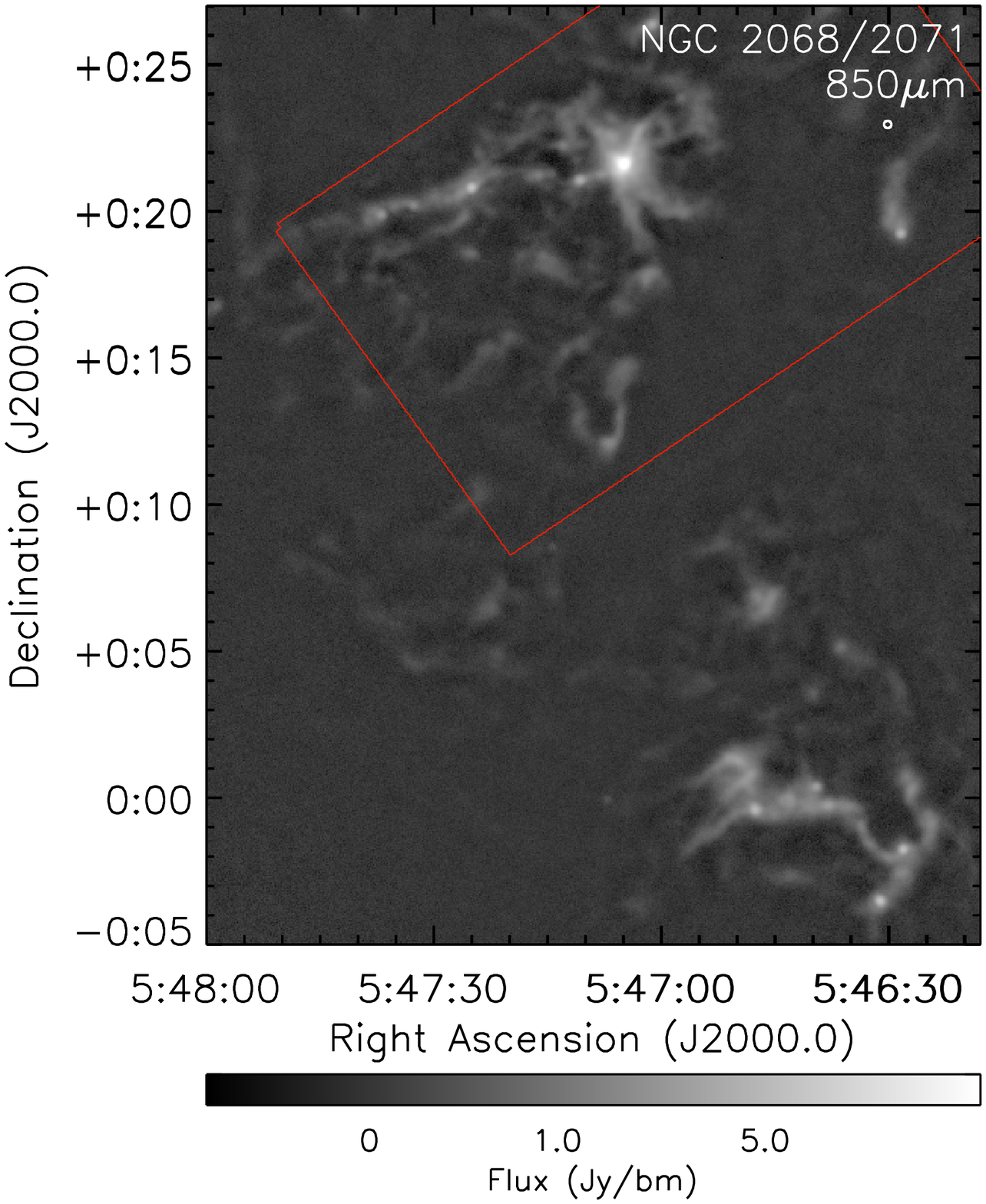} \\
\includegraphics[width=1.8in]{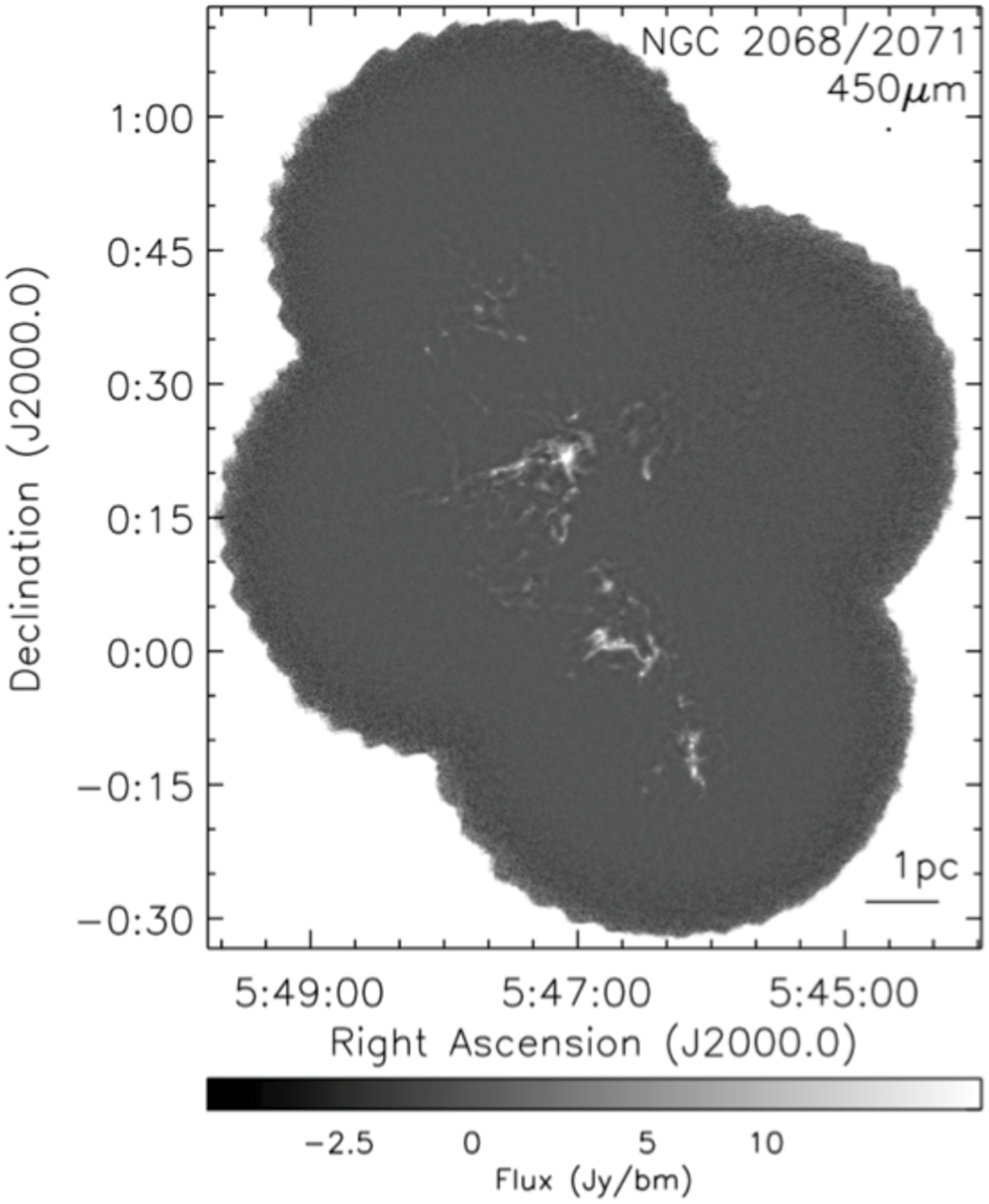} &
\includegraphics[width=1.8in]{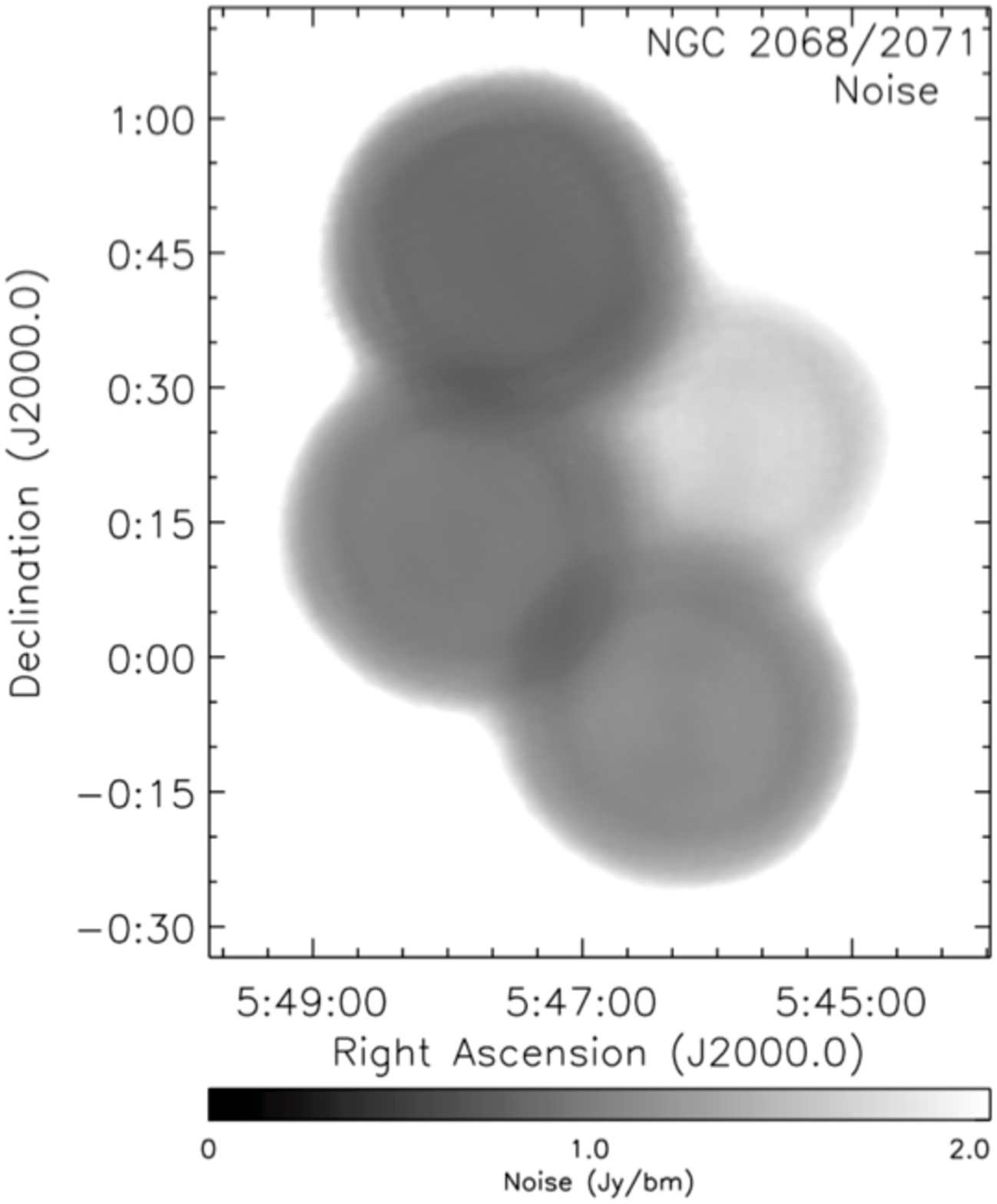} &
\includegraphics[width=1.8in]{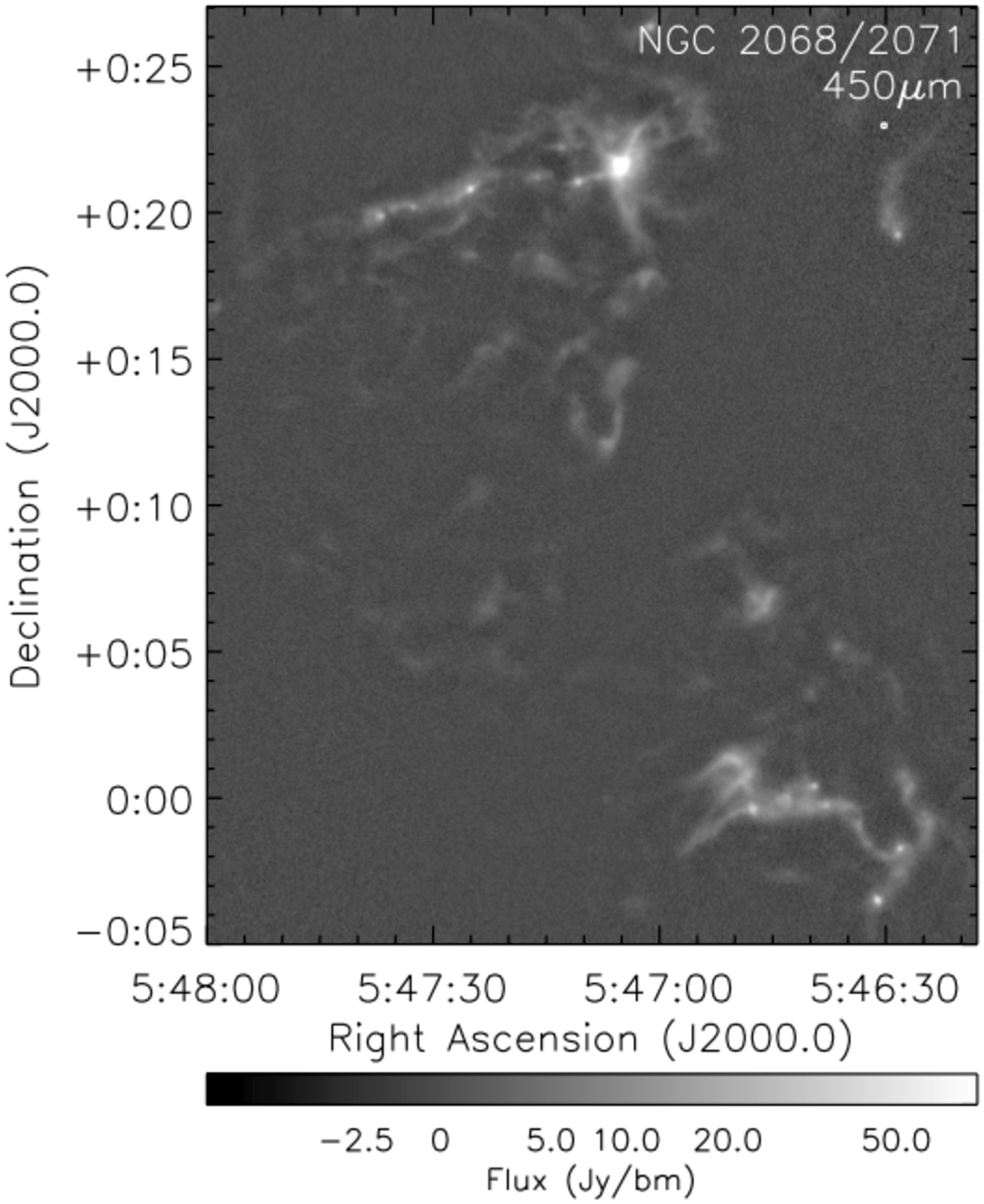} \\
\end{tabular}
\caption{The SCUBA-2 850~\mum (top) and 450~\mum (bottom) observations of \Ntwo\ in Orion B.
	See Figure~\ref{fig_N2023} for the plotting conventions used. } 
\label{fig_N2068}
\end{figure*}

Figures~\ref{fig_L1622} through \ref{fig_N2068} show the final reduced images, along with
their associated noise maps.  The external masks applied are indicated by the blue contours
on the 850~\mum noise map.  Note that the isolated pixels in the mask at the map edges
will have no effect on the scale of a dense core, since the contiguous area within 
those parts of the mask is too small.
Several of the brightest sources of emission in the maps are surrounded by 
negative (`bowl') features.  These features may slightly diminish the sizes and
total fluxes we derive for sources in Section~3, but based on artificial source-recovery
tests discussed in \citet{Mairs15} and our mask-making strategy, we expect our
results to be accurate to 20\% or better.


Portions of the \None\ and \Ntwo\ regions were also observed by the GBS in $^{12}$CO(3-2) with
HARP \citep{Buckle10} and reduced using ORAC-DR \citep{Jenness15}.  These areas are
indicated as contours in 
Figures~\ref{fig_overview}, \ref{fig_N2023}, and \ref{fig_N2068}.
The $^{12}$CO (3-2) emission line lies within the 850~\mum continuum band, and
therefore some fraction of the 850~\mum flux may in fact not be thermal dust
emission \citep[e.g.,][]{Johnstone03}.  
Observations of other star-forming regions \citep[e.g.,][]{Johnstone03,Drabek12,Sadavoy13,Hatchell13,Pattle15,Salji15a,Buckle15}
have shown that this `contamination' is generally
not a large effect ($< 20$\%), the main exception being regions with faint dust emission and
bright CO outflows, where the $^{12}$CO emission can dominate (up to 90\%).  
Over the regions where we have HARP CO observations, we estimate
the level of CO contamination on the observed 850~\mum flux.  Following the procedure
outlined in \citet{Drabek12}, 
we run an extra round of data reduction with the CO integrated intensity map included as
a negative source of emission in each raw datafile, scaled to the atmospheric transmission
of that evening.  These CO-subtracted maps are then mosaicked together, and compared with
the original 850~\mum mosaic.  This procedure ensures that the CO data are filtered and processed
identically to our 850~\mum data.  We calculate the fractional CO contamination level as
\begin{equation}
f_{CO} = \frac{S_{850,orig} - S_{850,noco}}{S_{850,orig}}
\end{equation}
where $S_{850,orig}$ is the flux in the original 850~\mum map and $S_{850,noco}$ is the 
flux in the CO-subtracted 850~\mum map.
Most of the area mapped has $f_{CO}$ below the (fractional) noise level at the same 
location,
implying an overall very small contamination level.  In \Ntwo, several small zones at the 
outskirts of the NGC~2071 cluster show $f_{CO}$ above the 20\% level, but these
are generally in areas of lower 850~\mum flux.  In \None, 
slightly off of the main NGC~2024 cluster, there are
several dense cores which show contamination levels above 50\% over part of their
extent (less than half of their full extent, and usually substantially less).
In general, however, the level of CO contamination is small.
Since most of the cores fall outside of the region with 
CO observations, we do not include the CO flux corrections in any of our subsequent
analysis.

\section{SOURCE IDENTIFICATION}
\label{sec_core_id}
We identify cores in the three 850~\mum Orion~B maps using FellWalker \citep{Berry15}, 
a source identification algorithm available as part of the 
{\sc CUPID}\footnote{\tt http://www.starlink.ac.uk/cupid} package \citep{Berry07} 
in Starlink.  The basic premise of FellWalker is to define the peaks and 
sizes of objects in images based
on local gradients, and the extent of pathways which lead to a given peak.
Like the more traditionally-used ClumpFind algorithm \citep{Williams94}, 
FellWalker does not assume a geometry when identifying cores. ClumpFind, however,
splits zones of complex emission into multiple cores based on user-selected contour
levels, whereas FellWalker relies on local gradients instead; \citet{Watson10} found
FellWalker generally produces superior results to ClumpFind, including a generally
better recovery of accurate peak and total fluxes of artifical cores inserted into maps.
FellWalker provides both a listing of the peak flux position for each
dense core and also a dense core footprint (i.e., a set of pixels all belonging to the core).
Appendix~\ref{app_cores} discusses the details of our source identification
process.  

\begin{figure*}[p]
\includegraphics[width=6.2in]{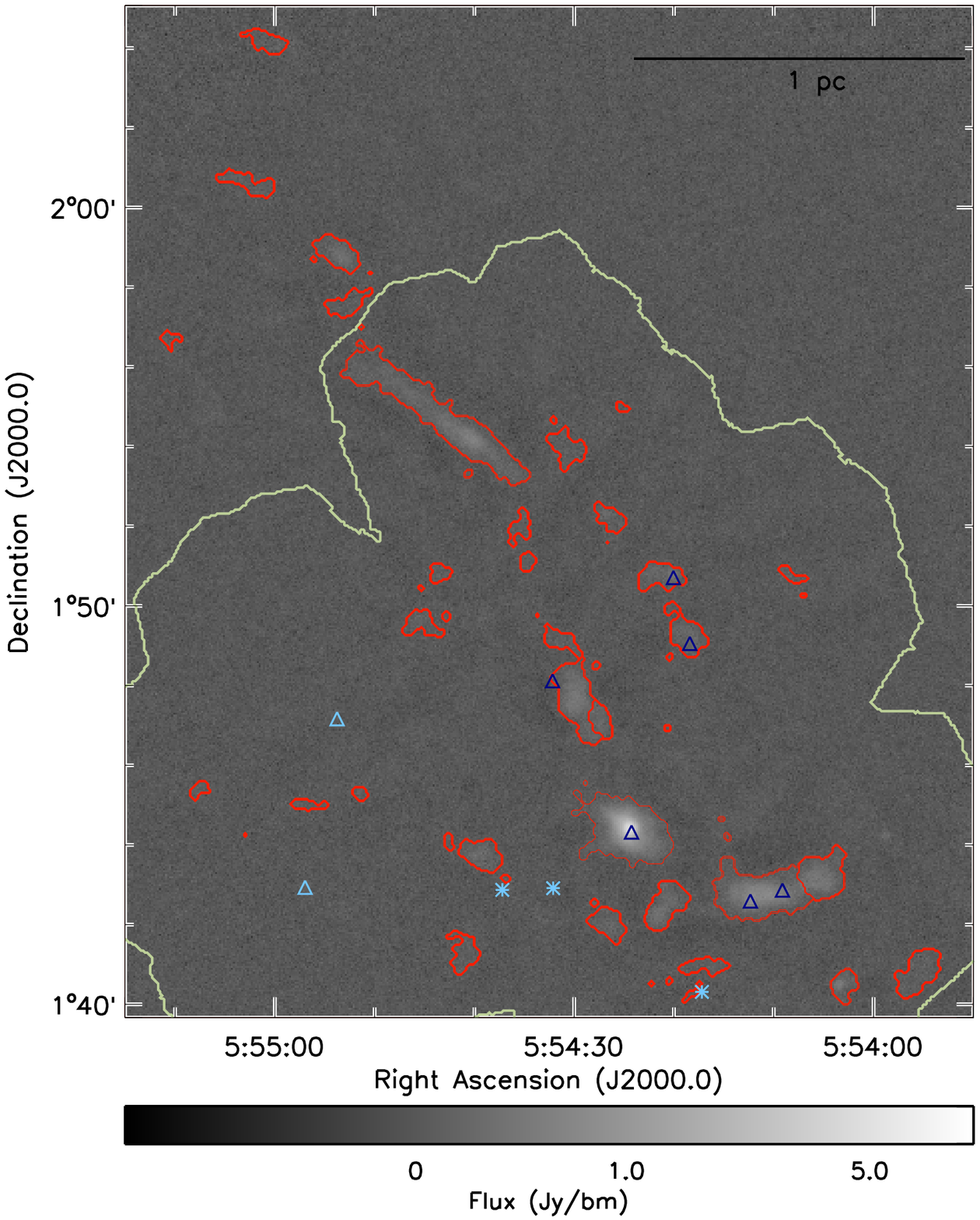}
\caption{Dense cores identified in L1622.  The background greyscale image shows the SCUBA-2
	850~\mum emission.  The red
	contours show the dense core boundaries, as identified with FellWalker, while the
	blue triangles show locations of protostars from the {\it Spitzer} YSO catalogue of
	\citet{Megeath12} and blue asterisks show the {\it Herschel} YSO catalogue 
	of \citet{Stutz13}.  Dark symbols indicate protostars associated with a 
	dense core, while light symbols indicate unassociated protostars.  The light
	yellow contour denotes the {\it Herschel} coverage for the \citet{Stutz13}
	catalogue (A. Stutz, priv. comm.). }
\label{fig_L1622_cores}
\end{figure*}

\begin{figure*}[p]
\includegraphics[width=6.5in]{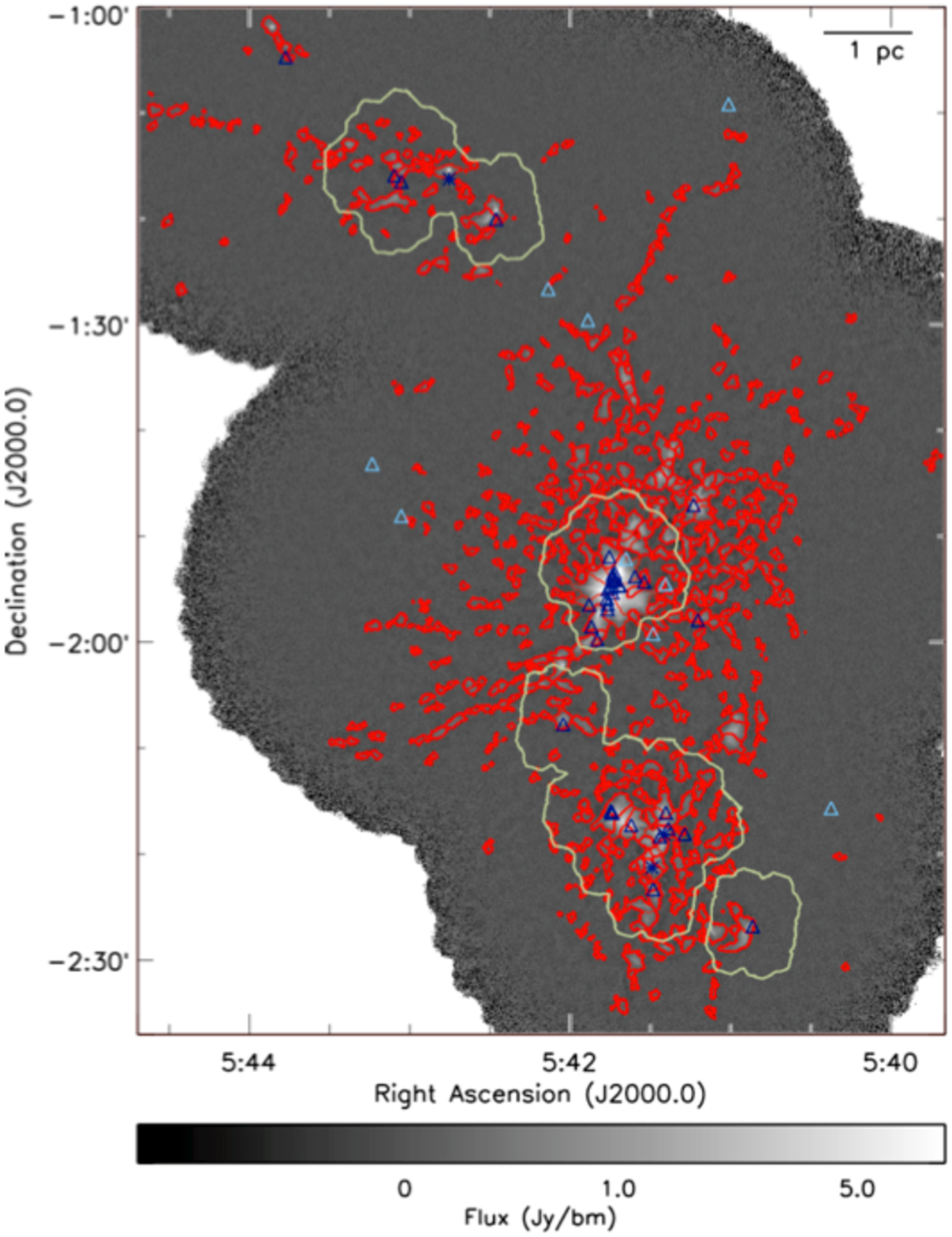}
\caption{Dense cores identified in \None.
	See Figure~\ref{fig_L1622_cores} for the plotting conventions used.
	}
\label{fig_N2023_cores}
\end{figure*}

\begin{figure*}[p]
\includegraphics[width=6.5in]{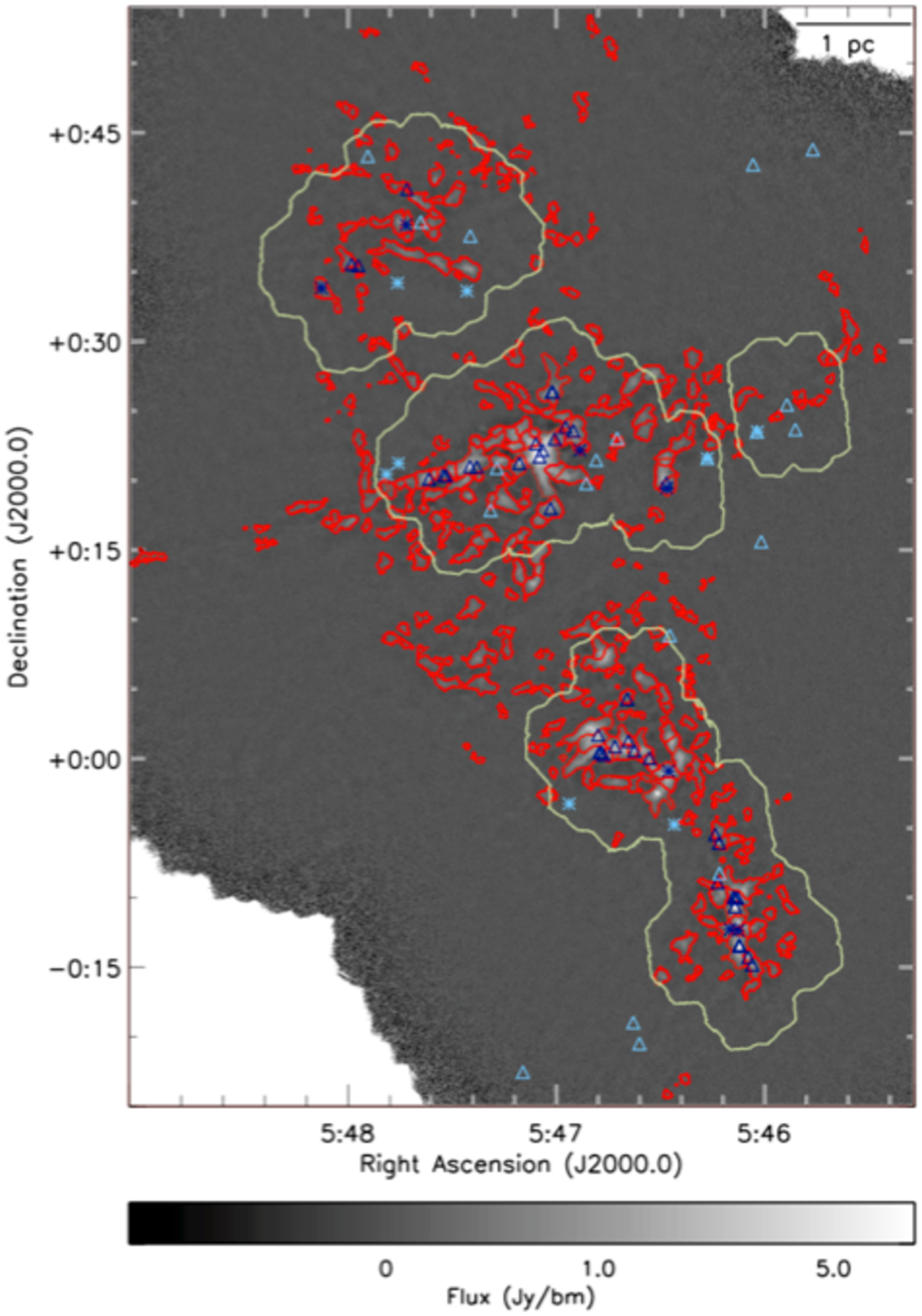}
\caption{Dense cores identified in \Ntwo.
	See Figure~\ref{fig_L1622_cores} for the plotting conventions used.
	}
\label{fig_N2068_cores}
\end{figure*}

We ran FellWalker with very relaxed settings, identifying 260, 1383, and 1020
potential sources, from which we then culled unreliable sources from.
After this subsequent elimination, we identified
29 reliable cores in L1622, 564 in \None, and 322 in \Ntwo.  
See Appendix~A for more details on our core identification strategy.
Our final core list includes cores with peaks potentially as low as twice the local
noise level.  While this is fainter than most core searches would be extended to,
a careful comparison of the 850~\mum data with
the {\it Herschel} 500~\mum data revealed that faint structures below the formal
3$\sigma$ typical cutoff were, in fact, real, and appear to have similar extents
at both wavelengths.
The dense cores we identify are shown in Figures~\ref{fig_L1622_cores} through 
\ref{fig_N2068_cores},
with the dense core footprints, and the 
{\it Spitzer}-identified protostars from \citet{Megeath12} 
and {\it Herschel}-identified protostars from \citet{Stutz13}
also shown (see next section for
more discussion on identifying protostellar cores).
We note that in Figure~\ref{fig_L1622_cores}, more closed contours are apparent
than the total number of cores identified.  {\it FellWalker} does not require that
cores have contiguous boundaries, and therefore sometimes near a core edge, some pixels
will be excluded, e.g., due to low flux, while other neighbouring pixels do satisfy all of
the core criterion and are included.  These isolated pixels represent a small fraction of
individual cores and by definition are located in low-flux areas of the map.  Therefore,
these isolated pixels have minimal influence on the properties we measure
(recall that the core size is based on the total area of the core footprint, not the maximal core 
extent).

Table~\ref{tab_cores}
provides a full list of the dense cores we identify within each
of the three regions.  In the table, core locations correspond to the position
of peak flux within the core.  The peak flux and total flux are calculated without any
background emission subtracted, but see Section~4 for further treatment of this issue.  
The peak flux is given in Jy~bm$^{-1}$, with the conversion
from mJy~arcsec$^{-2}$ made assuming an effective beam size of 14.6~\arcsec\ \citep{Dempsey13}.  
The core size 
is the effective radius, $R_{eff}$, calculated as the radius of a 
circle which spans the same area as the dense core (calculated using the
full dense core footprint).  For cores where HARP CO observations were made, we also
include the fraction of the core's area covered by the CO data, and the resulting core peak
fluxes and total fluxes at 850~\mum with the contribution from CO emission removed.  
We also calculate the peak and total 450~\mum flux using the
same dense core footprints as the 850~\mum data.  
Note that we do not make any attempt to account for the noise
level at 450~\mum within the dense core footprints.  In effect, 
cores with little to no 450~\mum
emission above the noise level may have a negative total flux within the 
core footprint.


The original SCUBA instrument at JCMT observed parts of \None\ and \Ntwo\ 
\citep[e.g.,][]{Motte01,Mitchell01,Johnstone01,Johnstone06,Nutter07}.  
\citet{Motte01} used a wavelet-based scheme to identify dense cores, which generally
identifies more compact regions of emission.  Other SCUBA analyses 
\citep{Mitchell01,Johnstone01,Johnstone06,Nutter07} used ClumpFind, which
tends to act more similarly to FellWalker, in identifying larger zones of emission
around each core.  We provide a detailed comparison of the dense cores identified
in \citet{Nutter07} as well as those published in the SCUBA Legacy Catalogue \citep{DiFrancesco08}
with our SCUBA-2 results in Appendix~\ref{app_scuba}.
We find generally good agreement between the cores identified in SCUBA and their 
corresponding match in the SCUBA-2 data.  Different core identification schemes, however,
can subdivide regions of complex emission differently, which generally leads to larger differences
in the total fluxes and sizes of the cores between the two measurements than
peak fluxes.
The SCUBA-2 observations are factors of four to six times more sensitive than the SCUBA
observations, with a median noise level of 3.7~mJy~bm$^{-1}$ compared to 16-23~mJy~bm$^{-1}$
in SCUBA \citep{Nutter07}.

\section{DENSE CORE PROPERTIES}
\label{sec_core_props}
We first classify all of the dense cores as starless or protostellar.  Our aim is to
make a conservative list of starless cores.  We start by
using the {\it Spitzer} catalogue from \citet{Megeath12} to identify protostars.
Specifically, any dense core which contained one or more protostars 
listed in the `all protostars' list from \citet{Megeath12} within the dense core's 
boundary was classified as protostellar.  We supplement our list of protostars by 
running a similar procedure on the full list of
candidate protostars from \citet{Stutz13} using {\it Herschel} data.
In other words, if any pixel of a core has a protostar lying within it
from either catalogue, we
classify the core as protostellar.
We note that the {\it Herschel} catalogue covers a smaller area within Orion~B, and focuses
exclusively on the most embedded YSOs.
This procedure allows us to identify five protostellar cores in L1622, 
25 in \None\ (of which 3 were {\it Herschel}-based) and 34 in \Ntwo\ (of which
6 were {\it Herschel}-based).  The number of starless cores in each region is
therefore 24, 539, and 288 in L1622, \None, and \Ntwo\ respectively.
Table~\ref{tab_cores}
denotes which dense cores we defined as protostellar.

\subsection{Masses}
\label{sec_coremass}
In addition to the dense core properties returned directly from FellWalker (size, peak
flux, and total flux), the core mass is an important property.  Using only the 
total 850~\mum flux measured for each core, we estimate the mass using the
equation
\begin{equation}
M = \frac{S_{\nu} D^2}{\kappa_{\nu} B_\nu(T)}
\end{equation}
from \citet{Hildebrand83}, where $S_\nu$ is the total flux at frequency $\nu$,
$\kappa$ is the dust opacity, and $B$ is the black body function at temperature $T$.
This simplifies to
\begin{equation}
M = 1.06 \times S_{850~\mu m} \times (e^{\frac{17~K}{T}} -1) 
	\times (\kappa_{850~\mu m}/0.0125~cm^2~g^{-1})^{-1} \times (D/415~pc)^2 
\end{equation}
with $M$ in solar masses and $S_{850~\mu m}$ in Jy~bm$^{-1}$.
We adopt a dust opacity of 
$\kappa_\nu = 0.1 \times (\nu/10^{12}Hz)^\beta$~cm$^{2}$~g$^{-1}$
with $\beta = 2$, i.e., 
$\kappa_{850\mu m} = 0.0125$~cm$^{2}$~g$^{-1}$, 
following \citet{Pattle15} and \citet{Salji15a} among others, 
and a distance of 415~pc following \citet{Buckle10}.
These two assumptions are similar to those used in previous SCUBA analyses.
Note, however, that
\citet{Motte01}, \citet{Johnstone06}, and \citet{Nutter07} assume a distance of 400~pc, while
\citet{Johnstone01} assumes 450~pc. 
Also, 
\citet{Johnstone01} and \citet{Nutter07} assume
$\kappa_{850} = 0.01$~cm$^2$~g$^{-1}$ while \citet{Motte01} and \citet{Johnstone06} assume
$\kappa_{850} = 0.02$~cm$^2$~g$^{-1}$.

The dense cores are likely to have a range of temperatures (both within
each core, and core-to-core), although the largest variation would
be expected for the protostellar cores.  \citet{Schneider13} find dust temperatures  
of $\sim20$~K or higher around the \None\ and \Ntwo\ clusters, where most of the
SCUBA-2 emission is observed\footnote{While the maps at each {\it Herschel} wavelength
analyzed by \citet{Schneider13} are publicly released, the derived temperature and
column density maps
are not similarly available at present.  A full re-derivation of the dust temperature
across Orion~B based on the {\it Herschel} data is beyond the scope of our present analysis.}.  
We therefore assume a constant temperature of 20~K,
consistent with \citet{Johnstone06} and \citet{Nutter07}, as well as \citet{Sadavoy10}; 
\citet{Motte01}, however, assumed a temperature of 15~K for starless cores
and 20-40~K for protostellar cores, while \citet{Johnstone01} assumed a constant value of 30~K.
\citet{Johnstone06} note the four most massive cores they
identified in Orion~B are known to harbour bright far infrared sources which have heated
them to above 50~K, which would lower their estimated masses considerably from the 
value measured assuming 20~K.  At 50~K, the masses would be a factor of 3.3 lower than
assuming a temperature of 20~K.  We expect high
temperatures to be most likely in some of the brightest protostellar cores, 
where the masses we estimate are largest.  
Uncertainties in the dust opacity and cloud distance also increase the uncertainty in
the dense core mass estimates.  The dust opacity at 850~\mum likely has some 
variation across the cloud, with some inter-core and core-to-core variations, as seen 
in the $\beta$ variations measured across the Perseus molecular cloud by 
Chen et al (2015, submitted) and \citet{Sadavoy13}.
We expect that the distance
will generally be relatively constant across the cloud, and is more likely to affect 
global population values (i.e., changes would increase / decrease all masses by 
the same factor), and 
should have a smaller effect on the relative masses estimated.

In addition to the uncertainties in the conversion factor between flux and mass, there is
one other important consideration.  Structures within molecular clouds are hierarchical
in nature, although our SCUBA-2 observations are insensitive to the largest
of these structures.  Source identification algorithms such as FellWalker associate 
zones of emission with a single source, whereas other types of algorithms such as
those based on dendrograms \citep[e.g.,][]{Rosolowsky08} treat emission as nested
levels in a hierarchy of emission.  Under the latter scheme, only a fraction of the
total emission at a given position would be associated with the top level of the
hierarchical structure (i.e., the dense core), while some fraction of the emission would
be associated with underlying larger structures.  We therefore make a second estimate
of the total flux associated with each core which accounts for some of this larger-scale
structure.  Conservatively, we take the median flux value of pixels along the boundary
of a core as representing the constant background level of underlying layers of structure,
and subtract that value from every pixel lying within the core.  We refer to this
as the background-subtracted total flux (and mass), and include the background-subtracted
flux 
in Table~\ref{tab_cores}.  
This background subtraction method will
overestimate the contribution of larger-scale emission, particularly in the more
clustered parts of the cloud, and therefore provides a strict lower limit to the dense core
masses.

\begin{figure*}[htbp]
\begin{tabular}{cc}
\includegraphics[width=3in]{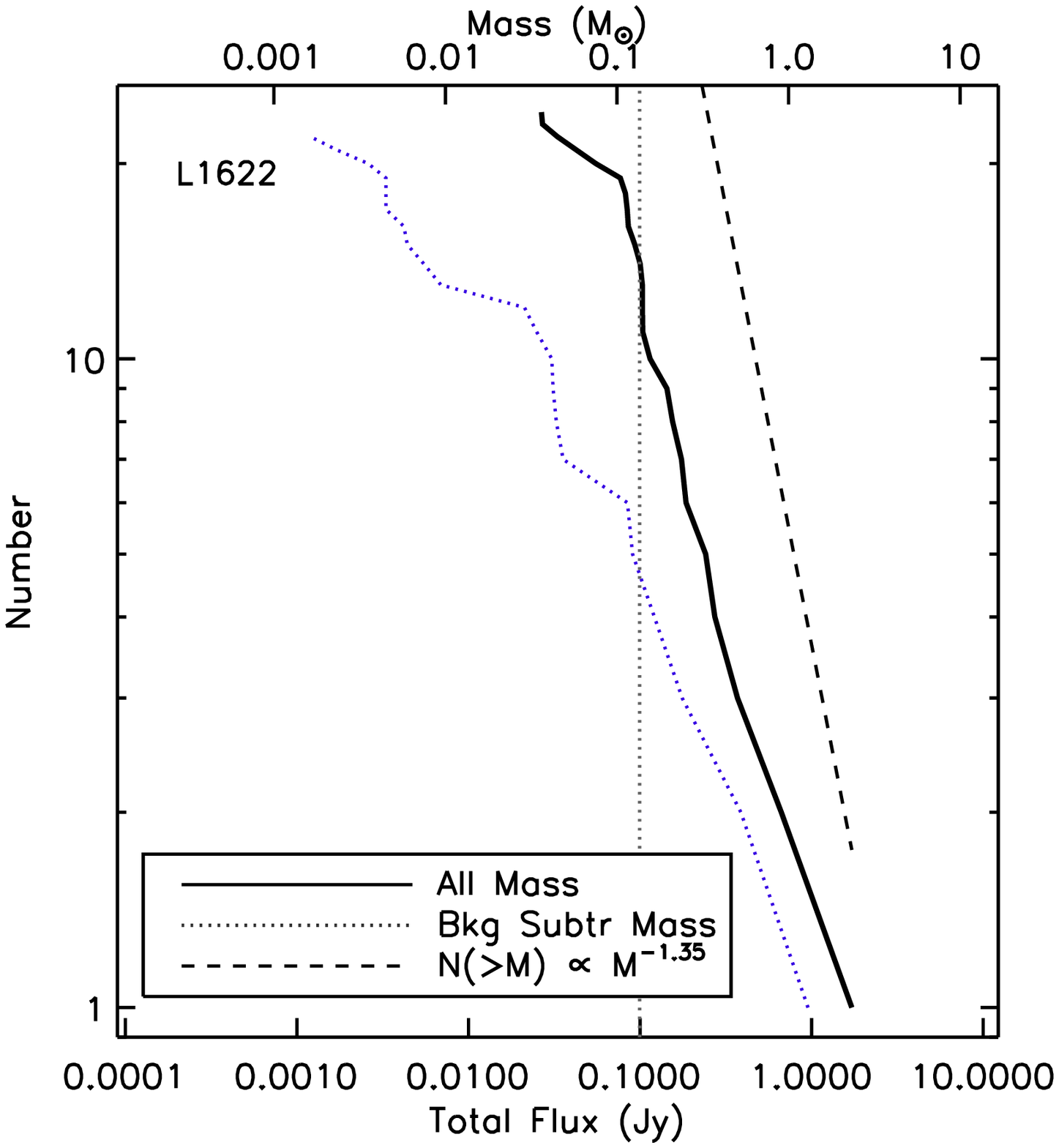} &
\includegraphics[width=3in]{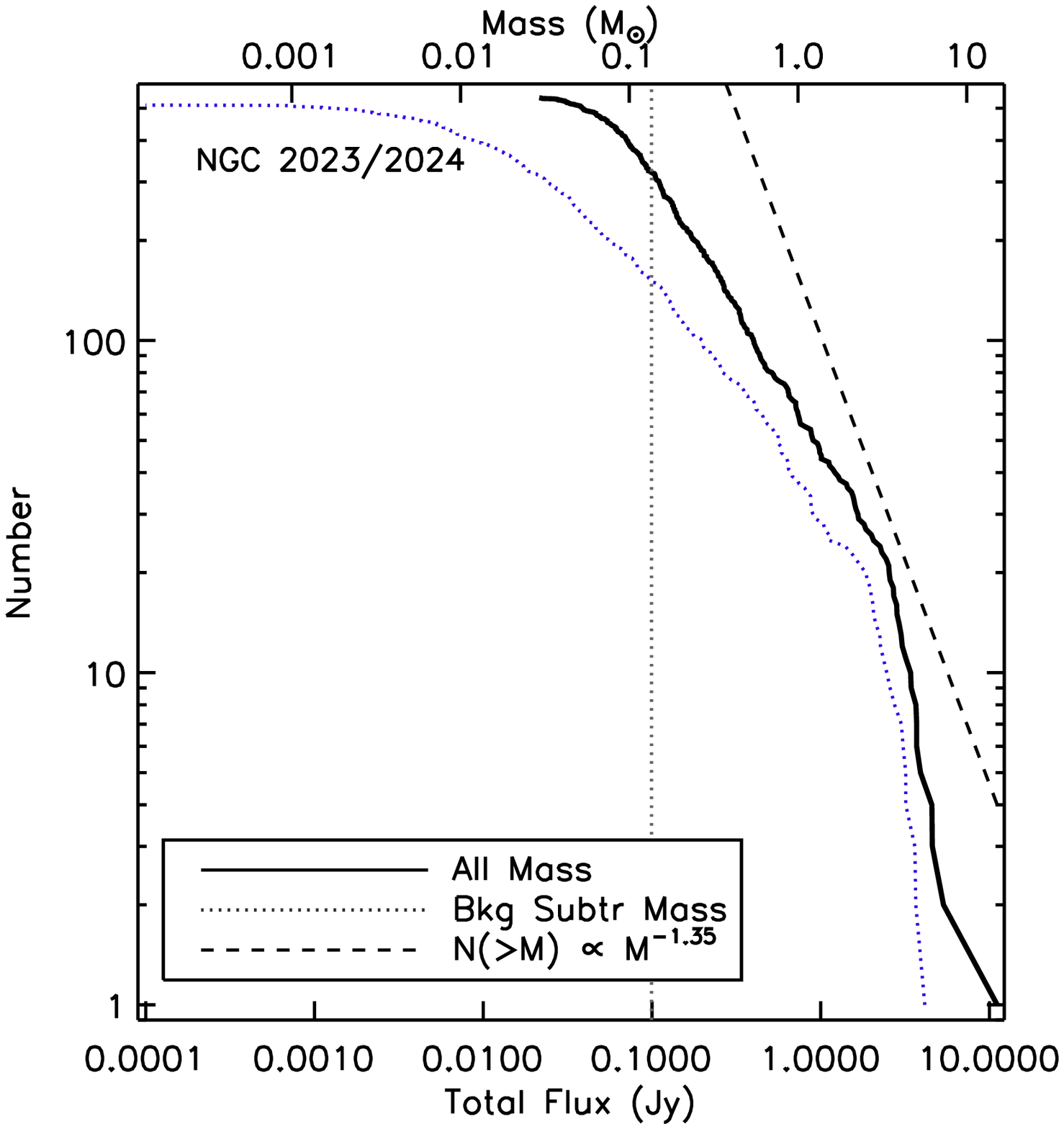} \\
\includegraphics[width=3in]{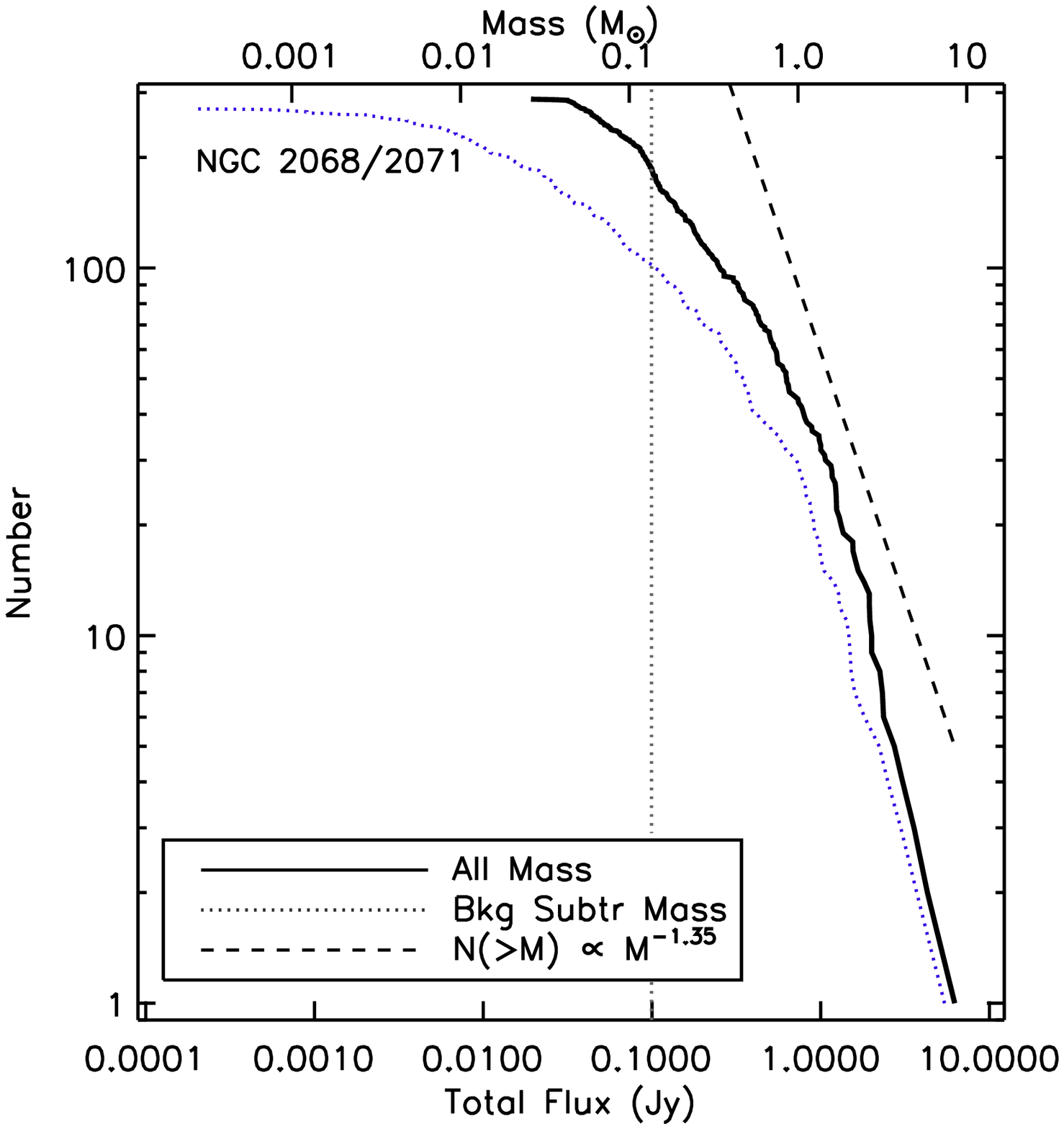} &
\includegraphics[width=3in]{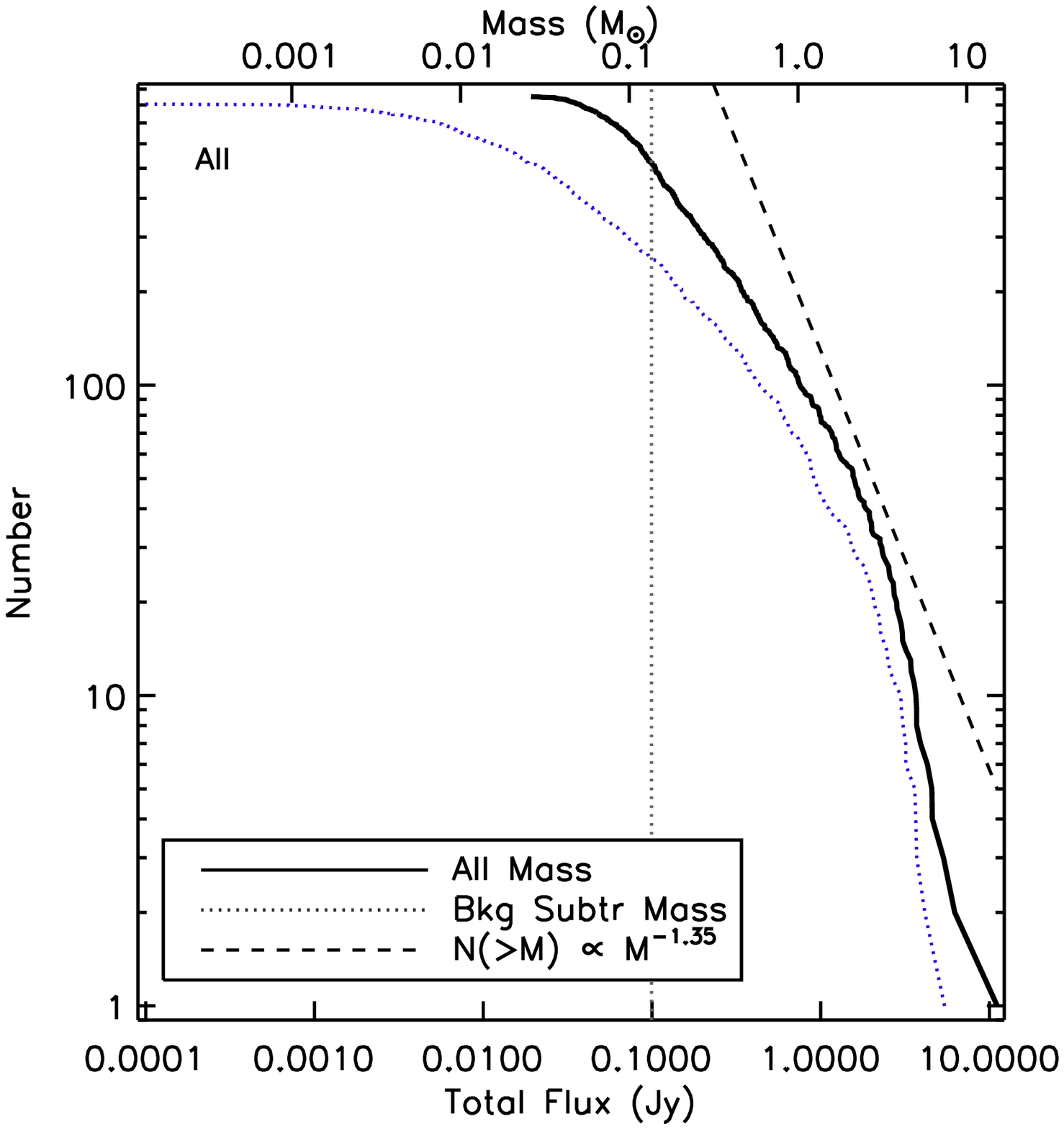} \\
\end{tabular}
\caption{Cumulative mass functions for starless dense cores in L1622 (top left), 
	\None\ (top right), \Ntwo\ (bottom left),
	and all three regions combined (bottom right).  The black solid line shows the 
	dense core masses using the full FellWalker-estimated masses, while the dotted
	blue line shows the background-subtracted masses.
	The dashed line shows a Salpeter slope of $N \propto M^{-1.35}$.
	The vertical grey line in the bottom right panel shows the completeness
	level. 
	The bottom horizontal axis
	in all plots shows the total flux measured, while the upper axis shows the approximate
	mass, as estimated by a simple constant conversion factor (see text for details).
}
\label{fig_mf}
\end{figure*}

Figure~\ref{fig_mf} shows the cumulative mass functions measured
from the total and background subtracted fluxes for the starless core
population in each of the three regions observed (top row and bottom left panel), 
and also as a combined sample (bottom right panel).  We omit the protostellar cores
on the basis that their masses are more likely to be over-estimated 
by assuming 
a constant temperature of 20~K.  
We estimate the completeness level from a flux level of 3~$\sigma$ across an
area equal to the median starless core size.
At the higher-mass end of the distribution, the slope is roughly consistent with the canonical 
Salpeter IMF \citep{Salpeter55}, for either estimate of the dense core masses.  
This similarity of the slope with the Salpeter IMF agrees with the original SCUBA analysis of 
\citet{Motte01}, and the combined SCUBA Orion A and B results of \citet{Nutter07},
among others.  Although the {\it Herschel} core mass distribution for Orion~B is not
yet available for a direct comparison 
\citep[see, however,][for the Orion~B column density PDF]{Schneider13},
other star-forming regions tend to follow a similar profile \citep[see, e.g.,][]{Andre14}.

At the very highest masses, we appear to have a slight deficit of starless cores relative
to a pure Salpeter distribution.  For example, extrapolating
the mass function shown in black from around 1~Jy (around 1~\Msol) up to 10~Jy using 
a Salpeter slope implies that there should be roughly three cores with total fluxes
above 10~Jy, whereas our sample contains only one.  The discrepancy 
between the Salpeter slope and the observed distribution of cores becomes even larger 
when the background
subtracted masses are used instead.  Both, however, are consistent within 3~$\sigma$
Poisson uncertainties.
An even larger sample of dense cores, ideally at least ten times more cores with high
masses, would be needed to confirm whether or not this result is statistically
significant.  An absence of massive dense starless cores might be partially attributable
to the tendency of object-identification algorithms to split large sources into multiple
components.  A real dearth of the most massive starless cores might also be partially
attributable to a slightly higher detection rate of protostars in the infrared; since
massive cores tend to have higher densities, it is possible that their natal protostars
would tend to have higher accretion rates, and therefore higher luminosities.  A larger
sample size, combined with a detailed consideration of the typical accretion rates derived
for detected protostars, would be necessary to test this scenario.

\subsection{Core Stability}

\begin{figure*}[htb!]
\begin{tabular}{cc}
\includegraphics[width=3.0in]{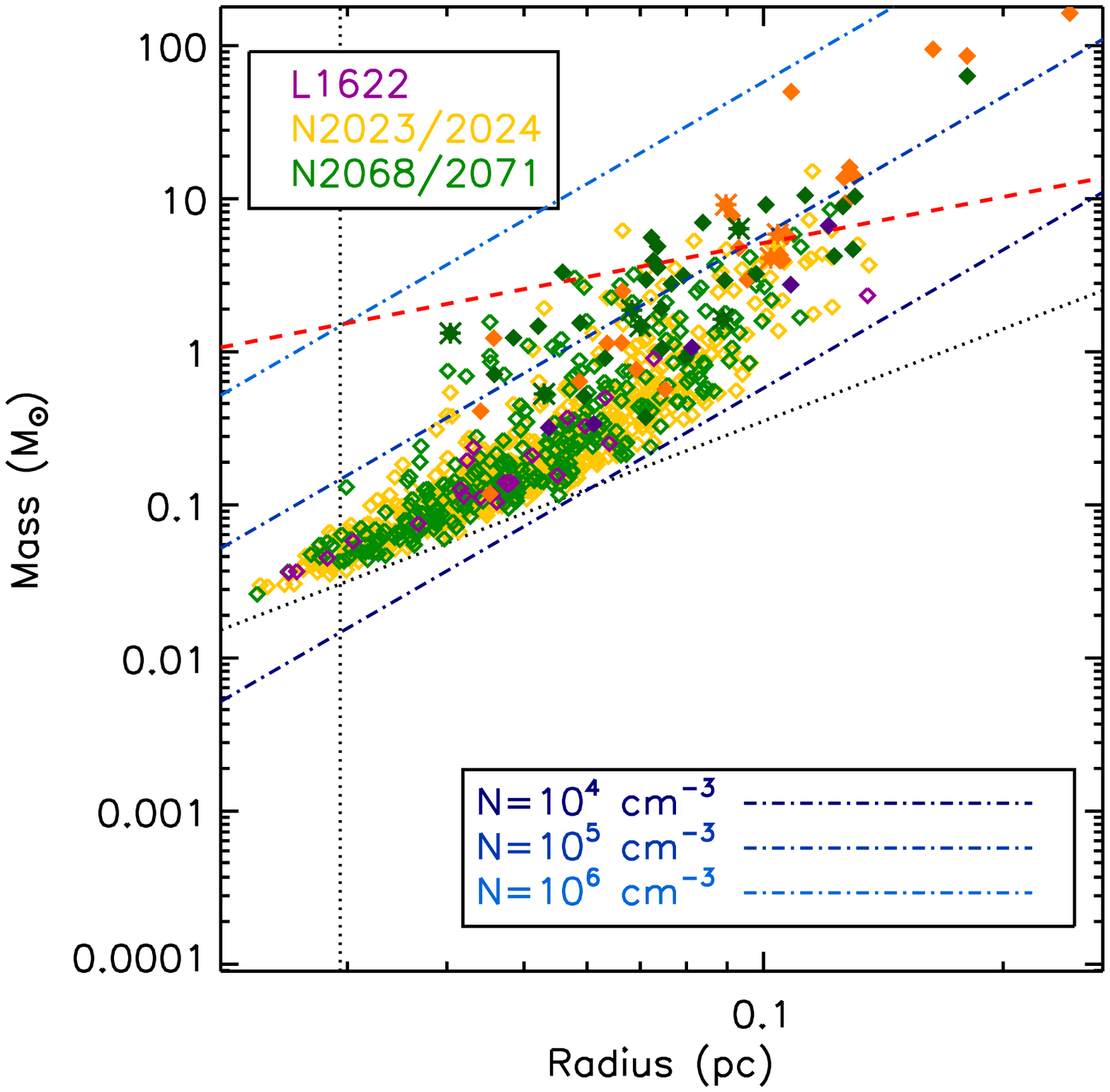} &
\includegraphics[width=3.0in]{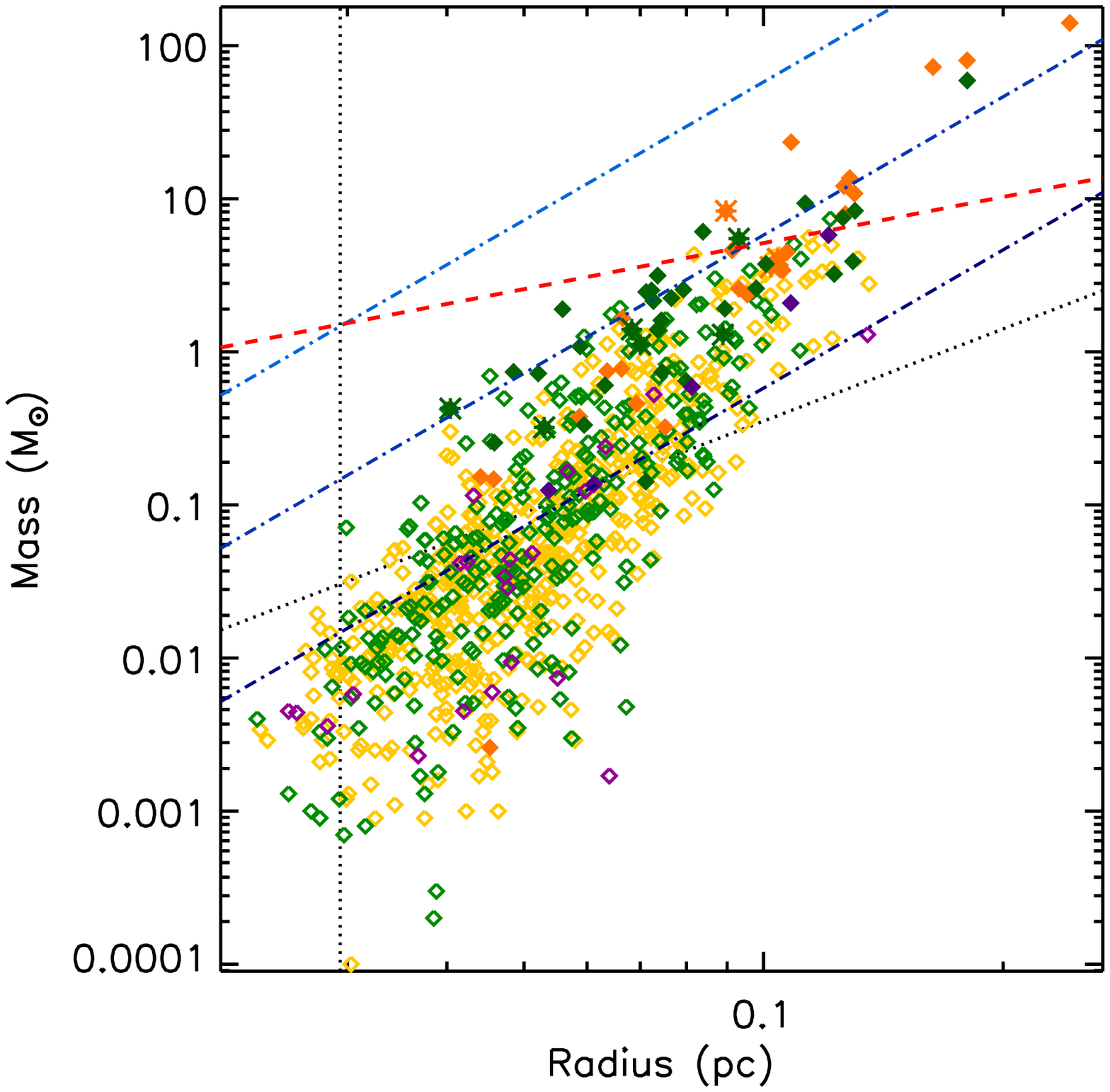} \\
\end{tabular}
\caption{Distribution of dense core masses and sizes for each region observed.  The colours
	indicate the region observed: L1622 (purple), \None\ (yellow), and \Ntwo\ (green),
	while the open diamonds indicate starless cores and the filled diamonds indicate
	protostellar
	cores.  The protostars have a slightly darker shading to enable better visibility,
	and the deeply embedded protostars from \citet{Stutz13} are denoted by asterisks.  
	The dotted lines denote the approximate sensitivity levels (the vertical
	line shows the beamwidth, while the diagonal line shows 3 times the typical rms
	integrated over a given radius).  The blue diagonal dash-dotted
	lines show the relationship expected for constant (3D) density objects, ranging
	from $10^{4}$~cm$^{-3}$ to $10^6$~cm$^{-3}$ from dark to light (bottom to top).  The dashed
	diagonal red line shows the locus of Jeans masses for a temperature of 20~K.
	The left panel indicates the total mass within each core, while the right
	panel indicates the background-subtracted mass within each core.  See text
	for details.
	}
\label{fig_m_r}
\end{figure*}

Using the sizes and estimated masses of all the dense cores, we can determine which
cores are stable to gravitational collapse.
Figure~\ref{fig_m_r} shows the core masses and radii for all three regions, as indicated
by the different colours.  
The JCMT effective beam width and the approximate
flux sensitivity, i.e., three times the median noise level integrated across a given area
are also shown.
Our core selection criterion is slightly more complex than can be captured by a 
single completeness level.  In particular, we removed sources that failed several local
signal-to-noise ratio criteria, which are more stringent than the global level indicated.
In Figure~\ref{fig_m_r}, we show masses derived from the
total fluxes of all cores (left panel), as well as masses derived from background-
subtracted total fluxes (right panel).  The background-subtracted mass estimates tend to
be smaller (as expected), and can be significantly smaller than our nominal total 
mass completeness level.  We emphasize that our simple method for estimating the 
background level overestimates the true core background, likely by a significant amount
for cores in crowded regions, and therefore those results should be treated
with caution.

Assuming a spherical geometry for the dense cores allows us to estimate their mean
densities.  Lines of constant density of $10^4$~cm$^{-3}$, $10^5$~cm$^{-3}$, 
and $10^6$~cm$^{-3}$ are
plotted in Figure~\ref{fig_m_r}.
Most of the cores in the left panel lie between $10^4$~cm$^{-3}$ and several $10^5$~cm$^{-3}$.
Although not explicitly calculated there, the range spanned by our more massive dense
cores is similar to that inferred from Figure~5 of \citet{Johnstone01} and Figure~7 of 
\citet{Johnstone06}.  \citet{Motte01} use a wavelet-based source-finder and include 
deconvolution of the telescope beam in their final size measurement, which tends to lead 
to smaller sizes (and therefore higher mean densities) than we report.
We also compared our results to those we would obtain using core sizes deconvolved
by the telescope beam.  Deconvolution had little effect most cores, since the majority of 
cores that we identify are significantly larger than the beam.

In Figure~\ref{fig_m_r}, we also plot the
locus of Jeans mass for each radius for an assumed temperature of 20~K.  
Dense cores above the Jeans mass locus are expected to be unstable to 
collapse if thermal pressure
provides the only avenue of support against gravitational collapse, and indeed, 
the majority of cores in this regime are associated with a protostar
(of 33 cores in the unstable regime, 24 are protostellar), although we 
caution that the protostellar masses may be overestimated.
Using
instead the background-subtracted mass decreases the already small number
of cores which lie above the Jeans instability line (17 unstable cores, of which 15
are protostellar).
\citet{Johnstone01} and \citet{Johnstone06} similarly found that most dense cores lie
within the range of stable Bonnor-Ebert sphere models \citep[an equilibrium isothermal 
sphere model;][]{Ebert55,Bonnor56}.  In addition, cores above this range
tended to have high central concentrations, which are correlated with the presence
of protostars (see discussion in the following section).
\citet{Motte01}, however,
argued that most of their identified dense cores were gravitationally unstable,
with this difference being directly attributable to their smaller core size measurements
obtained using a wavelet-based technique. 

The inclusion of velocity information from a dense gas tracer is important to determine
the role of turbulent motions in offsetting gravitational instability.  While 
primarily sensitive
to more diffuse gas than our SCUBA-2 observations, HARP $^{13}$CO and C$^{18}$O observations
of \None\ and \Ntwo\ show typical line widths of 1 -- 3~km~s$^{-1}$ \citep{Buckle10}, suggesting
that some level of non-thermal support is likely present in the Orion~B dense cores.  With
observations of a dense gas tracer such as N$_2$H$^+$, a more detailed consideration can
be made of the level of non-thermal support present for each dense core 
\citep[e.g.,][]{Kirk07,Pattle15}.
While non-thermal support mechanisms can explain the presence of starless cores
lying above the Jeans stability line, it is harder to understand the presence of protostellar
cores which appear to be Jeans stable.  
The most likely possibility is that the core boundaries we use in our analysis encompass both
a smaller-scale unstable region where the protostar has formed and a larger-scale zone
around it which is still stable, therefore making the core as a whole to appear to be stable.

\subsection{Concentration}
We measure the central concentration of each dense core as: \begin{equation}
C = 1 - \frac{1.13B^2S_{850}}{\pi R_{eff}^2F_{850}}
\end{equation}
following \citet{Johnstone01}, where $B$ is the effective beam width (in arcsec), 
$S_{850}$ is the total
flux (in Jy), $R_{eff}$ is the effective radius (in arcsec; see Section~3 for the definition
of $R_{eff}$), and $F_{850}$ is the 
peak flux (in Jy~bm$^{-1}$).  For dense cores that are well-approximated by the 
Bonnor-Ebert sphere model, those
having concentrations above 0.72 would be unstable to gravitational collapse
\citep{Johnstone01}.  
Previous work \citep{Jorgensen07,Jorgensen08,vanKempen09} has also shown that highly concentrated
dense cores tend to be associated with protostars.  

\begin{figure*}[htbp]
\includegraphics[width=5in]{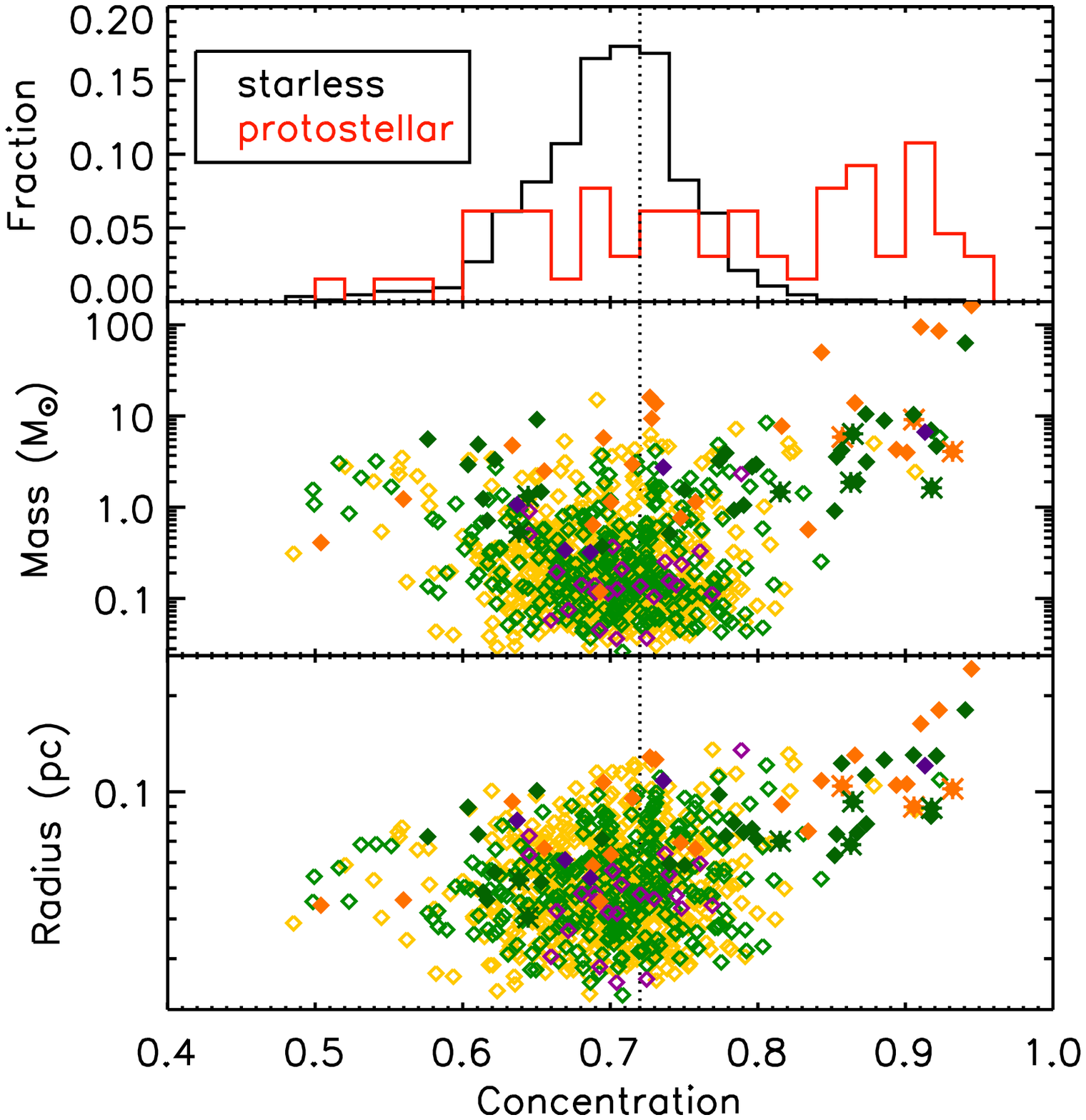}
\caption{The distribution of dense core concentrations (top panel) compared with their
	estimated masses (middle panel),
	and effective radii (bottom panel).  The red line in the top panel and filled
	diamonds in the bottom two panels indicate dense cores associated
	with a protostar, with the deeply embedded protostars from \citet{Stutz13} shown
	in asterisks, while the black line in the top panel and the open
	diamonds in the bottom two panels indicate starless cores.  
	Cores with concentrations above 0.72 (vertical dotted line) would be
	gravitationally unstable under a Bonnor-Ebert sphere model.
	}
\label{fig_concs}
\end{figure*}

In Figure~\ref{fig_concs},
we show the concentration of the dense cores compared with their masses and effective
radii.
The top panel shows that the majority of protostellar
cores have high concentrations that are normally taken to indicate gravitational
instability (42 protostellar cores, versus 23 at lower concentrations).
The starless cores have a much tighter distribution of concentrations
around a value of $\sim 0.72$, which a two-sided KS test shows is statistically
distinct, with a probability of $3\times10^{-10}$ that the protostellar and starless core
concentrations were drawn from the same parent sample.
We note that some of the cores are elongated, complicating both the application of
the Bonnor-Ebert sphere model and the interpretation of the concentration measurement.
FellWalker does not calculate core elongations, since it does
not fit any pre-determined shape to the cores.
We use the ratio of the `size' of the core along the horizontal and vertical axes, each defined as 
the flux-weighted standard deviation of core pixel values from the flux-weighted 
centre position, as a rough proxy for core elongation.  
With this measure, only 12\% of the cores are elongated (ratios of 2 or higher),
and they have a similar distribution of concentrations and effective radii to the 
other cores, so they do not bias the global distribution.  

Lower concentrations for protostellar cores could indicate more evolved sources
\citep[c.f.][]{vanKempen09}.
SCUBA-2 is insensitive to the mass contained within the central protostar
itself, so a protostar which has accreted much of the mass in its envelope would
tend to have a lower concentration \citep[see][for a discussion of protostellar mass versus
`envelope' mass in the
context of comparisons with numerical simulations]{Mairs14}.
The protostellar cores which lie the furthest below the thermal Jeans line, and
those with smaller total masses both tend to
have lower concentrations as well, which supports this hypothesis.

In contrast to prior work, we find that a significant number of starless dense cores have
high concentrations that would nominally indicate instability (299 starless cores
have concentrations above 0.72 while 551 have lower concentrations).
At least some of these higher values of concentrations are likely attributable to the 
increased sensitivity of SCUBA-2 compared with SCUBA.
\citet{Johnstone03} and \citet{Johnstone06} find a range of concentrations from about 0.3 to 
0.9 for dense cores in Orion B using SCUBA data, whereas our
concentration measurements range between roughly 0.5 and 0.95.  
Since the resolution of SCUBA and SCUBA-2 are identical, these differences must
be attributable to the improved sensitivity of SCUBA-2 data and possibly
also the core identification algorithm used (ClumpFind versus FellWalker). 
FellWalker, like ClumpFind,
tends to include lower flux material around peaks within the boundary of a core.  
Thus, cores identified in higher sensitivity observations will tend to have
larger sizes and total fluxes, with the core area increasing faster than the total flux
(since only faint pixels are being added).  
We therefore expect that the increased
sensitivity of the SCUBA-2 observations, coupled with the improved recovery of
larger-scale emission, will increase the concentrations of our cores
relative to similar analyses of SCUBA observations.
At the same time, cores with larger areas relative to their fluxes (or masses) will
appear more stable in the
mass versus radius analysis shown in Figure~\ref{fig_m_r}.

\subsection{Dense Cores and Ambient Cloud Pressure}

\citet{Lombardi14} used a combination of {\it Planck} and {\it Herschel} data across
Orion (A and B clouds) to estimate the total column density of material down to a 
resolution of 36\arcsec\ in areas with {\it Herschel} coverage.  We use this
map, including \citeauthor{Lombardi14}'s recommended scalings between optical depth 
at 850~\mum and total column density, to compare with the SCUBA-2 dense cores.  
L1622 falls outside of the \citet{Lombardi14} column density map, and so is not
included in this analysis.  
Previous analyses
\citep[e.g.,][]{Onishi98,Johnstone04,Hatchell05,Kirk06,Enoch06,Enoch07,Konyves13} have shown that
dense cores tend to be found in regions of higher overall column density, although historically
these analyses have relied on much lower resolution measurements of the overall cloud
column density.

\begin{figure*}[htb]
\begin{tabular}{cc}
\includegraphics[width=3in]{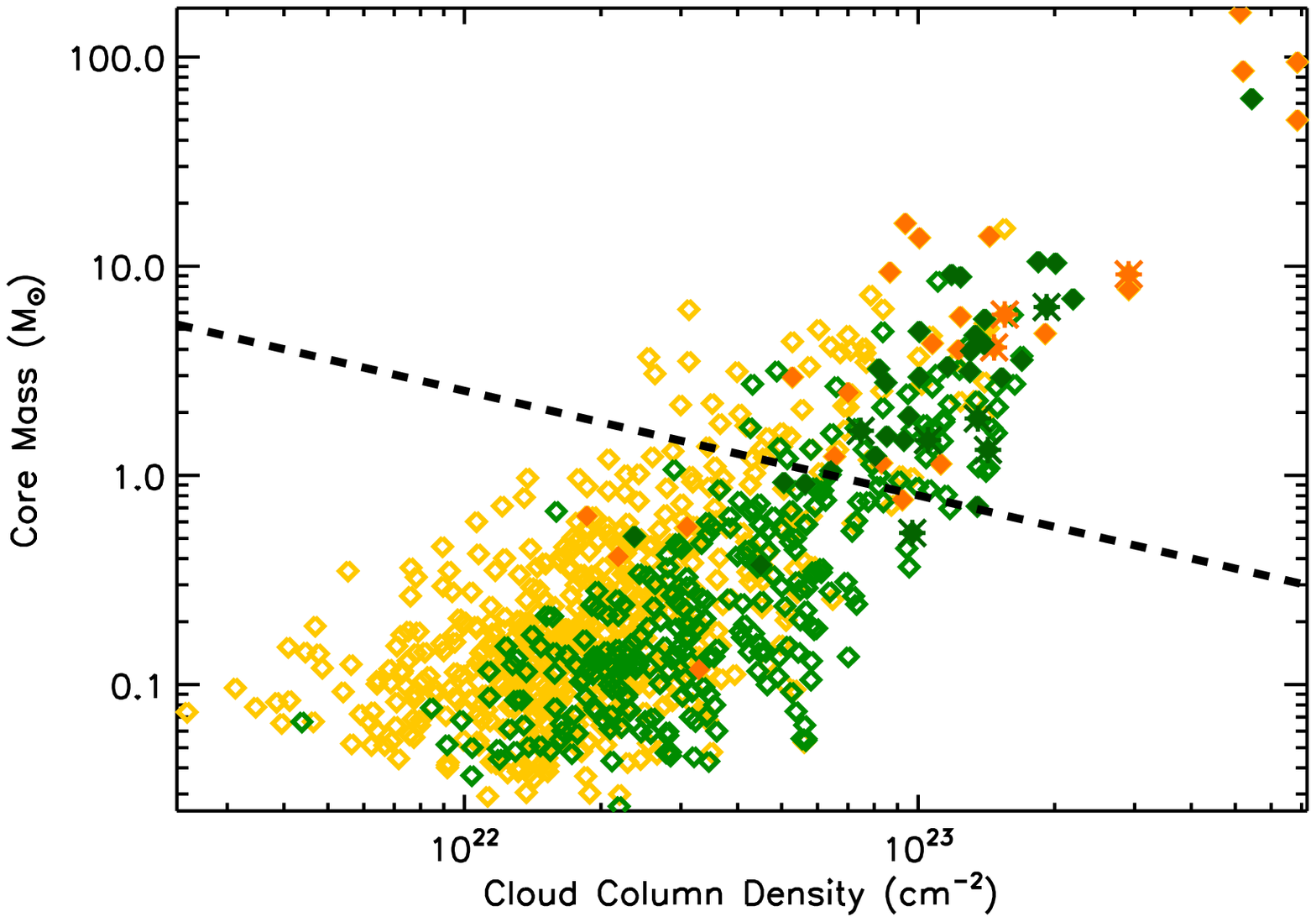} &
\includegraphics[width=3in]{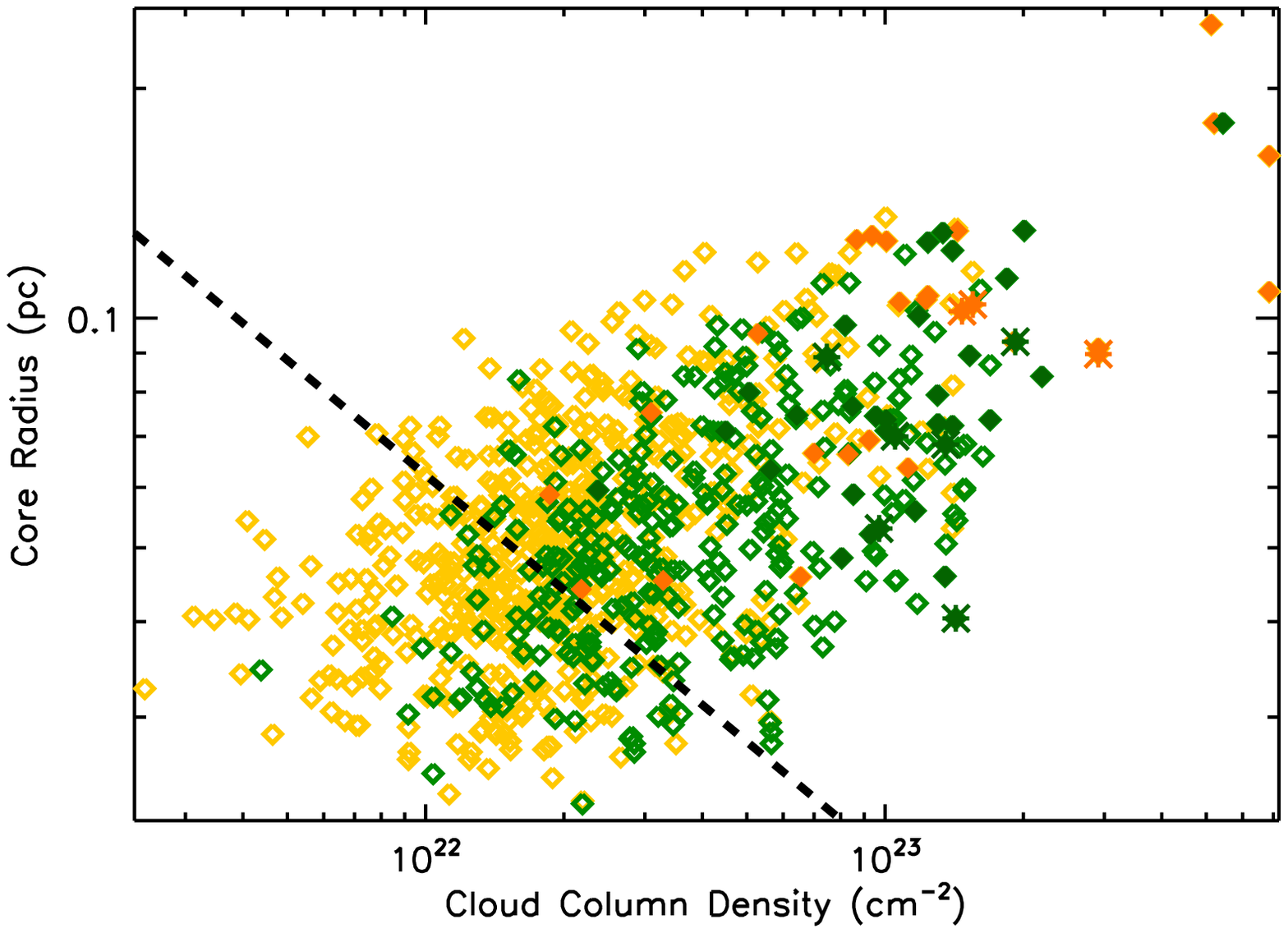} \\
\end{tabular}
\caption{Dense core sizes (left) and masses (right) compared to the local cloud column density
	from \citet{Lombardi14}.  The colour scheme follows Figure~\ref{fig_m_r}, i.e.,
	\None\ cores are plotted as yellow/orange diamonds, while \Ntwo\ cores are plotted
	as green / dark green diamonds.  The darker, filled diamonds indicate protostellar cores,
	and the asterisks denote the protostars from \citet{Stutz13}.
	In both panels, the dashed black line shows the critical radius (left) and mass (right)
	for an isothermal sphere model at 20~K with an external pressure derived from the
	local cloud column density.  Most of the protostellar cores lie above the line of
	critical stability.}
\label{fig_cores_pressure}
\end{figure*}

Under the assumption that a molecular cloud is a sphere, the column density at 
a given location within the cloud can be used as a proxy for the external pressure due
to the overlying weight of the cloud.  In this simple model, a higher local column
density implies a three dimensional position closer to the cloud centre, and hence
a larger weight of overlying cloud material.  While the \citet{Lombardi14} column
density map clearly shows that the Orion~B cloud is more complex than a sphere,
the spherical assumption provides a practical method to estimate the bounding pressure on
dense cores due to the ambient cloud material.  Furthermore, the model's implication that
sources in higher column density zones are likely surrounded by more material than
those in lower column density zones seems generally reasonable. 
Following \citet{McKee89}, and the
implementation in \citet{Kirk06}, the pressure at depth $r$ in a cloud is
given by 
\begin{equation}
P(r) \simeq \pi G \bar{\Sigma} \Sigma(r)
\end{equation} where $\bar{\Sigma}$ is the mean
column density and $\Sigma(r)$ is the column density measured at cloud depth $r$.  For cores
near the cloud centre, the column density along the line of sight to the core is
roughly twice this value, i.e., $\Sigma_{obs} = 2 \times \Sigma(r)$.  In both
\None\ and \Ntwo, the mean cloud column density over the area observed by SCUBA-2 is
$9 \times 10^{21}$~cm$^{-2}$.  For each core, we measure the local cloud column density as the
maximum value of the \citet{Lombardi14} column density map within the core's footprint 
(there are typically only a few resolution elements within each core footprint).
If we make the assumption that the cores can be well represented by an isothermal
sphere model, the critical radius and mass of each core can be written as
\begin{equation}
R_{crit} = 0.49\frac{c_s^2}{\sqrt{G P}}
\end{equation}
and
\begin{equation}
M_{crit} = c_s^4 \sqrt{\frac{1.4}{G^3 P}}
\end{equation}
where $c_s$ is the sound speed and $G$ the gravitational constant 
\citep[equations adapted from][]{Hartmann98}.
In Figure~\ref{fig_cores_pressure}, we show the relationship between core sizes and masses
and the total cloud column density at the core positions.  
We see a strong correlation between the cloud column 
density and the core masses, and a weak correlation between the cloud column density
and the core sizes, in contrast with \citet{Sadavoy10}, who compared 
SCUBA-based dense core properties with extinction-based column density measures in five 
nearby molecular clouds, including Orion.  Given the large scatter in the relationships that
we observe, we expect the discrepancy with \citet{Sadavoy10} is the result of the
much larger number of
dense cores in our present analysis, and the larger parameter space that they occupy.

With the pressure of the overlying cloud material estimated using the spherical-cloud
assumption discussed above, many of the cores have sizes and masses larger than can
be thermally supported given this external weight of the cloud.  By size, all protostars
and 522 of 826 starless cores lie above the critical value, while by mass, 50 of 60 protostars
and 101 of 826 starless cores lie above the critical value (note that cores in L1622 are 
not included in this analysis).
The fraction of cores deemed unstable by this simple pressure analysis
is a significant change from the apparent
thermal stability of the dense cores seen in Figure~\ref{fig_m_r}
(24 of 60 protostellar cores lie above the thermal Jeans mass compared to 9 of 826
starless cores; see Section~4.2) and shows that
the pressure from the ambient molecular cloud plays a strong role dense core stability.
A similar result has been seen in other dense core analyses 
\citep[e.g.,][]{Kirk07,Lada08,Pattle15}.
Beyond stability considerations from a hydrostatic equilibrium model, non-thermal forces
may be contributing significantly to the pressure on individual cores, which might help to
explain the large scatter apparent in Figure~\ref{fig_cores_pressure}.

While we identify dense cores inhabiting a wide range of cloud column densities, we note
that the correlation between the cores' size or mass with the cloud column density
also implies that there is a minimum column density value at which pressure-unstable cores
are found.  This minimum column density is approximately 10$^{22}$~cm$^{-2}$, which is
somewhat higher than the column density threshold observed in nearby star-forming
regions, which is usually around 5-7$\times 10^{21}$~cm$^{-2}$ 
\citep[e.g.,][]{Onishi98,Johnstone04,Kirk06,Enoch06,Enoch07,Konyves13}.

\subsection{Cloud Structure and Core Lifetimes}

\begin{figure*}[htb]
\begin{tabular}{cc}
\includegraphics[width=3in]{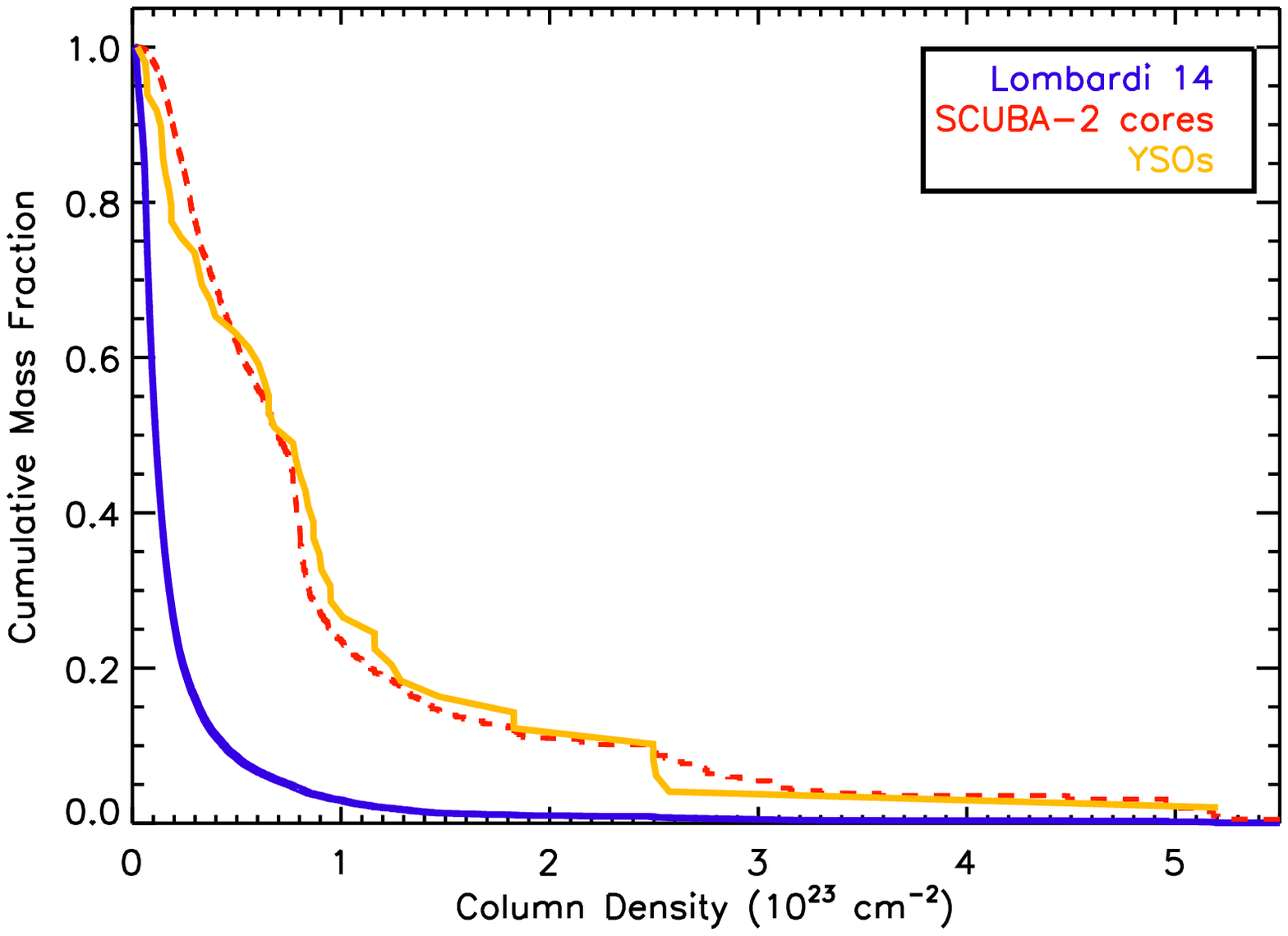} &
\includegraphics[width=3in]{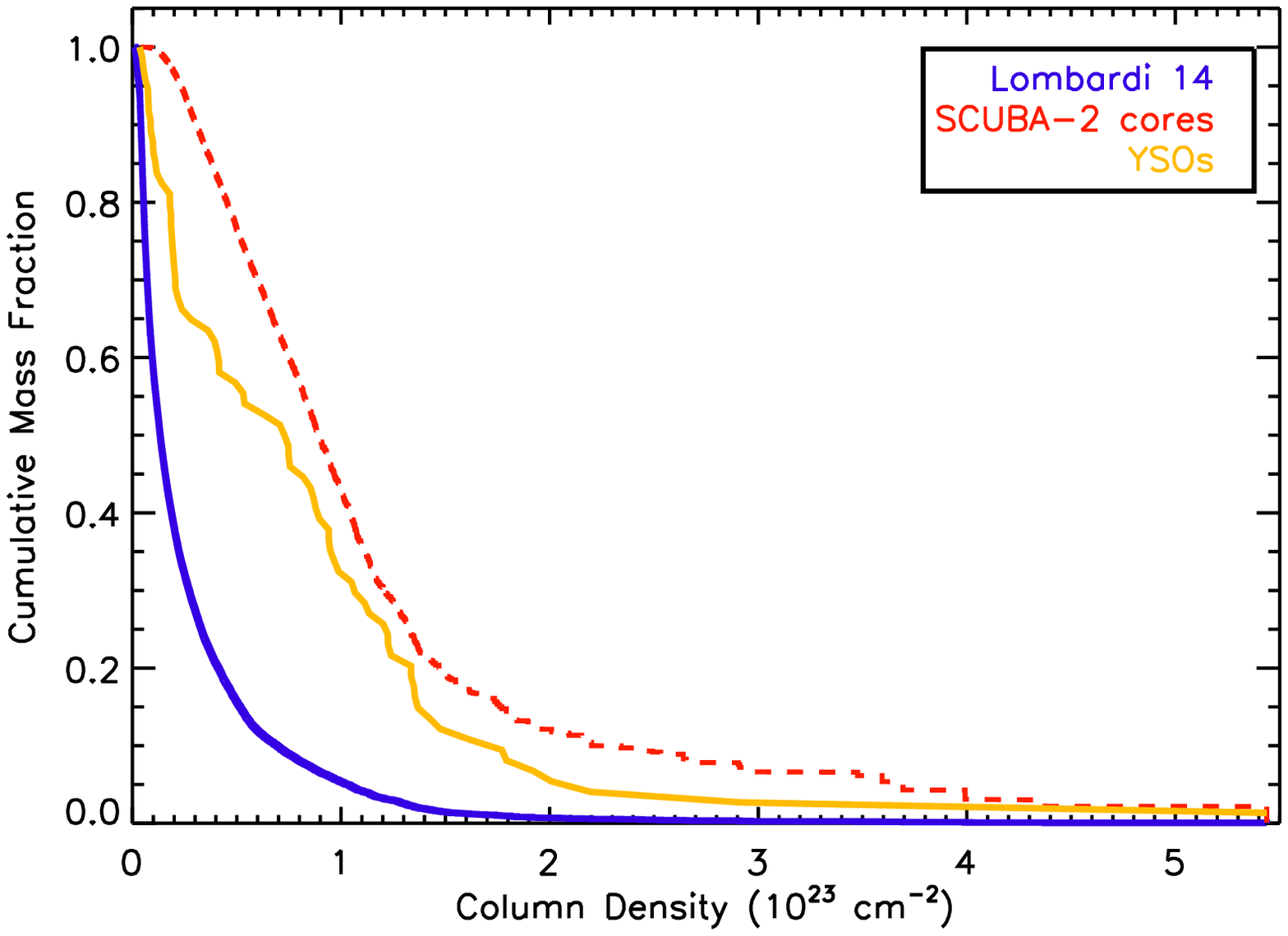} \\
\end{tabular}
\caption{A comparison of the cumulative mass fraction within SCUBA-2 cores and the entire cloud of 
	gas and dust as measured by \citet{Lombardi14} across \None\ (left panel) and 
	\Ntwo\ (right panel).  In each panel, the solid blue line shows the fraction of mass
	at a given column density or higher within the area observed
	by SCUBA-2, while the dashed red line shows the fraction of mass in dense cores. 
	The yellow line shows the fraction of protostars
	from \citet{Megeath12} and \citet{Stutz13} at the same column density or above.}
\label{fig_cores_vs_col}
\end{figure*}

In Figure~\ref{fig_cores_vs_col}, we compare
the cumulative mass fractions of dense cores (using the total mass estimated) and 
cloud mass as functions of the 
cloud column density from \citet{Lombardi14} for \None\ and \Ntwo. 
For a fair comparison between the 
dense core mass and cloud mass fractions, we consider only areas observed with SCUBA-2.
For the dense cores, we take the column density at each pixel that lies
within a dense core footprint (excluding other pixels as noise).
Figure~\ref{fig_cores_vs_col} shows that the dense cores seen by SCUBA-2 are 
associated with the highest column density material. 
For example, roughly 24\% and 43\% of the mass in SCUBA-2 cores in \None\ and \Ntwo\ 
respectively
is associated with total column densities above 
$10^{23}$~cm$^{-2}$, whereas only a small fraction of the cloud material 
(3\% and 5\% respectively) 
lies within this range.  
We also compare in Figure~\ref{fig_cores_vs_col} the protostar number fraction with 
the column density at the locations of
protostars from \citet{Megeath12} and \citet{Stutz13}.  Assuming that all protostars
have similar masses, the fractional number of protostars within a column density contour 
is equivalent to their fractional mass within that same contour.  
We find that the protostars are also
concentrated in regions of high column density, although slightly less so than
the dense cores in \Ntwo.  Since the protostellar list from
\citet{Megeath12} likely includes some slightly older protostars that have drifted
from their birthsites, it is not surprising
for a YSO population to have a slightly wider range of column densities than
the dense cores.

\begin{figure*}[htb]
\begin{tabular}{cc}
\includegraphics[width=3in]{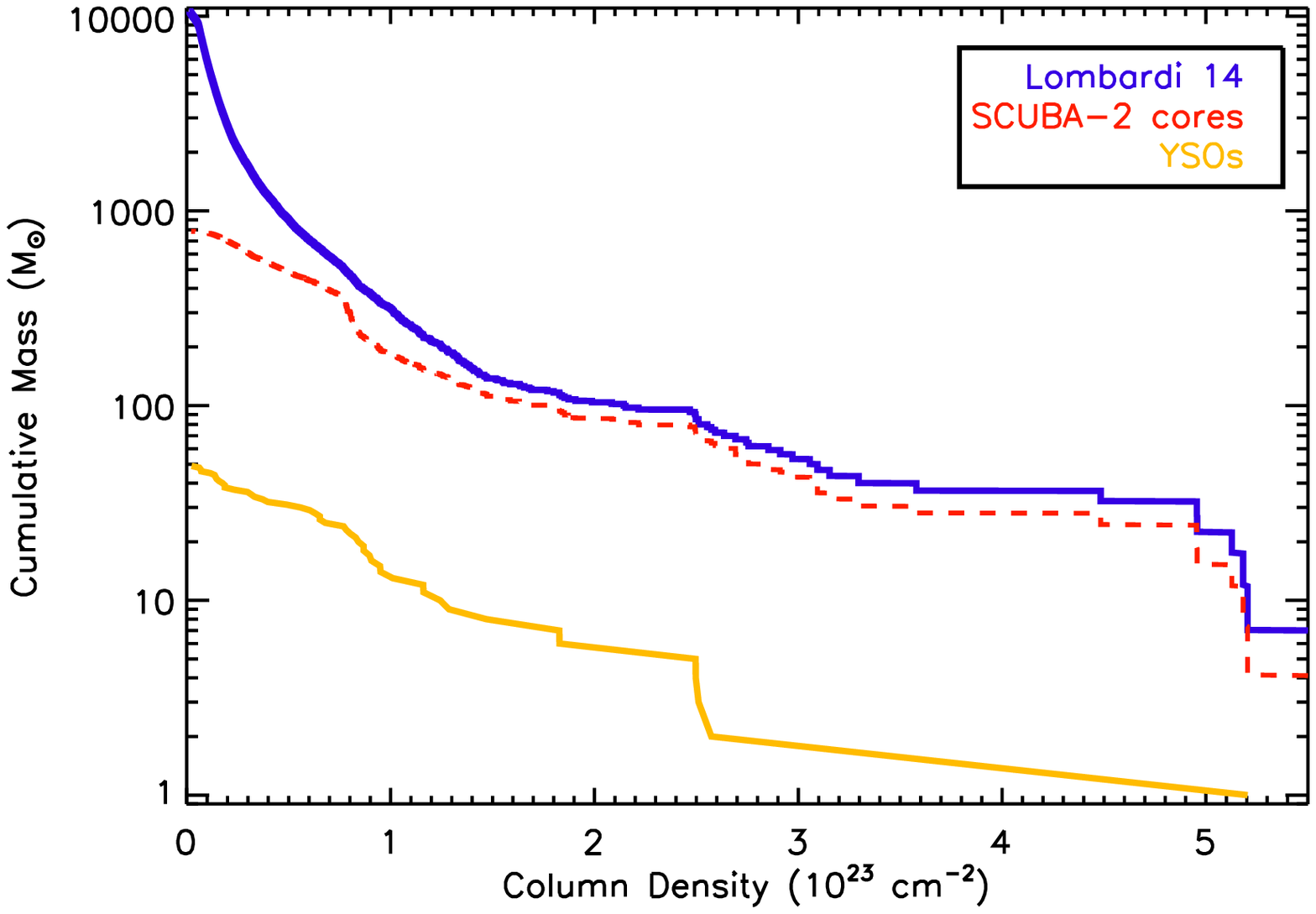} &
\includegraphics[width=3in]{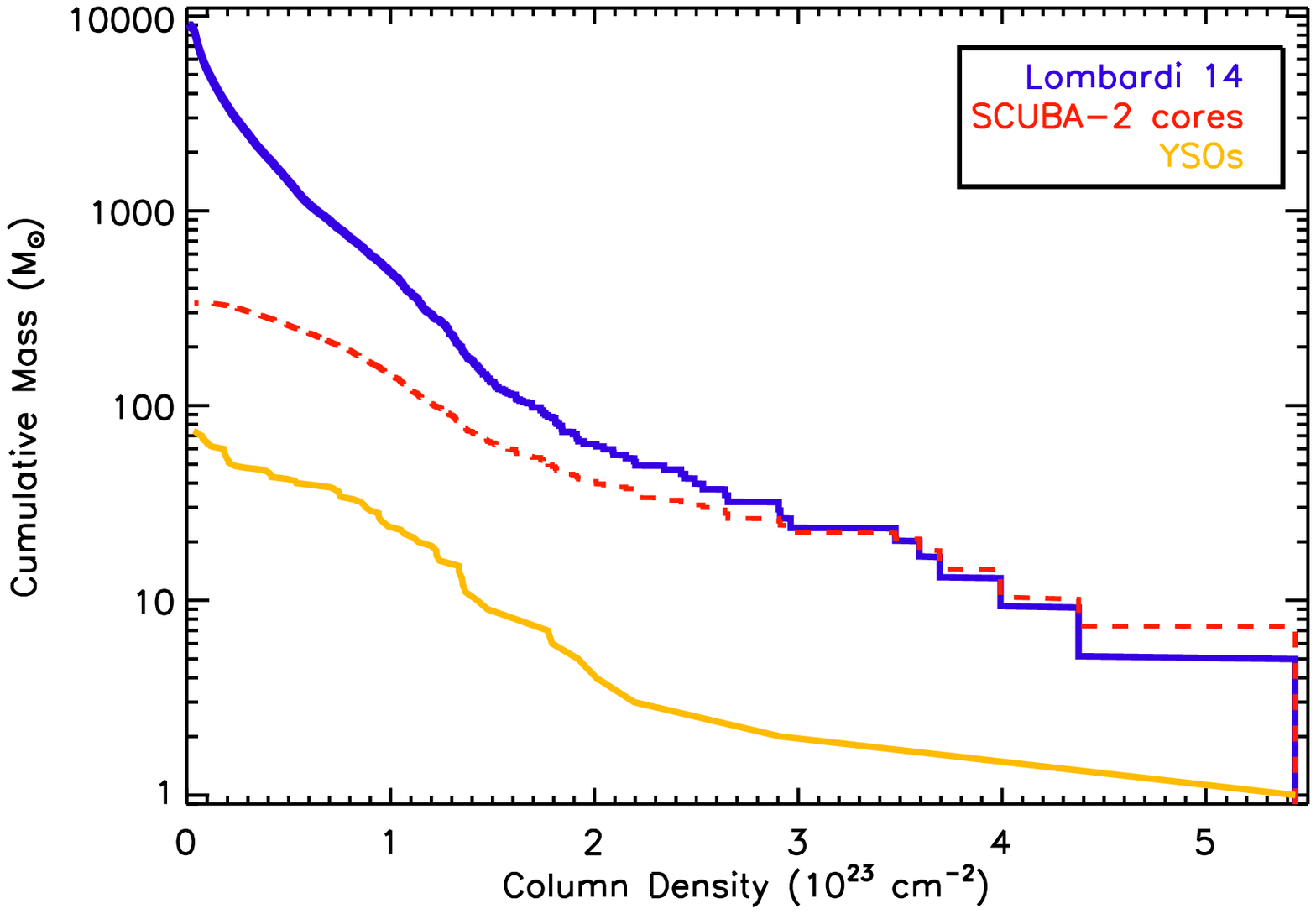} \\
\end{tabular}
\caption{A comparison of the cumulative mass within SCUBA-2 cores and the entire 
	gas and dust mass as measured by \citet{Lombardi14} across \None\ (left panel)
	and \Ntwo\ (right panel).  In each panel, the solid blue line shows the total
	mass at a given column density or higher within the area observed by SCUBA-2,
	while the dashed red line shows the total mass in dense cores (see 
	Section~\ref{sec_coremass} for the assumptions used).  The yellow line shows
	the mass in protostars, assuming they each have a mass of 0.5~\Msol.}
\label{fig_mass_vs_col}
\end{figure*}

We can also consider the above distributions in terms of total masses, as shown in
Figure~\ref{fig_mass_vs_col}.  The total mass in \None\ and \Ntwo\ within the areas
observed by SCUBA-2 is 10600~\Msol\ and 9000~\Msol, respectively, while the total
mass in dense cores is 780~\Msol\ and 340~\Msol, respectively.  Again, we
emphasize that the dense core masses are estimated assuming a constant temperature of
20~K.  Both \None\ and \Ntwo\ have several large and massive protostellar dense cores
for which this assumption will cause the mass to be overestimated.  
These particular dense cores are
coincident with the highest total column densities in the \citet{Lombardi14} map,
which is responsible for making the total dense core mass strangely appear larger than the
total gas and dust mass at the highest column densities in Figure~\ref{fig_mass_vs_col}.
It is also likely that, due to their slightly lower resolution 
(compared to SCUBA-2), \citet{Lombardi14} may
slightly underestimate the total mass in the highest column density and smallest
scale structures.
Even with these caveats, it is interesting to note that 
above $1-2\times10^{23}$~cm$^{-2}$,
nearly all of the high column density material is already in dense cores.  Below this
column density, the dense cores represent an ever-decreasing fraction of the total
mass.

At the highest column densities, 10-20\% of
the dense core material is already located within protostars,
if we make the assumption that each protostar has a mass of 0.5~\Msol. 
At lower column
densities, the mass in YSOs is only around 6\% of the dense core mass in \None\ while it
is 16 -- 22\% in \Ntwo.  In both regions, the mass within YSOs is less than 1\% of the total
cloud mass.  There is no indication of a strong relationship between the total
column density and the ratio of YSO mass to dense core mass, though it is possible that
systematic biases in our simple mass estimations which hide
such a trend (e.g., if YSOs tend to be more massive in high column density environments).
At lower column densities, the ratio of YSO mass to dense core mass in
\None\ is roughly a factor of 4 lower than in \Ntwo.  
This result could imply that \None\ is younger, and that
the protostars there have only started to form recently.  Although the total numbers
of sources are small, 
\citet{Stutz13} also found a higher proportion of the youngest protostellar candidates
(``PBRs")
in \None\ than in \Ntwo\ relative to YSOs found in the {\it Spitzer}-based 
catalogue of \citet{Megeath12}.
This result also supports the scenario of \None\ being younger, as does the relatively
larger percentage of YSOs that we see at lower column densities in \Ntwo\ 
(Figure~\ref{fig_cores_vs_col}).

The ratio of starless cores to protostellar cores has also been used as a tool to
estimate the relative lifetimes of the two stages, with the estimated protostellar
lifetime then used as an anchor to obtain absolute lifetimes.
Previous analyses of dense cores detected with SCUBA and similar instruments have
suggested lifetimes of both to be several tenths of a Myr, with a similar 
number of protostellar
and starless cores identified \citep[e.g.,][]{Enoch08,Hatchell07,Kirk05}, while
earlier analyses, such as that of \citet{Jessop00} suggested the starless 
core lifetime decreases with the core's volume density.
With our more sensitive census of cores detected with SCUBA-2, we identify a much larger population
of starless dense cores, and can re-visit this question, although we caution that
examining only cores within a single cloud may introduce some bias.
Furthermore, some of the dense cores in our sample may be transient features which
never evolve to form a star.
In our full sample, we have 851 starless cores and 64 protostellar cores, 
i.e., a ratio of 13:1.  If we sub-divide the dense cores into bins of varying 
mean density, we find a roughly 1:1 ratio for starless to protostellar cores above mean
densities of $10^5$~cm$^{-3}$, and a rapidly increasing ratio beyond that, as shown
in Table~\ref{tab_proto_ratio}.  In Table~\ref{tab_proto_ratio}, we include the ratio
of protostellar cores for both the full dense core sample, as well as when the cores
are restricted to those more massive than 0.1~\Msol\ (i.e., those which presently
have sufficient mass to form a star and may therefore be less likely to be 
transient features).  While the protostellar ratios differ in the
lower density bins, depending on which sample is examined, both show the same trend
of a protostellar ratio which decreases rapidly with protostar density. 
This result is qualitatively in agreement with 
\citet{Jessop00} in that dense core lifetimes do indeed appear to be longer 
for cores with lower
mean density.  We caution, however, that our assumption of a constant temperature
tends to bias the protostellar core masses (and hence mean densities) to higher
values, which would therefore serve to increase the fraction of protostellar cores
in the higher density bins from their true value.  
Similarly, if some starless cores were colder than
20~K, their masses and densities would be {\it under}estimated which would increase
the number of starless cores in the higher density bins.

The concentration measured for each core is likely to be less biased by a non-constant
temperature than density / mass estimates are.  Separating the dense cores into those
with high and low concentrations (above and below the nominal maximum stable value of 0.72)
shows that more concentrated dense cores are more likely to be protostellar.  The
starless to protostellar core ratio for high concentrations is 7:1 (299 versus 42)
while the ratio for low concentrations is 24:1 (551 versus 23).
We note that these ratios are very similar when only dense cores more massive than
0.1~\Msol\ are considered: there, the ratios are 6:1 and 22:1 respectively. 

Both the concentration and mean density results support the simple picture that
as dense cores evolve, they tend to become denser and more centrally concentrated
before they are able to form a protostar.  

\input{tab2}

\section{DISCUSSION}
\label{sec_disc}

\citet{Lada91} identified roughly 300 YSOs in each of 
the \None\ and \Ntwo\ regions, corresponding to an additional 150~\Msol\ for each
region beyond the YSO masses discussed in the previous section.  
With an efficiency of converting
dense core mass into YSOs of 30\%, approximately 235~\Msol\ and 100~\Msol\
of YSOs in \None\ and \Ntwo, respectively, may be created from the current population of 
dense cores.  This number would roughly double the existing stellar populations in both regions,
and is several times larger than the existing YSO population in either region.  The
total amount of mass at lower densities in each region is around 10000~\Msol; if even
1\% of this mass ends up also contributing to future stars, it would contribute
about the same amount of stars again.  Both of these regions therefore may one day
harbour stellar clusters containing many hundreds of stars.
At the present star formation rate, it will take several million years to deplete the
current population of dense cores.  Since the most massive dense starless cores 
present reach only about 10~\Msol, it is likely that B stars will be the most massive
that can eventually form and help to drive the dissipation of the remaining cloud material.

L1622 appears to have less material available to form additional YSOs with a total dense core
mass of 18~\Msol\ and roughly 6~\Msol\ presently in YSOs.  The total cloud mass cannot
be estimated to the same precision as \None\ and \Ntwo\ since a full column density map 
is not presently available.  Based on the CO maps of \citet{Maddalena86}, however,
L1622 appears
to be a factor of at least several less massive than \None\ or \Ntwo.  This too suggests
that a limited amount of star formation may occur in the future in L1622.
As outlined in the introduction, the distance to L1622 is less certain, and some
observations suggest a distance of $<200$~pc \citep[see discussion in][]{Reipurth08}.  
If this closer distance is indeed
correct, then L1622 would be an even more quiescent region than our analysis
here suggests.  For example, all of the core sizes would increase by a factor of $\sim 2$, while the
masses would decrease by a factor of $\sim 4$. Also, the shorter distance
would push all of the cores
below the thermal Jeans line in Figure~\ref{fig_m_r}, while the cores' concentrations
would remain unchanged.  Since L1622 cores represent a small fraction of the total
core population analyzed here, there would be minimal impact on our overall conclusions.

\section{CONCLUSION}
\label{sec_conc}
We have presented 
a first-look analysis of SCUBA-2 observations of the Orion B molecular cloud taken
as part of the JCMT Gould Belt Survey.  The improved sensitivity and larger detector size of
SCUBA-2 compared to SCUBA has allowed for significantly larger and more sensitive maps, 
with these SCUBA-2 observations reaching an rms of 3.7~mJy~bm$^{-1}$,
four to six times lower
than previous SCUBA observations.  Approximately 0.6, 2.1, and 1.7 square degrees were
mapped in L1622, \None, and \Ntwo, respectively.
In addition to the catalogues presented here, all of the 
reduced datasets analyzed in this paper (850~\mum and 450~\mum emission maps, the 
CO-subtracted 850~\mum map, and the 850~\micron-based FellWalker core footprint, 
along with maps of the variance per pixel, and
the external mask applied) are available at 
{\tt https://doi.org/10.11570/16.0003}.

We used the FellWalker algorithm to identify 915 dense cores within the 850~\mum map, and
analyzed their basic properties.  Protostellar dense cores are identified through association
with a protostar in the {\it Spitzer} \citep{Megeath12} or {\it Herschel} \citep{Stutz13}
catalogues.  Assuming a constant temperature of 20~K yields a starless core mass 
function similar to that derived in other studies, with the high-mass end following 
a roughly Salpeter slope.  Comparing the core masses and radii showed that most cores
have mean densities between $10^4$~cm$^{-3}$ and several $10^5$~cm$^{-3}$.  Dense cores with
masses above the thermal Jeans mass for the assumed temperature of 20~K tend to be
protostellar, although there are both starless cores and protostars on both sides
of this relationship.  A larger number of cores appear to be unstable when the
bounding pressure due to the weight of the overlying cloud material is accounted for.  
We measure a range of central concentrations for the dense cores
which tends to have larger values than previous SCUBA analyses \citep{Johnstone03,Johnstone06},
which we speculate is due to our deeper sensitivity.  At the highest mean densities,
the lifetimes of the starless and protostellar stages of dense cores appear to be fairly
similar, consistent with previous observations, while the least dense cores in our sample
may be longer-lived entities, if they are destined to form stars at all.  
Comparison of the distribution of dense cores we identified
to the overall cloud column densities in \None\ and \Ntwo\ measured by \citet{Lombardi14} shows
that at high column densities, above $1-2\times 10^{23}$~cm$^{-2}$, nearly all of the 
material is contained in the dense cores, while at lower cloud column densities, 
dense cores comprise a much smaller fraction of the material.  Based on the amount of
dense gas available, we predict that each of \None\ and \Ntwo\ will form at least as
many stars as are currently present, while L1622 has little dense material available to
supplement the present-day small protostellar population.  
We will present an in-depth analysis of the clustering properties of the dense cores
in Kirk et al (2015, in prep).

\appendix
\section{Dense Core Identification}
\label{app_cores}
Here, we describe our identification of dense cores, which was based on Starlink's
Fellwalker algorithm \citep{Berry15}.
The FellWalker algorithm is based on the idea of hiking through a set of hills: peaks are
defined as local maxima, with their extents based on the routes that a hiker could ascend 
to reach the top of each peak.  Users can set parameters in the algorithm such as the
minimum peak size, minimum dip between neighbouring peaks, minimum ascent slope, etc.
Prior to running FellWalker, we made a qualitative comparison between our 850~\mum maps
and the publicly available {\it Herschel} 500~\mum map \citep{Schneider13}.
This comparison revealed that surprisingly faint structures
in the 850~\mum map all have counterparts in the {\it Herschel} 500~\mum data. 
Given this correspondence,
we adopted less stringent FellWalker parameters than the default recommended values.
In particular, the parameters we modified from the default values are:
\begin{itemize} 
\item{ {\tt FellWalker.AllowEdge = 0} {\it (eliminate objects touching a map edge)} }
\item{ {\tt FellWalker.Noise = 0.5*RMS} {\it (extend object search deeper into the noise)} } 
\item{ {\tt FellWalker.MaxJump = 2.5} {\it (reduce area for identifying shared peaks)} }
\item{ {\tt FellWalker.MinPix = 5} {\it (allow smaller objects to be identified)} }
\item{ {\tt FellWalker.CleanIter = 5} {\it (tidy up jagged source edges)} }
\end{itemize}

We then ran the FellWalker source list through a second program to eliminate spurious sources,
which were numerous with the relaxed criteria above.
Since the noise at the edges of the maps and mosaics is larger than in the centre, a large
number of spurious sources were identified with FellWalker, which assumes a constant noise
level across the map.  We tested a variety of criteria to weed out spurious sources, and
found the following set of criteria to be the most effective at removing noise sources while
retaining real sources.  We removed: 1) sources smaller than the effective beamsize, 
2) sources which had fewer than three pixels above twice the local noise level, 
and 3) sources which, when slightly smoothed, had fewer than fifteen pixels above the local noise
level.  In addition, sources identified very near to the map edge (where the noise was highest)
were eliminated if they had fewer than 22 pixels above the local noise level, when the image
was slightly smoothed (i.e., 50\% more pixels than the third criterion).  All of the remaining
sources passed our visual inspection.  In general, the 3rd criterion eliminated the most sources.
The first criterion almost never rejected any sources since FellWalker itself eliminates
sources smaller than the beam.  FellWalker identified 260, 1383, and 1020 sources in L1622, 
\None, and \Ntwo\ respectively, which reduced to 29, 564, and 322 reliable cores after the
cuts described above.

\section{Comparison to SCUBA}
\label{app_scuba}
Parts of both the \None\ and \Ntwo\ regions were observed with the original
SCUBA instrument, and analyses of these observations which include independently-derived
core catalogues are given in 
\citet{Motte01}, \citet{Mitchell01}, \citet{Johnstone01}, \citet{Johnstone06},
\citet{Nutter07}, and \citet{DiFrancesco08}.  Both 
\citet{Nutter07} (hereafter NWT07) and the SCUBA Legacy Catalog of \citet{DiFrancesco08} 
(hereafter SLC) 
include all observations taken during SCUBA's
operation of Orion~B, and so we use these two works to compare the sensitivity of SCUBA
and SCUBA-2 in Orion~B\footnote{For completeness, we note that one single small `jiggle
map' (pointed observation) was taken in the L1622 region, but given the small quantity of data,
we do not make comparisons in this region.}. 

\begin{figure*}[htb!]
\begin{tabular}{cc}
\includegraphics[width=3.0in]{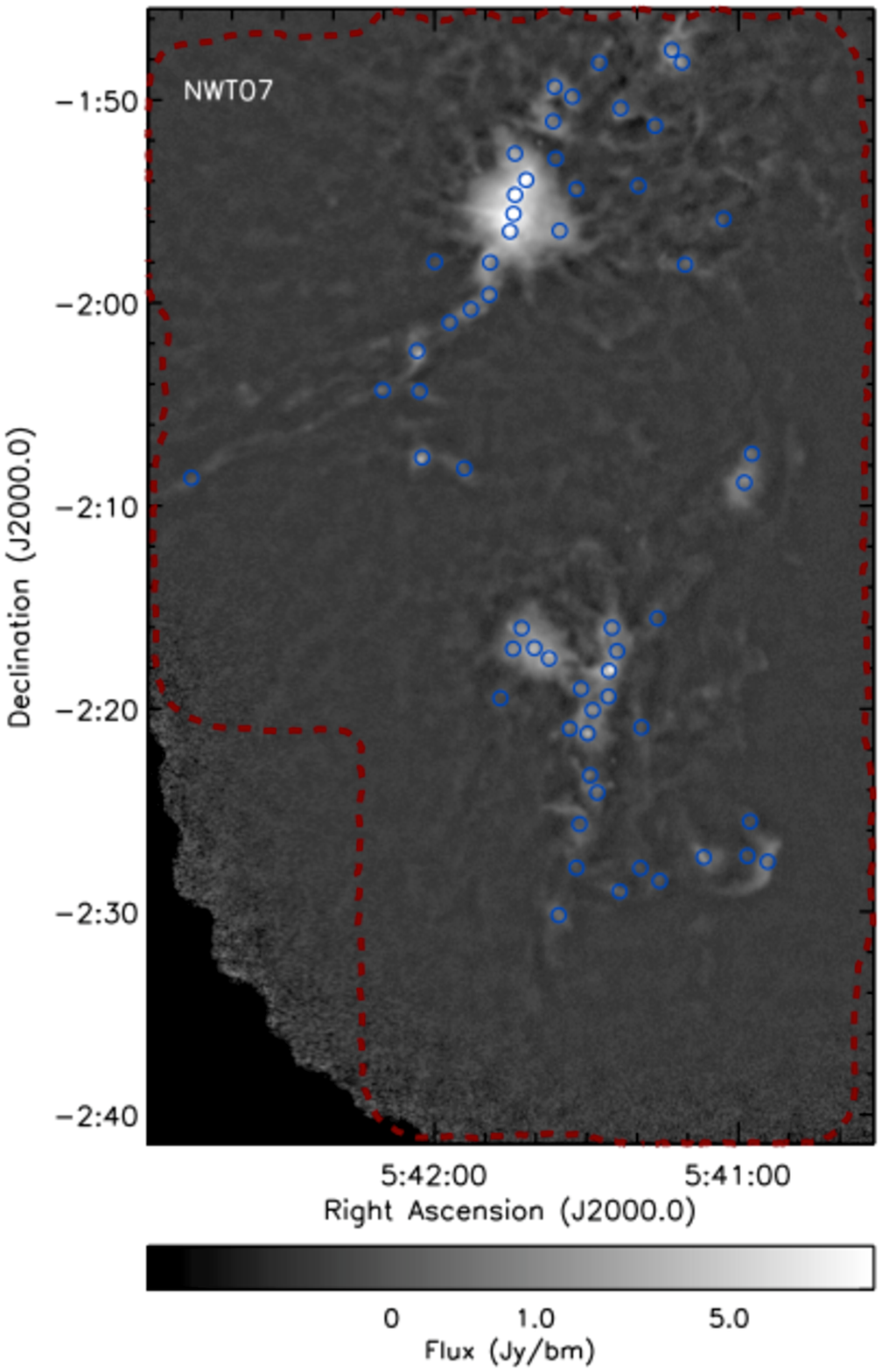} &
\includegraphics[width=3.0in]{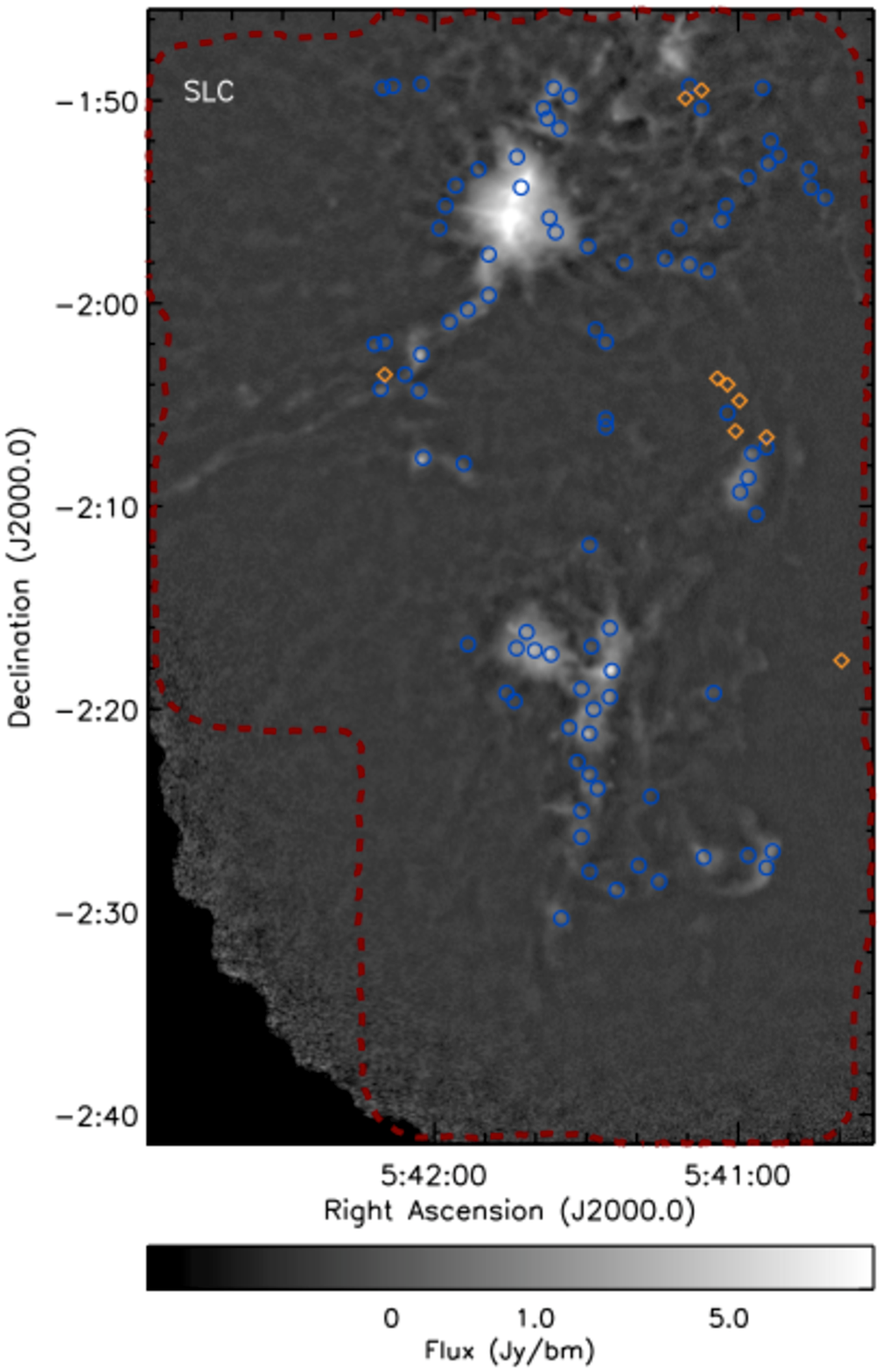} \\
\end{tabular}
\caption{A comparison of cores identified with SCUBA and the present SCUBA-2
	850~\mum map in \None.  The greyscale in both panels shows the SCUBA-2 
	observations;
	the red dashed line indicates the area observed with SCUBA.
	The left panel shows the peak positions of NWT07 cores, while
	the right panel shows the SLC cores. 
	In both panels, blue circles indicates cores which had a match in our
	SCUBA-2 catalogue, while orange diamonds 
	indicate cores with no match in our
	catalogue (i.e., peak position outside of all SCUBA-2 FellWalker core footprints).
	All of the NWT07 cores had a match in \None.
}
\label{fig_compl_im_2023}
\end{figure*}

\begin{figure*}[htb!]
\begin{tabular}{cc}
\includegraphics[width=3.0in]{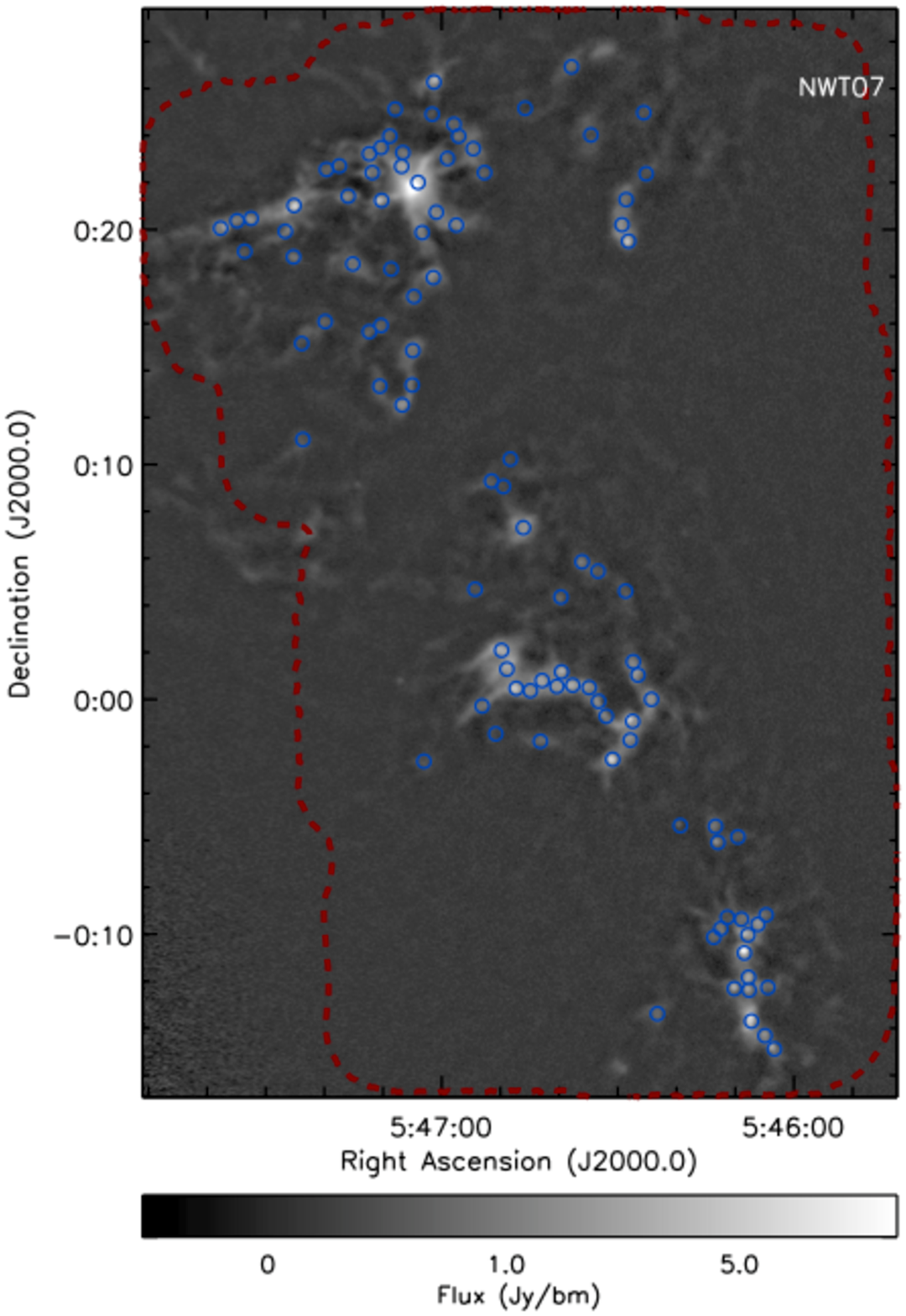} &
\includegraphics[width=3.0in]{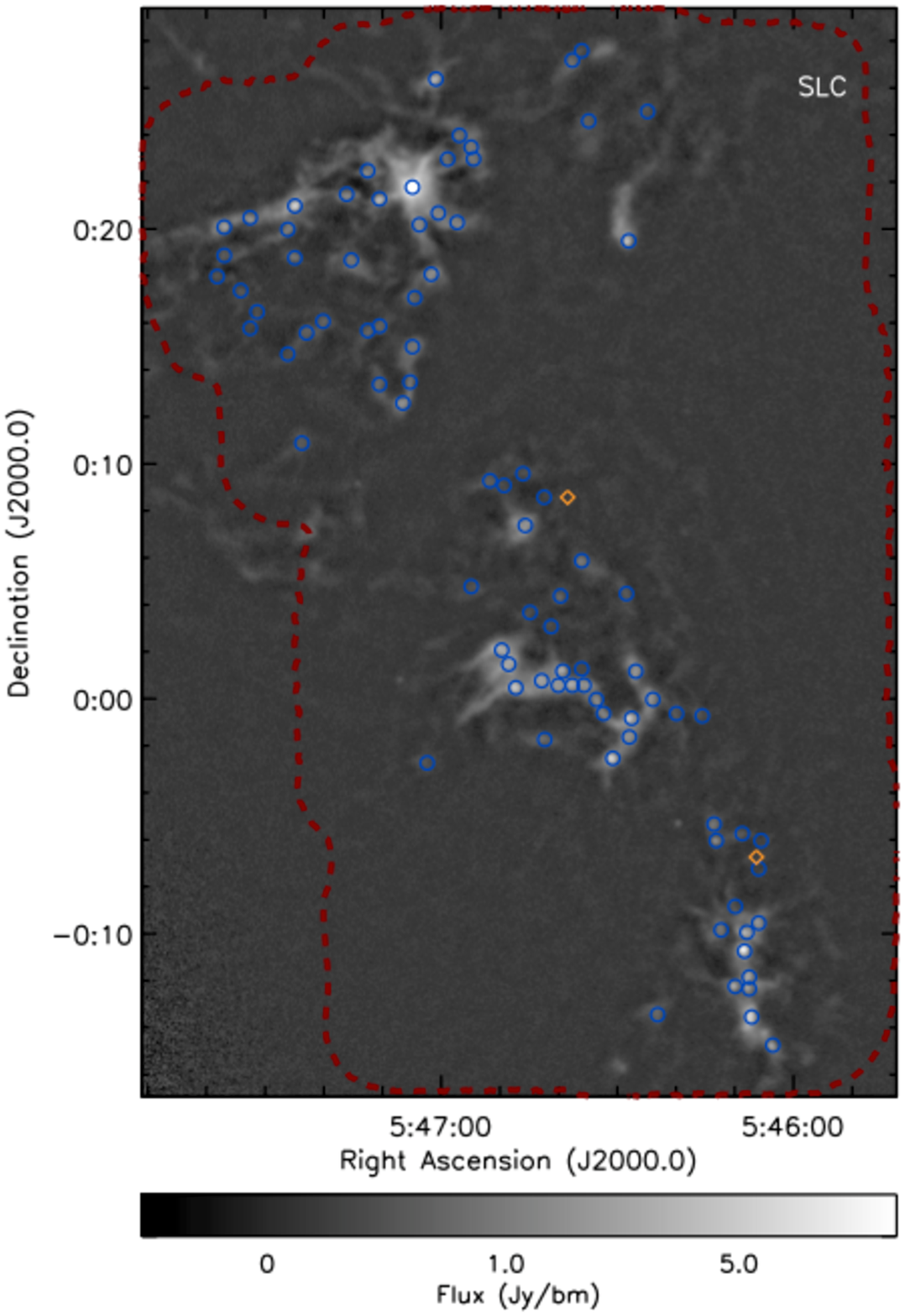} \\
\end{tabular}
\caption{A comparison of cores identified with SCUBA and the present SCUBA-2
	850~\mum map in \Ntwo.  See Figure~\ref{fig_compl_im_2023} for the 
	plotting conventions used.}
\label{fig_compl_im_2068}
\end{figure*}

Figures~\ref{fig_compl_im_2023} and \ref{fig_compl_im_2068} show the SCUBA-2 850~\mum 
images of the portions of \None\ and \Ntwo\ that were covered by SCUBA,
based on
the `coverage maps' from the SLC\footnote{ 
{\tt http://www4.cadc-ccda.hia-iha.nrc-cnrc.gc.ca/\\
community/scubalegacy/}}.  The left panels
of Figures~\ref{fig_compl_im_2023} and \ref{fig_compl_im_2068} show the cores 
listed in NWT07, while the
right panels show the SLC cores.
Since both \None\ and \Ntwo\ have clustered and complex emission, we expect  
differences to arise in core boundaries and the level of fragmentation.
To compare the core catalogues quantitatively, we associate the peak position of each 
SCUBA core with the FellWalker core whose boundary it lies within. 
Indeed, Figures~\ref{fig_compl_im_2023} and \ref{fig_compl_im_2068} show that
all of the cores in NWT07 have a match in our SCUBA-2 core catalogue,
while several of the SLC cores do not have a match.
Table~\ref{tab_core_match} shows the number of cores matched for each
catalogue.  Comparing the number of cores in the two SCUBA-based catalogues which
do and do not have a match in the SCUBA-2 catalogue suggests that the SLC probes deeper
than NWT07 in \None\ while the reverse is the case in \Ntwo.  In both cases, the
SLC is more susceptible to falsely identifying cores.  The SCUBA-2 catalogue
includes several times more cores than seen with SCUBA over a comparable area: roughly 300
additional cores were found in the SCUBA-2 map of the part of \None\ observed with 
SCUBA (compared to 60-90 cores identified with SCUBA), while roughly 150 additional cores
were found in \Ntwo\ (compared to 90-100 identified with SCUBA).

\input{tab3}


\begin{figure*}[htbp]
\begin{tabular}{cc}
\includegraphics[width=2.8in]{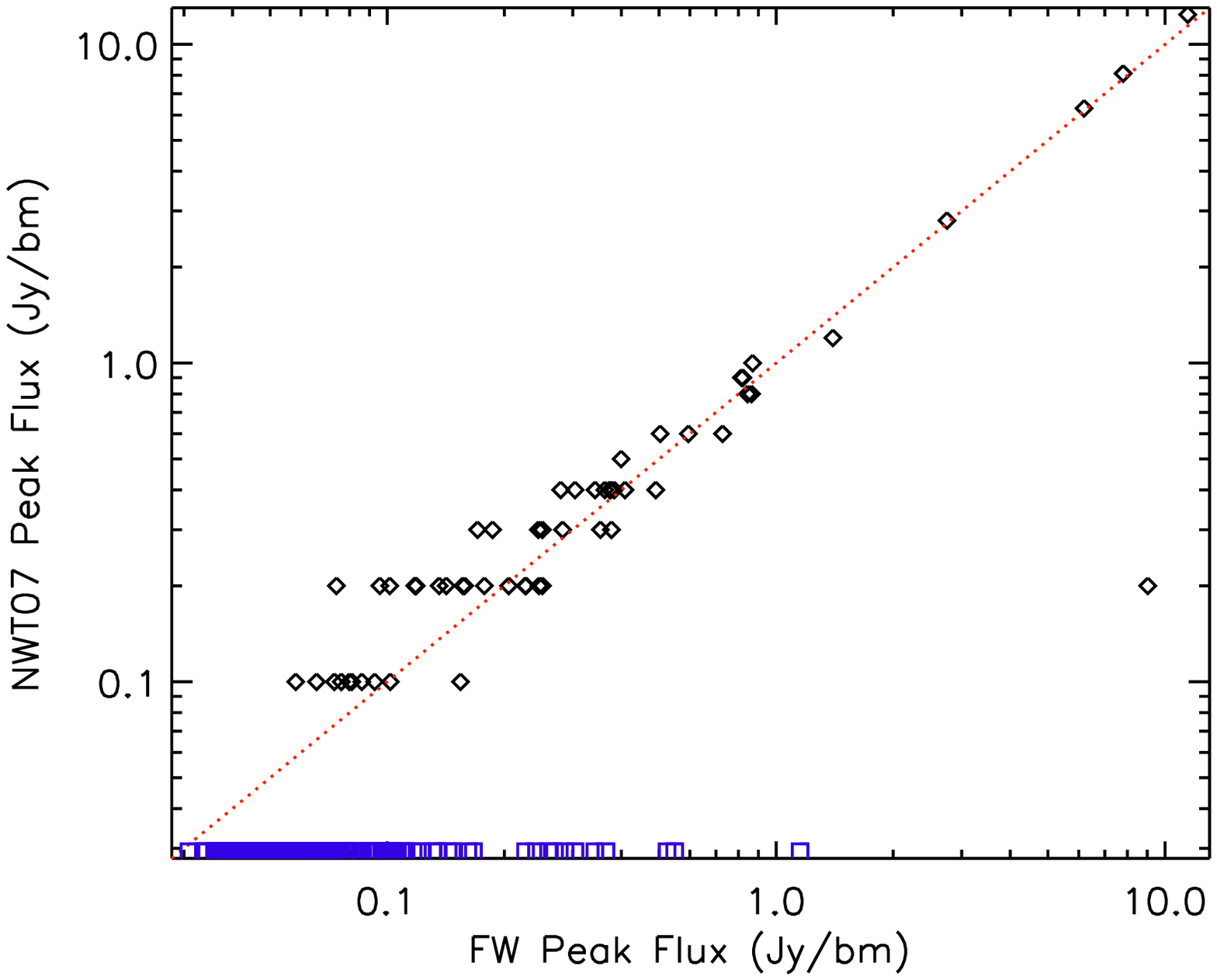} &
\includegraphics[width=2.8in]{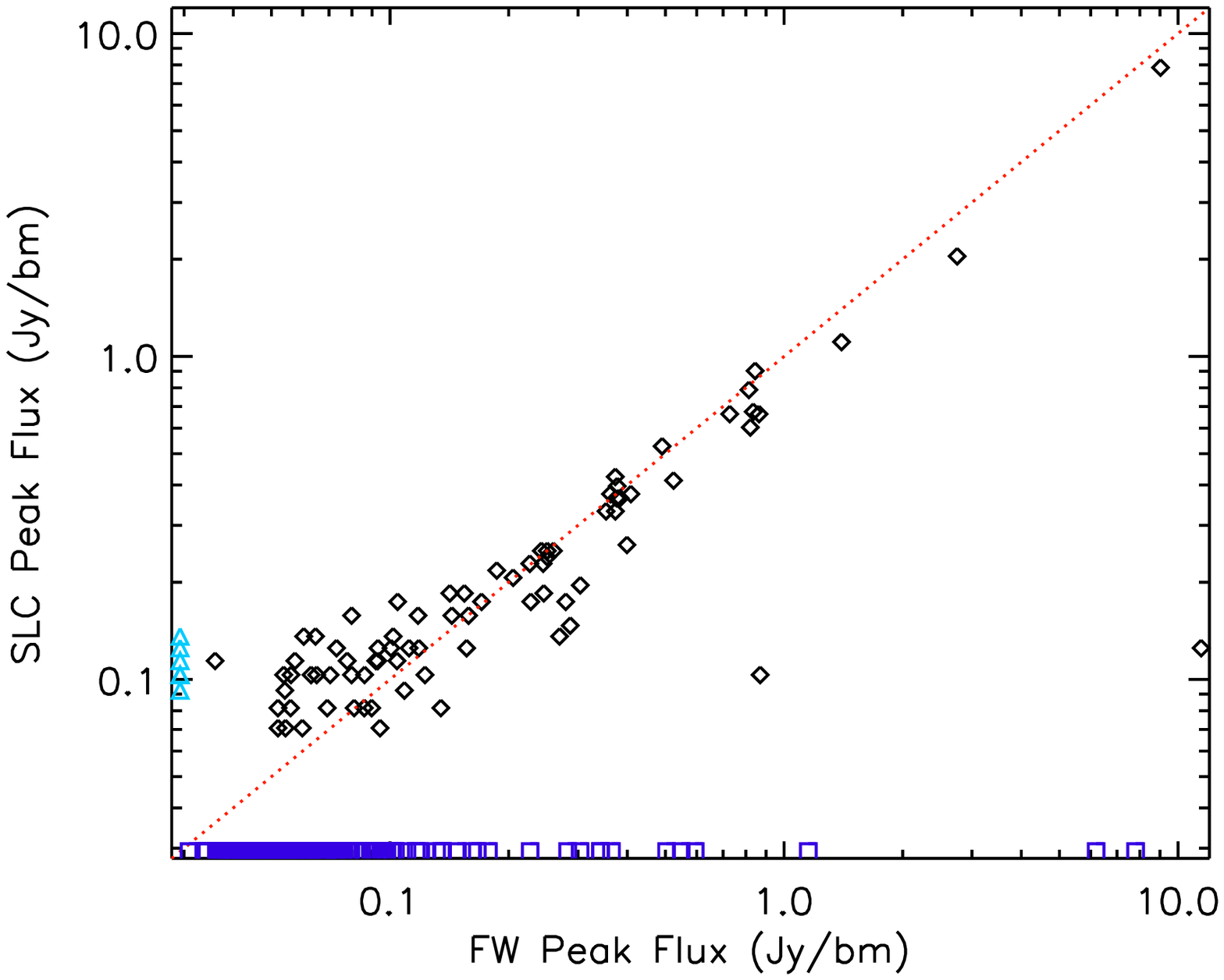} \\
\includegraphics[width=2.8in]{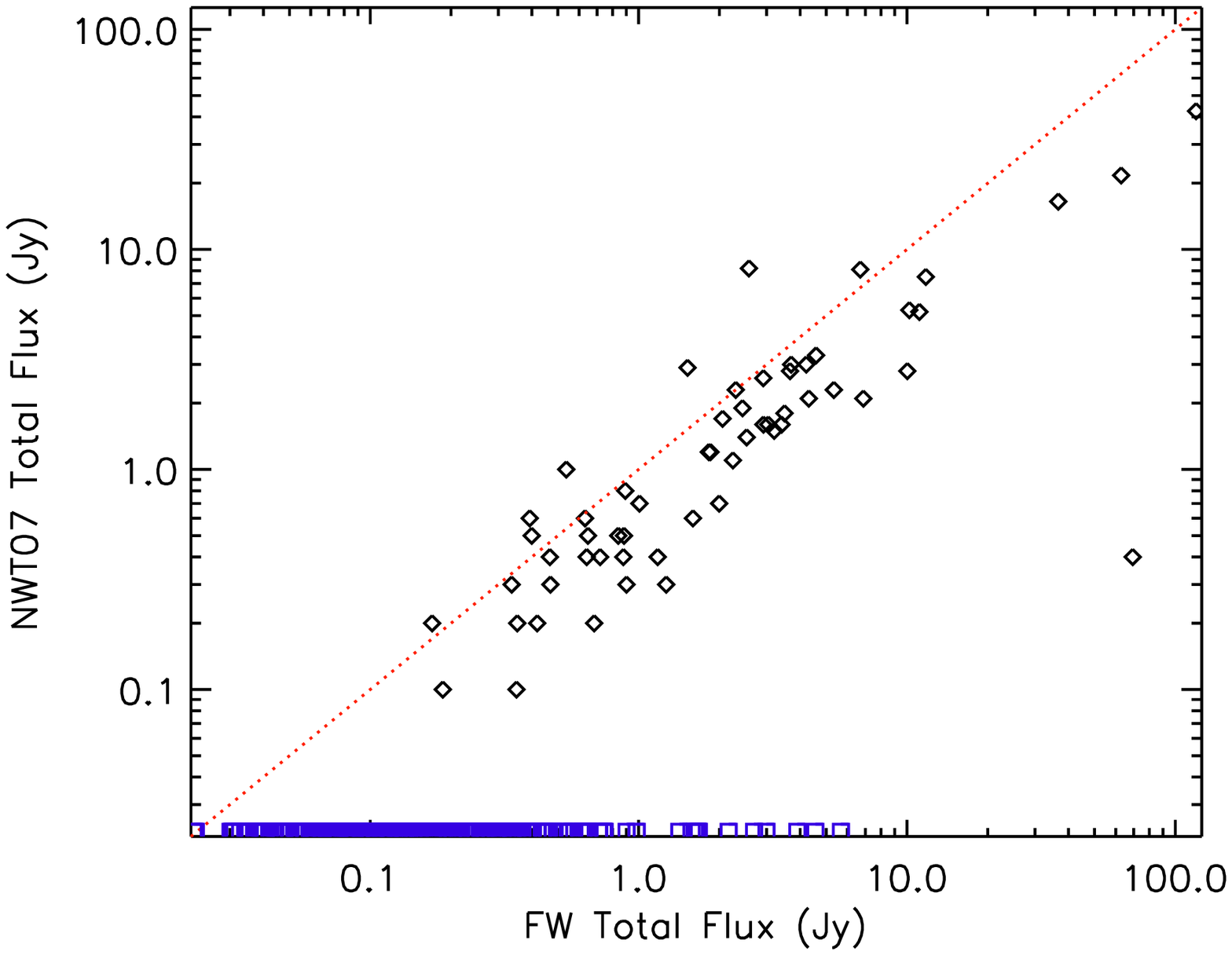} &
\includegraphics[width=2.8in]{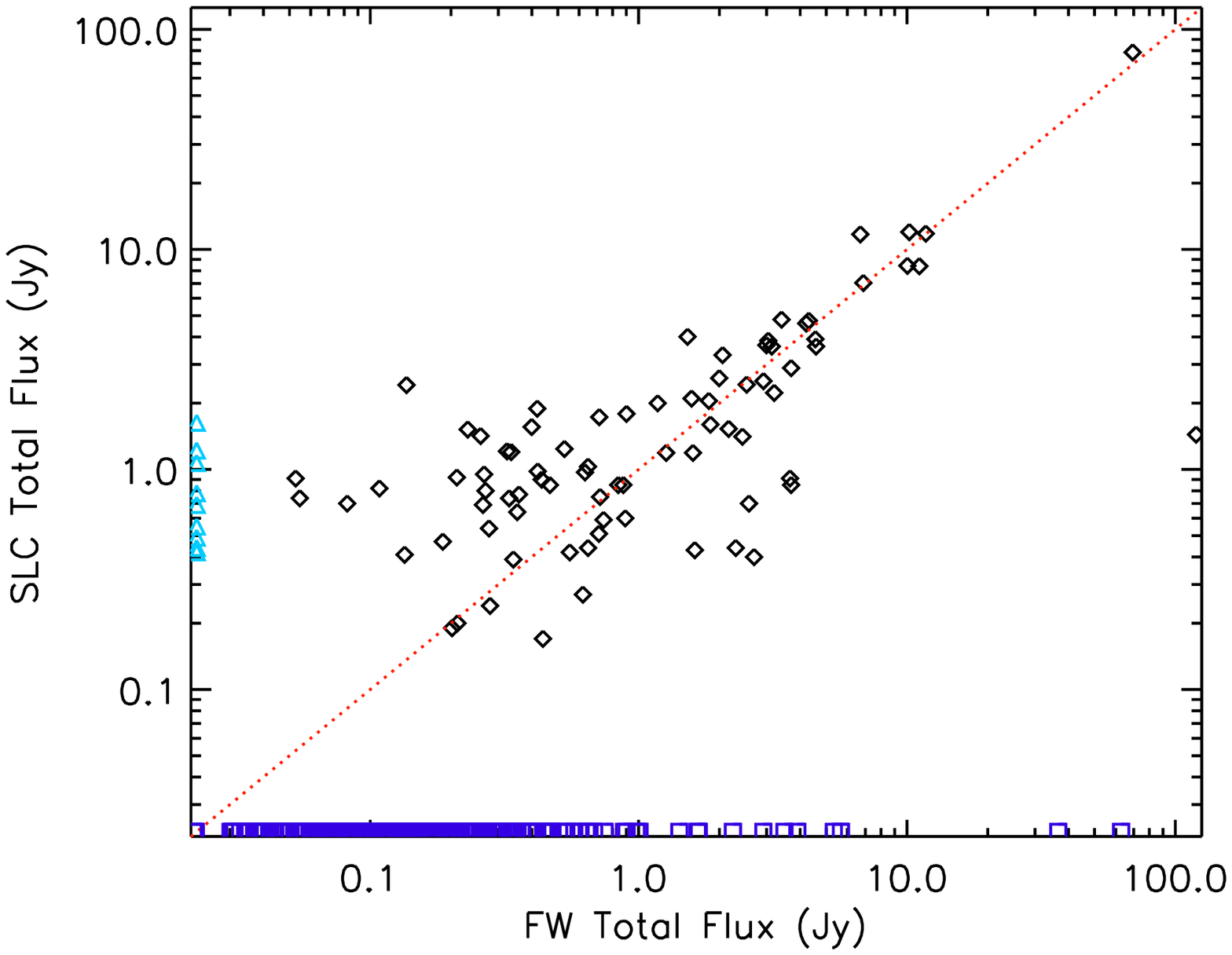} \\
\includegraphics[width=2.8in]{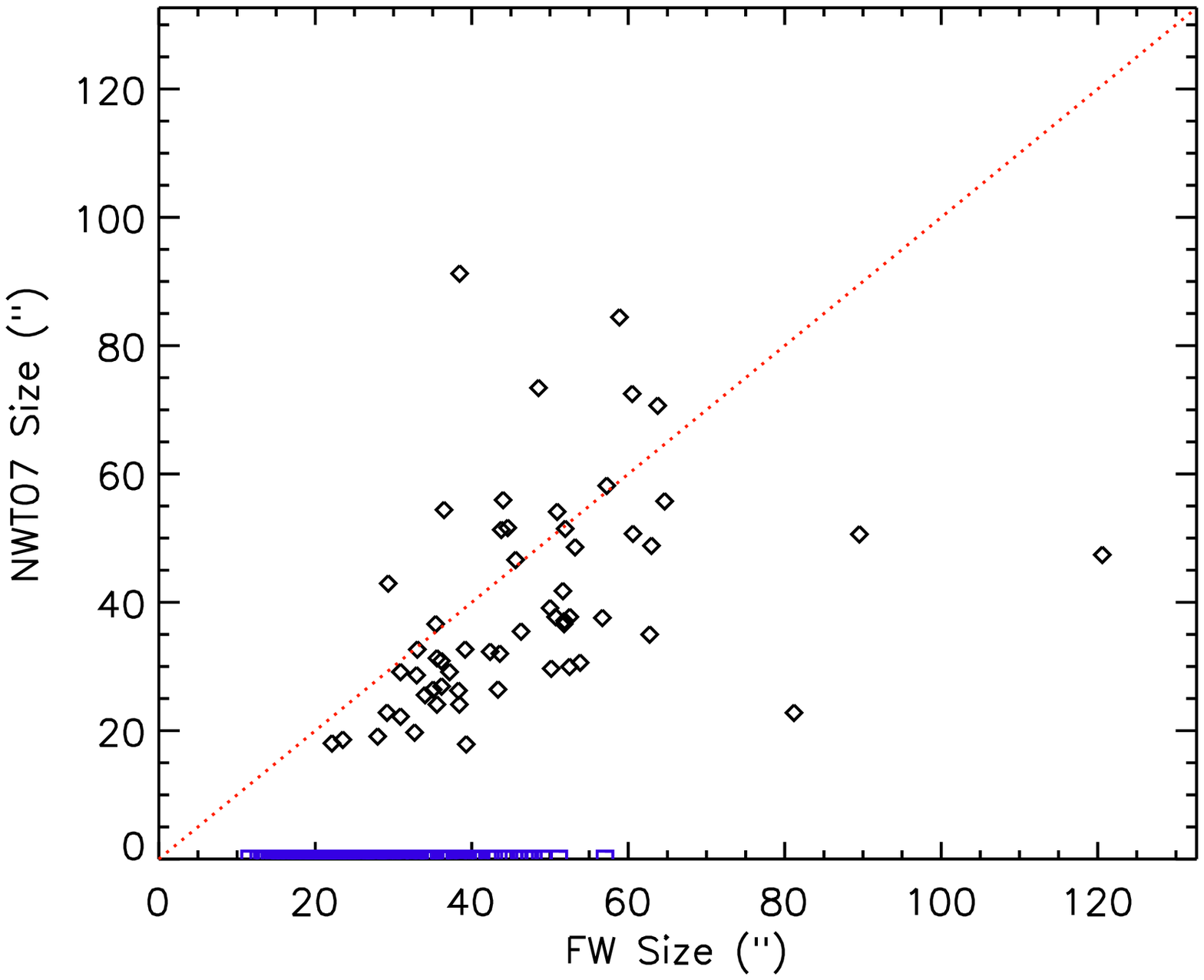} &
\includegraphics[width=2.8in]{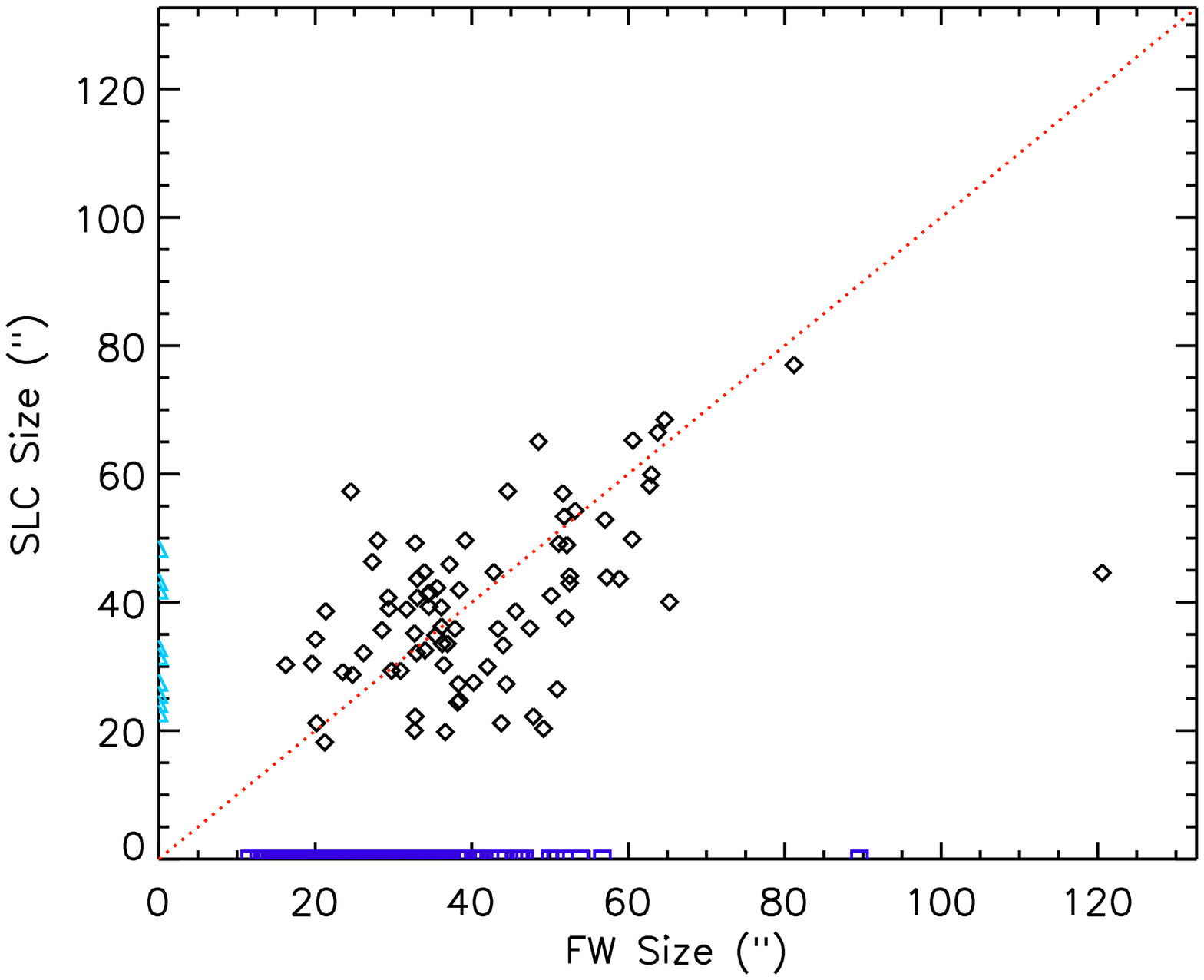} \\
\end{tabular}
\caption{A comparison of cores identified in \None\ with SCUBA-2 data and
	SCUBA data by NWT07 (left) and the SLC (right).
	In each plot, the SCUBA-2 FellWalker core 
	catalogue value is given on the horizontal axis, while the SCUBA catalogue value
	is given on the vertical axis.
	The top row shows the peak flux, while the middle row shows the total
	flux, and the bottom row shows the size (radius).  A radius of 60\arcsec\
	corresponds to 0.12~pc at the distance of Orion~B.
	Black diamonds show cores which had a match in both catalogues, while dark blue
	squares show cores only identified in the SCUBA-2 data and light blue triangles
	show cores only identified in the SCUBA data.
	The red dotted line indicates a one-to-one relationship.}
\label{fig_compl_flux_2023}
\end{figure*}

\begin{figure*}[htbp]
\begin{tabular}{cc}
\includegraphics[width=3in]{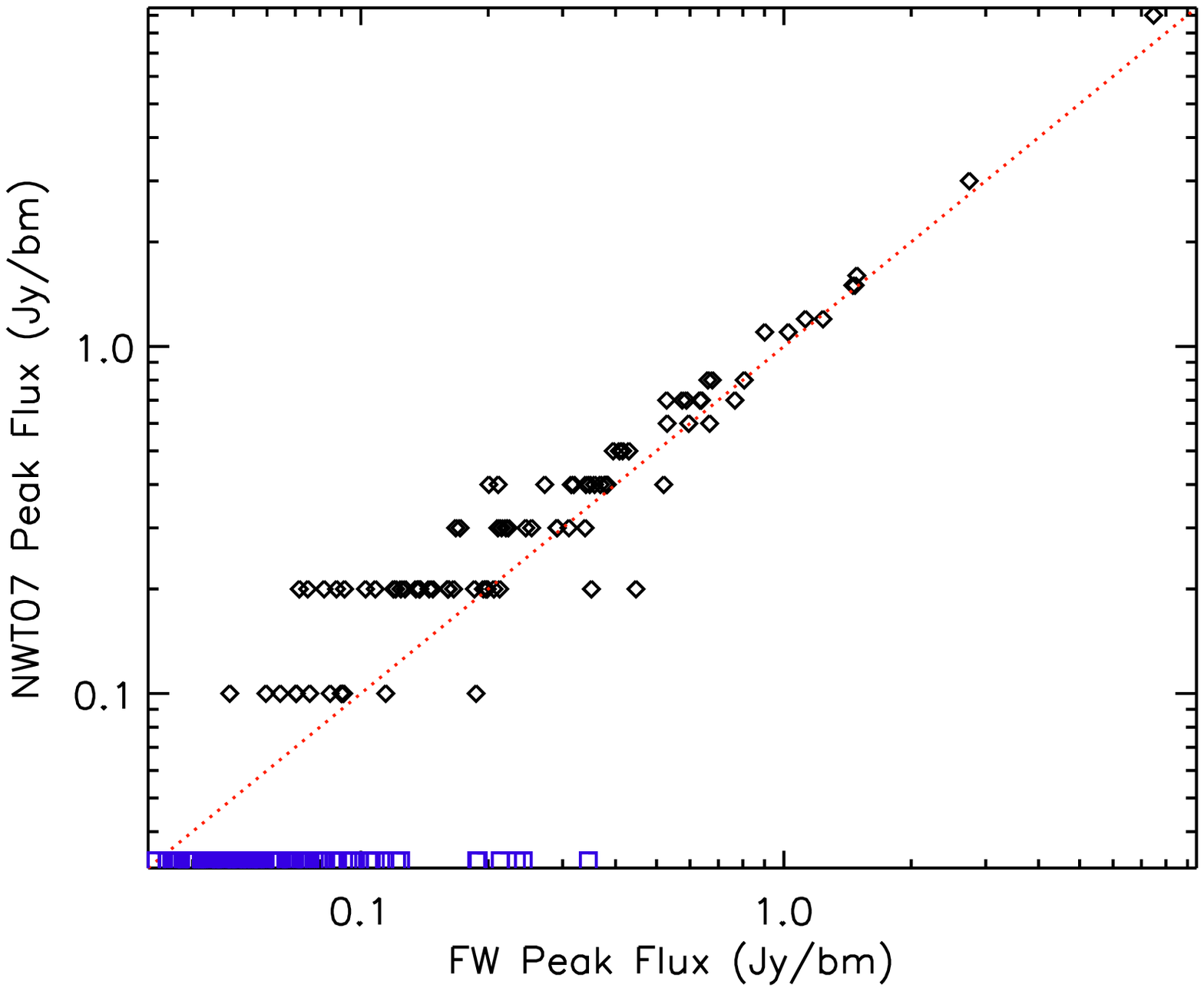} &
\includegraphics[width=3in]{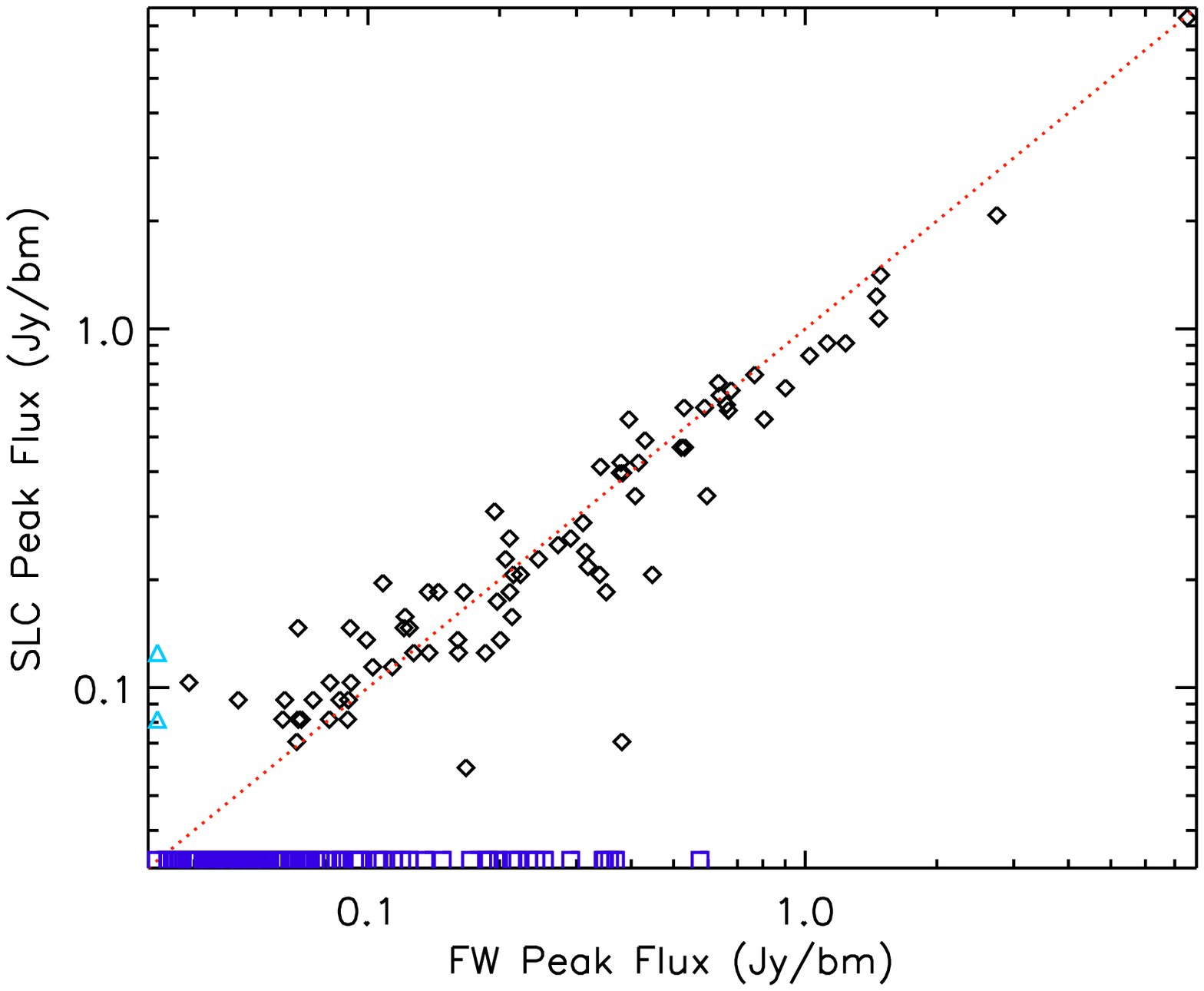} \\
\includegraphics[width=3in]{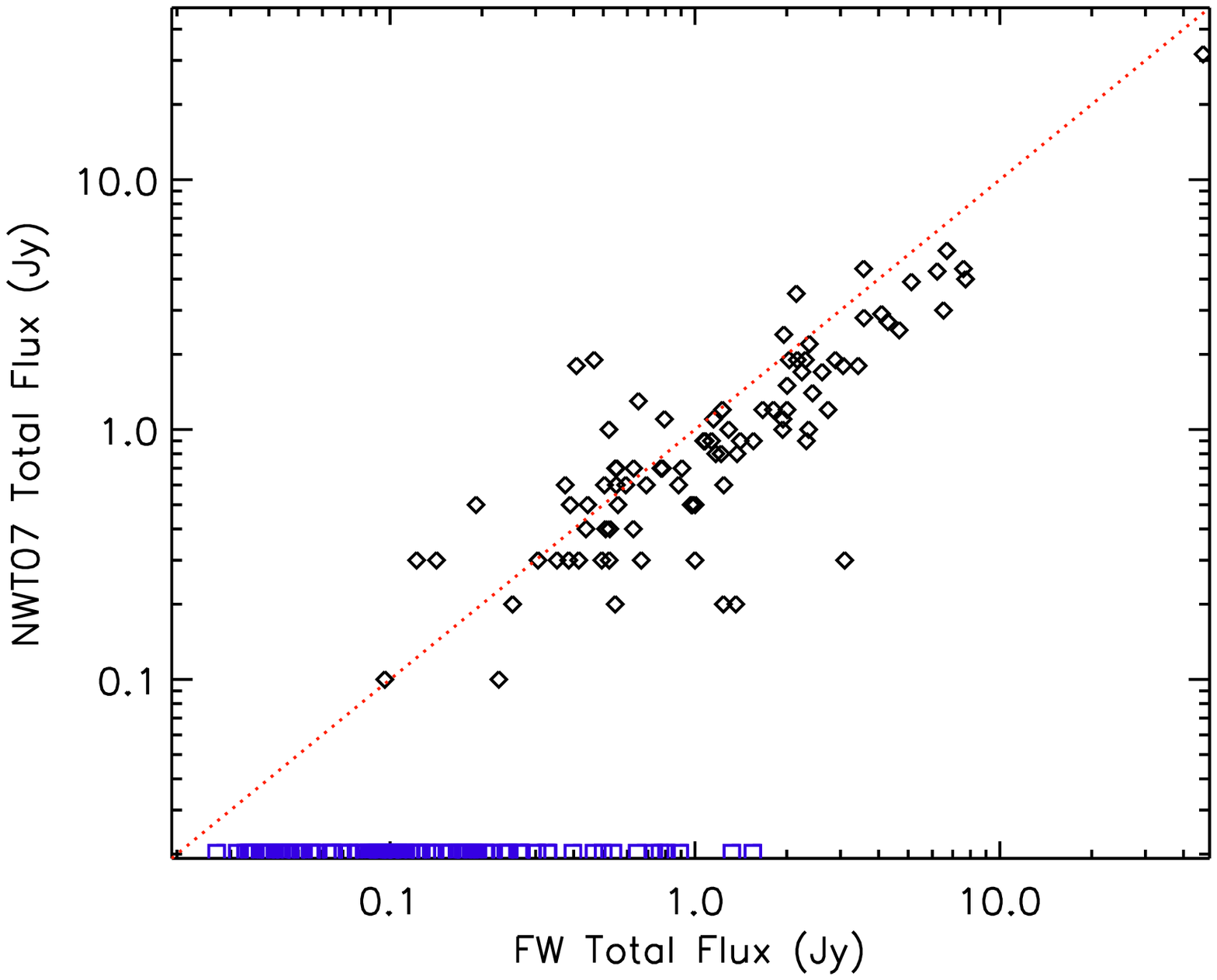} &
\includegraphics[width=3in]{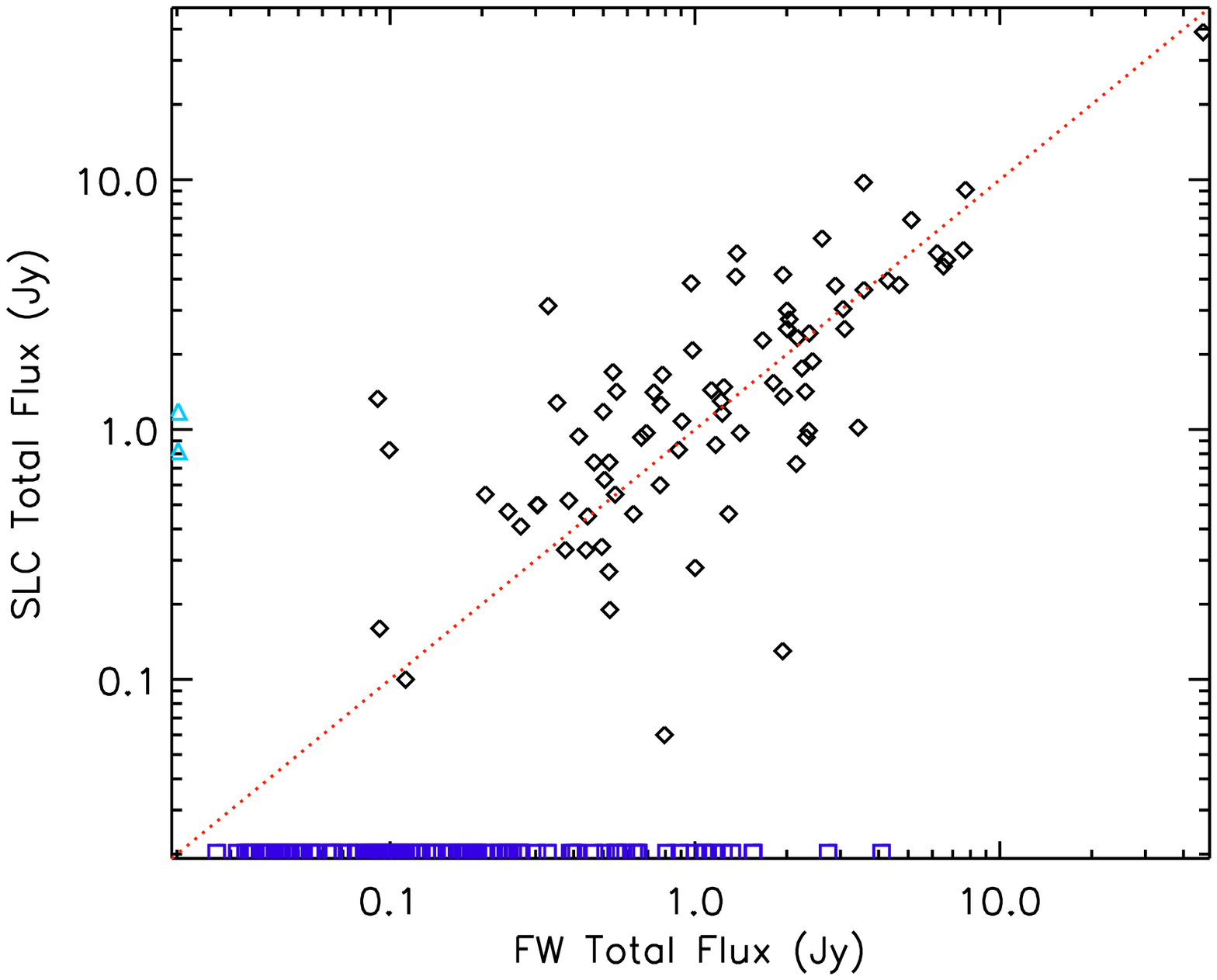} \\
\includegraphics[width=3in]{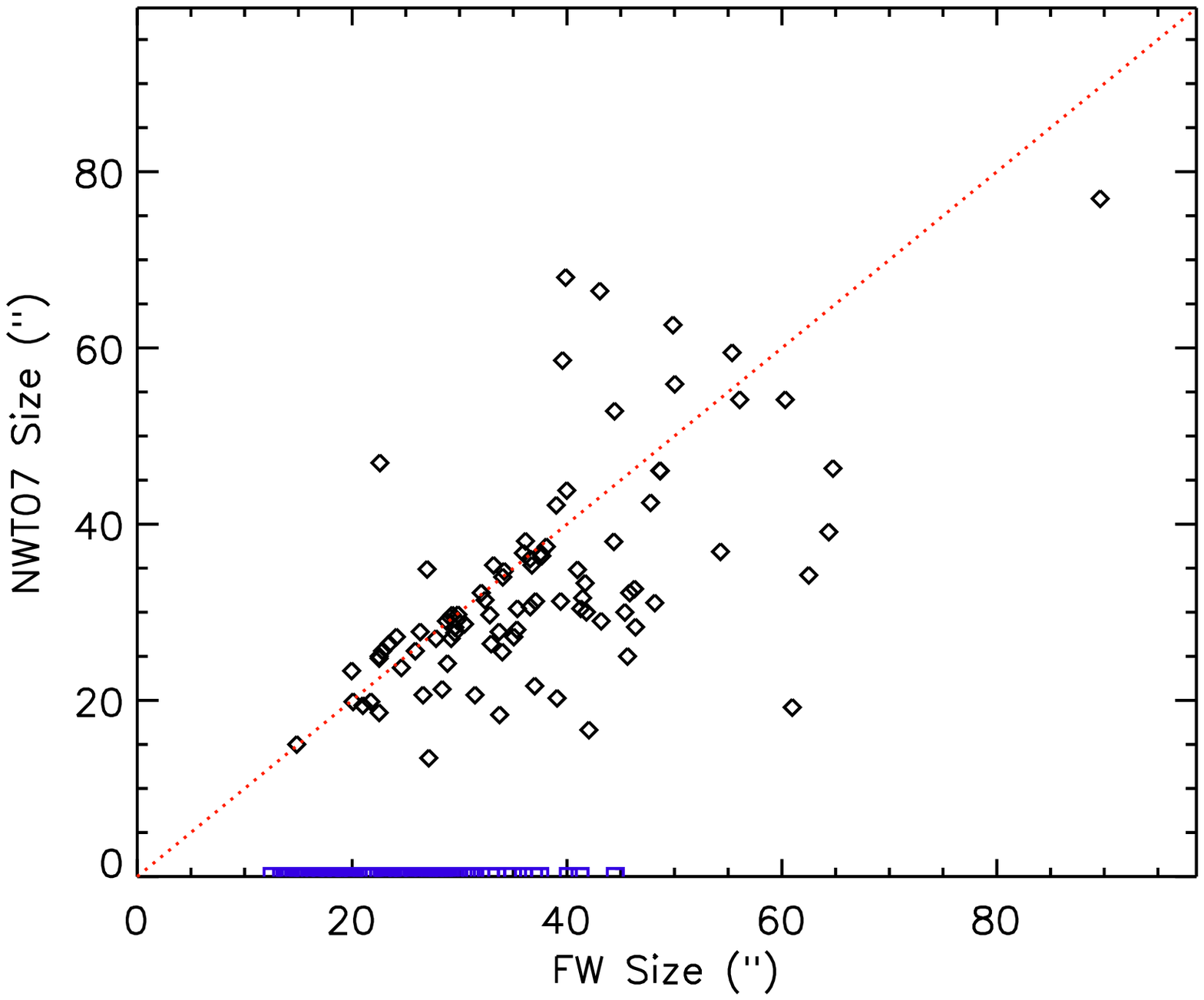} &
\includegraphics[width=3in]{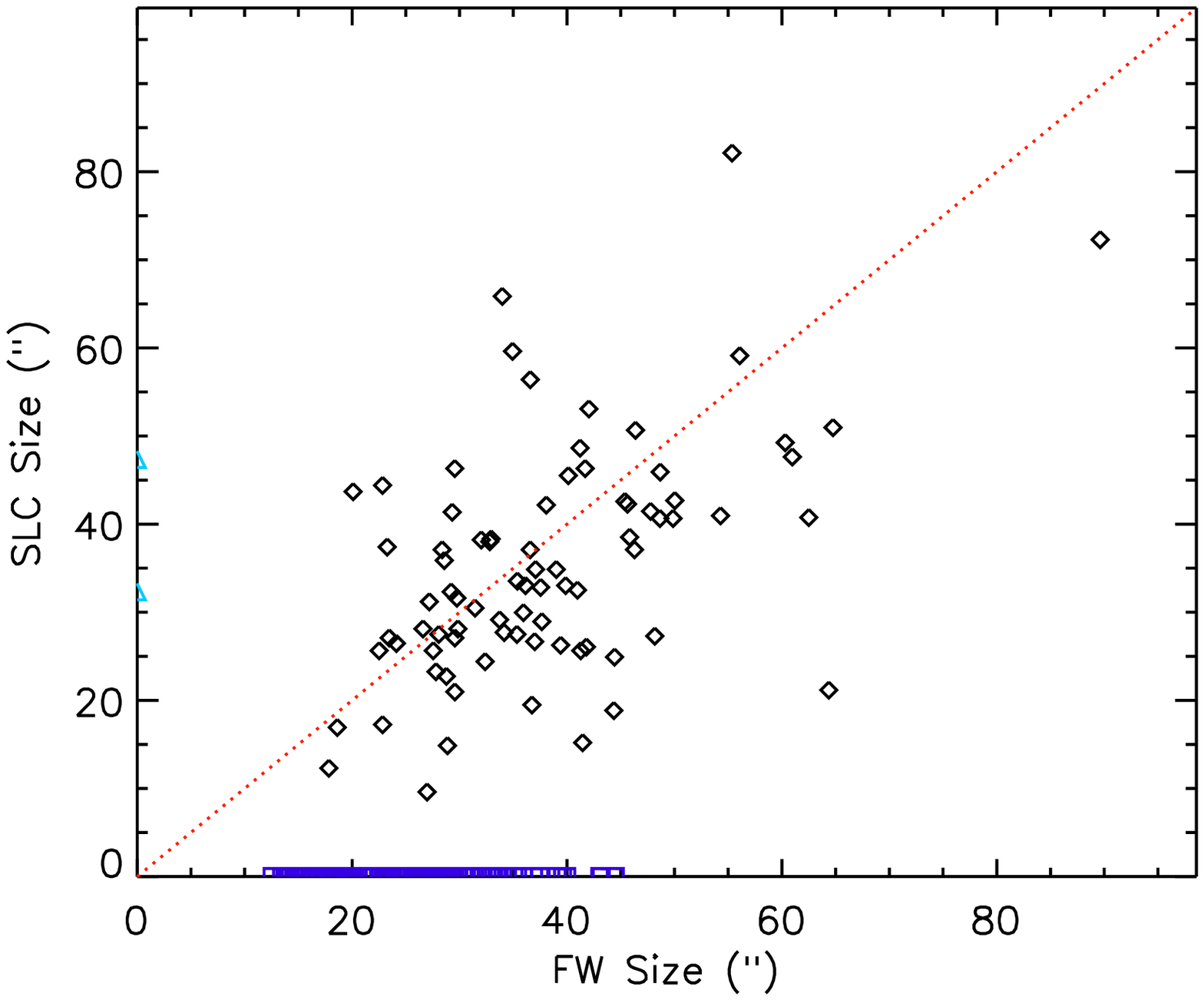} \\
\end{tabular}
\caption{A comparison of cores identified in \Ntwo\ with SCUBA-2 data and
	SCUBA data by NWT07 (left) and the SLC (right).
	See Figure~\ref{fig_compl_flux_2023} for
	the plotting conventions used.}
\label{fig_compl_flux_2068}
\end{figure*}

Figures~\ref{fig_compl_flux_2023} and \ref{fig_compl_flux_2068} show 
comparisons of the peak fluxes, total fluxes, and
sizes measured for cores in \None\ and \Ntwo\, respectively.  
The SCUBA Legacy maps were created with a resolution of 19.8\arcsec 
\citep{DiFrancesco08}, degraded from the nominal best value to better handle noise
features in their processing of multiple datasets of differing quality.  We roughly
correct for differences expected in the SLC values caused by this lower resolution:
we decrease the SLC peak flux by the ratio of beam sizes, and deconvolve the SLC
radius with the 19.8\arcsec\ beam, and reconvolve it with the SCUBA-2 14.6\arcsec\ beam
(given typical core sizes, the radius correction tends to be very small).  The 
NWT07 catalogue had a similar effective radius to our SCUBA-2 map, so
corrections are not needed for that comparison.
Most of the core catalogue values agree reasonably 
well between the two SCUBA measurements and the SCUBA-2 measurement, although there
is significant scatter around a perfect one-to-one relationship.  We expect the peak
flux to show the best agreement, and indeed that is generally true.  Note that the 
scatter in the comparison with the NWT07 peak fluxes is likely due to the 
lower precision of their published catalogue values, which were only given in tenths of
a Jy~bm$^{-1}$.  Since generally cores become more
numerous at lower flux levels (when above the detection limit), we expect that most of the 
cores shown at each of the several tenths of a Jy~bm$^{-1}$ level in peak flux would in 
fact have slightly lower true values, lowering the apparent scatter.  The total flux
and size plots tend to show a larger scatter between the SCUBA and SCUBA-2 values,
as these measures are more sensitive to precisely how the dense core boundaries are
defined, which tends to vary more between core identification methods in regions of
complex emission.
We do not see any indication of systematic calibration issues between the
SCUBA and SCUBA-2 maps: the median ratio of peak or total flux in SCUBA and SCUBA-2
cores is not consistently higher or lower than one across both regions in the
NWT07 or SLC catalogues.

Barring the problem with the lower precision in the peak flux values given in
NWT07, we find generally good agreement between the measured core properties 
from SCUBA-2 and those in NWT07 and the SLC.  The
SCUBA Legacy processed images have some negative features (``bowls'') around bright
cores.  Also, the \None\ and \Ntwo\ regions happened to be split between
two `tiles' in the SLC dataset, which appears to have caused a particularly poor
automated reduction of these regions.  The resulting large scale
bowls and pedestals make the resulting core catalogue less accurate.  
This difference likely acounts both for the large scatter away from the
one-to-one relationship in the total flux comparison, as well the presence of SLC cores
which have no match in our SCUBA-2 catalogue.  Visual inspection of the SLC cores
which are unmatched show they are either caused by noise spikes at map edges,
or are small-scale noise features coincident with a large-scale pedestal that raised
them above the global flux cutoff used.


\begin{figure*}[htb]
\begin{tabular}{cc}
\includegraphics[width=3in]{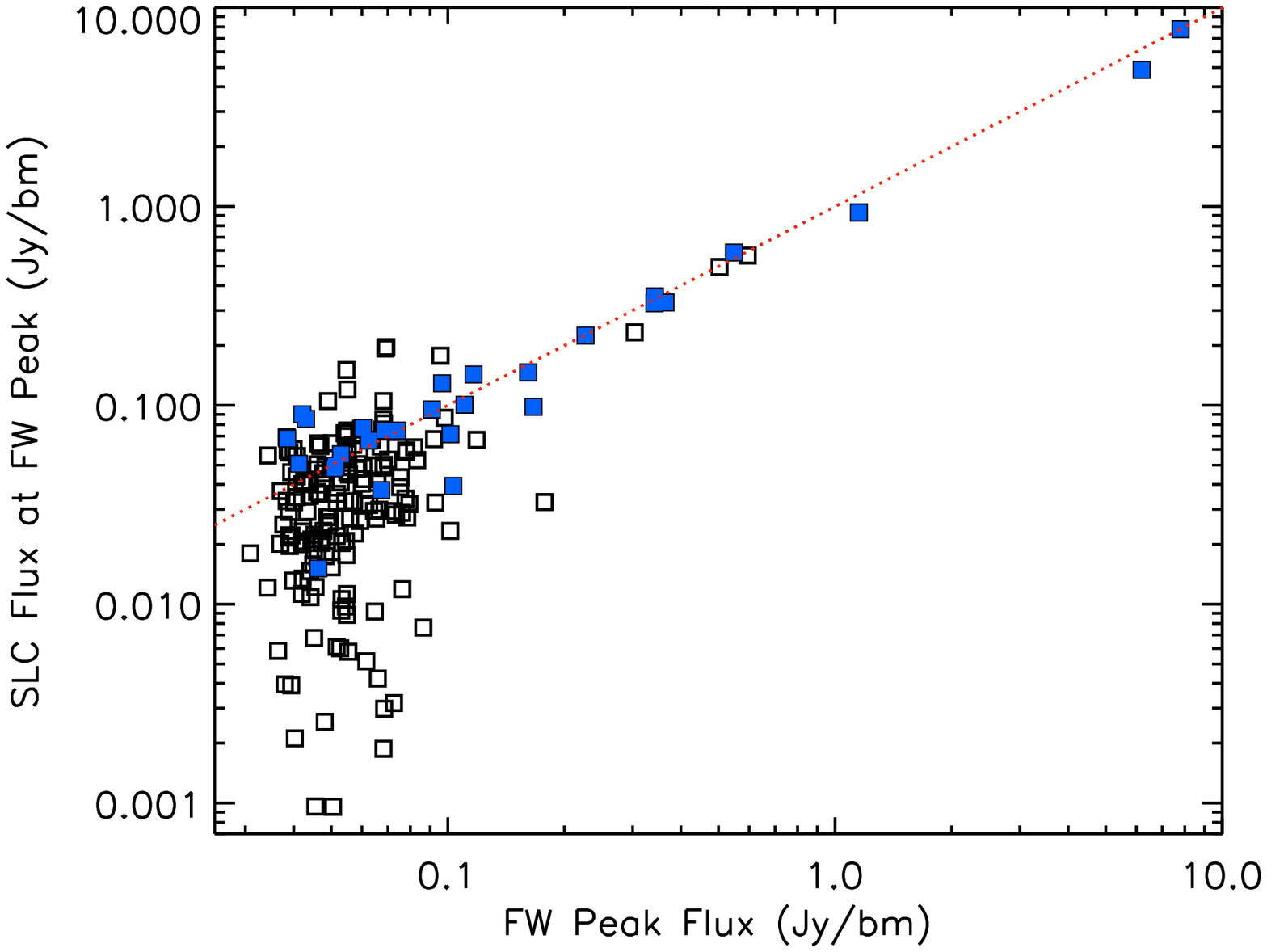} & 
\includegraphics[width=3in]{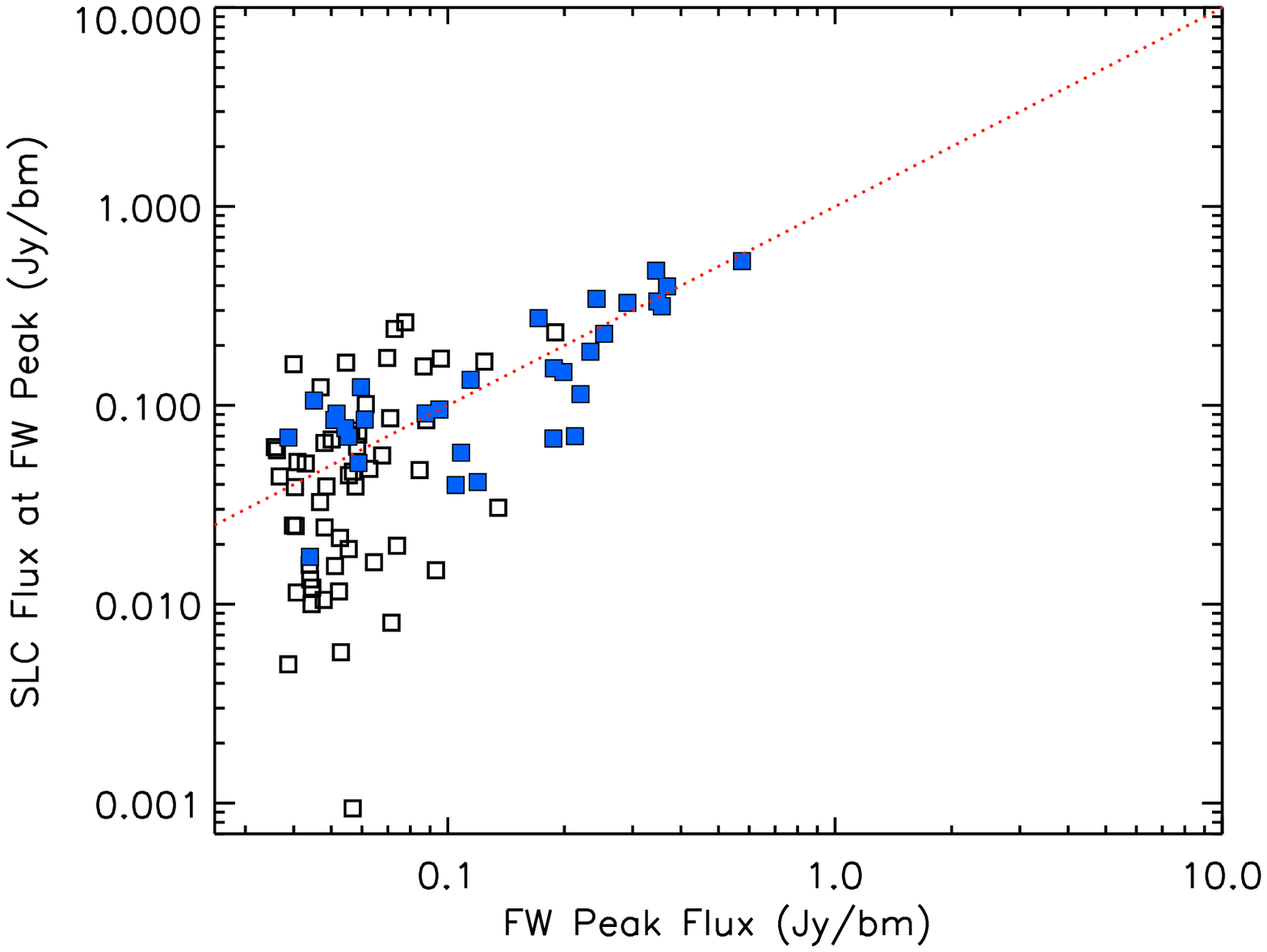} \\ 
\end{tabular}
\caption{A comparison of flux values for FellWalker-identified cores not explicitly matched
	to an SLC entry.  The horizontal axis shows the peak flux measured in the SCUBA-2
	map for each un-matched FellWalker core.  The vertical axis shows the flux in the
	SCUBA Legacy map at the location of the FellWalker core peak, while the red dotted line
	shows a one-to-one relationship.  The filled blue squares show pixels in the SCUBA
	Legacy map which are associated with an SLC source (i.e., the FellWalker core is
	classified as `unmatched' due to differences in core boundaries used by the two
	algorithms), while the black empty squares show pixels in the SCUBA Legacy map which
	are not associated with an SLC source.  See text for details.}
\label{fig_FW_unmatched}
\end{figure*}

For our comparison with the SLC, since the 850~\mum maps and source boundaries are 
publicly available, we can explicitly also test the cause of the un-matched FellWalker
sources.  In Figure~\ref{fig_FW_unmatched}, we show the 
SCUBA Legacy flux at the position of each unmatched FellWalker core peak position.
As can be seen in the figure, there is generally a good correspondence between the 
FellWalker core peak flux and the SCUBA Legacy flux at the same location.
We additionally find that the majority of the brighter unmatched
FellWalker cores have peak positions which lie within an SLC source boundary.
In these cases, it is clear that the reason that the FellWalker core was unmatched is 
differences in source boundaries in regions of complex emission: the SLC source which
the FellWalker core peak lies within was matched to a different FellWalker core.
We also examined the brightest
unmatched FellWalker cores whose peak positions did not lie within any SLC source boundaries,
and in all cases we found that there were sources visible in the SCUBA Legacy map that
were excluded from the final SCUBA Legacy catalogue.  There were two causes for this exclusion.
First, there were several bright emission peaks lying very close to a SCUBA Legacy map
edge (these correspond to the brightest few open squares in each panel of 
Figure~\ref{fig_FW_unmatched}).  
Since many SCUBA Legacy maps suffer from significant edge artefacts, the SLC 
used strong criteria to eliminate potentially spurious objects identified near a map edge, 
and evidently on
occasion those criteria also removed several real sources from their catalogue.  Second,
some peaks of emission were located within strong negative bowls in the SCUBA
Legacy map: while the peaks themselves were bright enough to be easily discernable, their
extents were truncated by the surrounding negative bowl to such an extent that these 
sources were likely eliminated by minimum size criteria imposed on the SLC.  The apparent
presence of bright unmatched FellWalker cores in Figure~\ref{fig_compl_flux_2023} and 
\ref{fig_compl_flux_2068} are therefore not indicative of major inconsistencies between
the SCUBA Legacy map and the SCUBA-2 map.

\begin{figure*}[htbp]
\begin{tabular}{cc}
\includegraphics[width=3in]{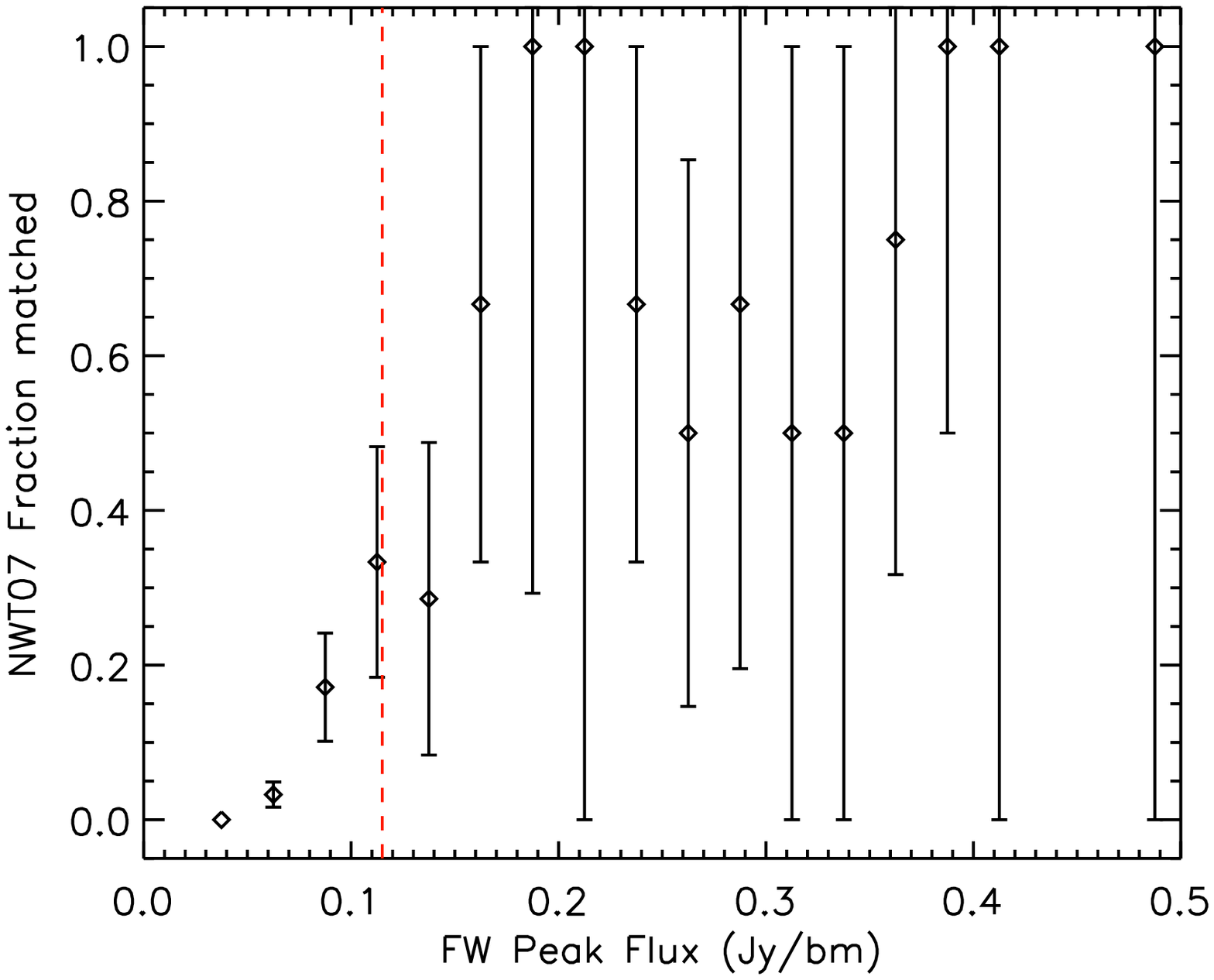} &
\includegraphics[width=3in]{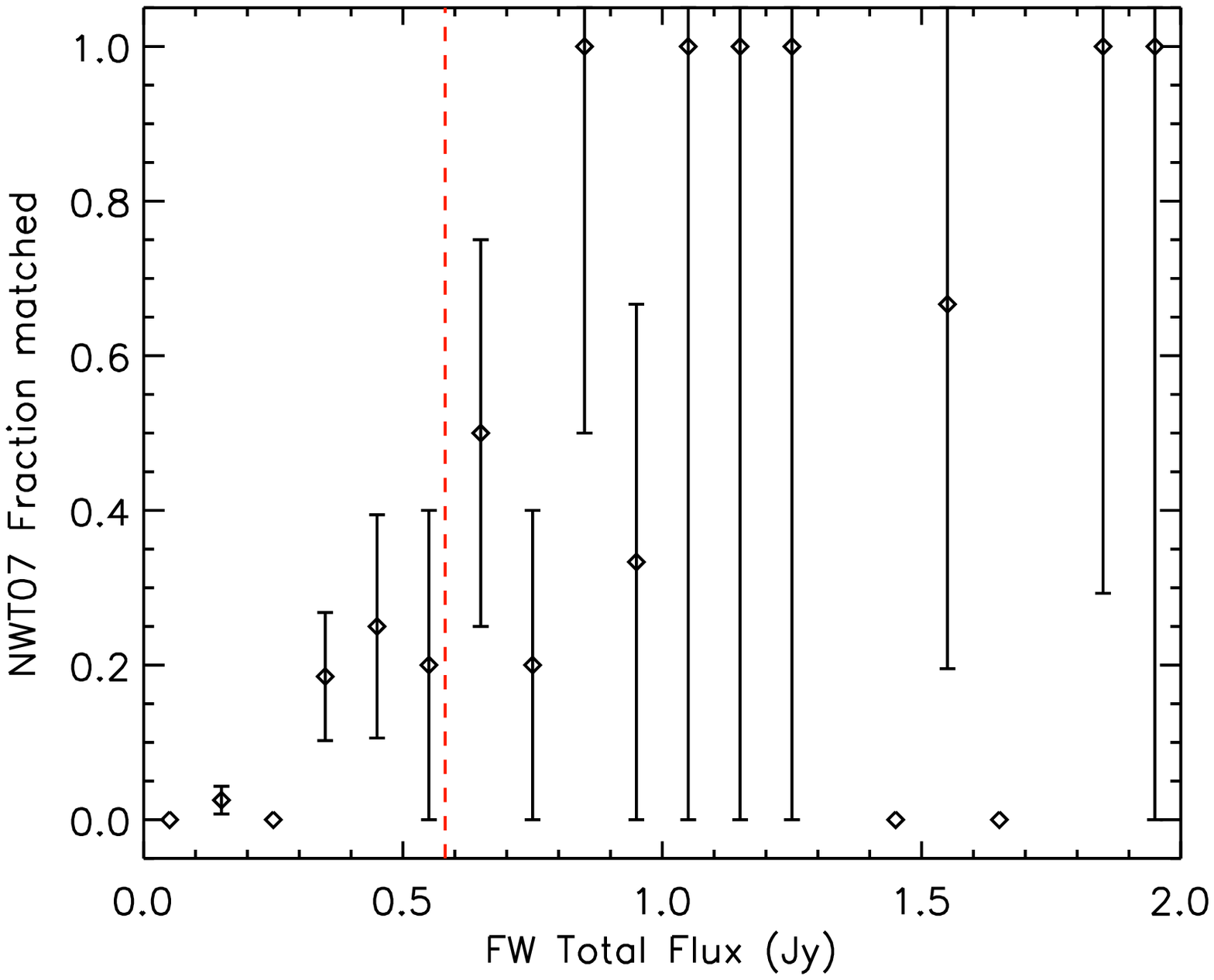} \\
\includegraphics[width=3in]{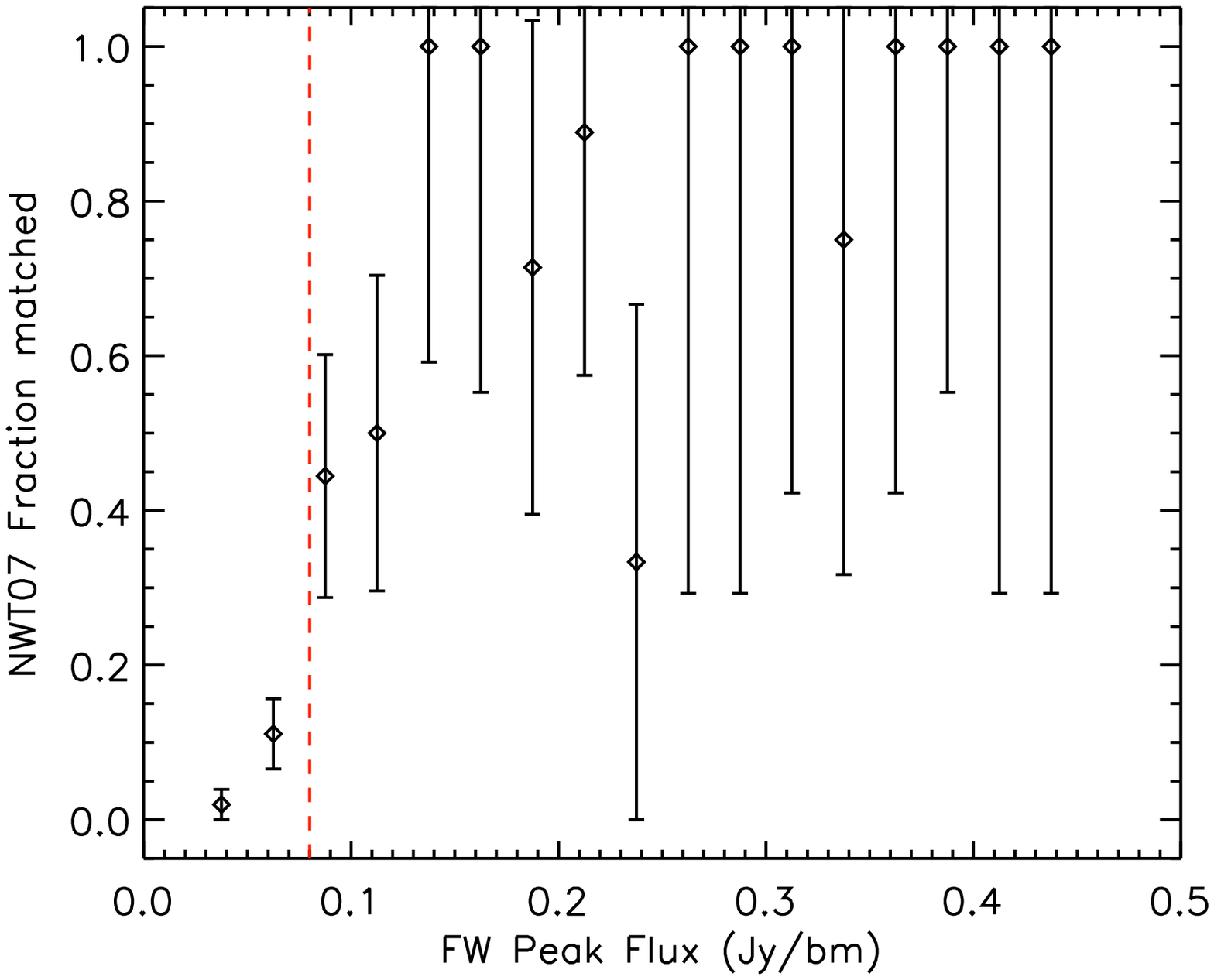} &
\includegraphics[width=3in]{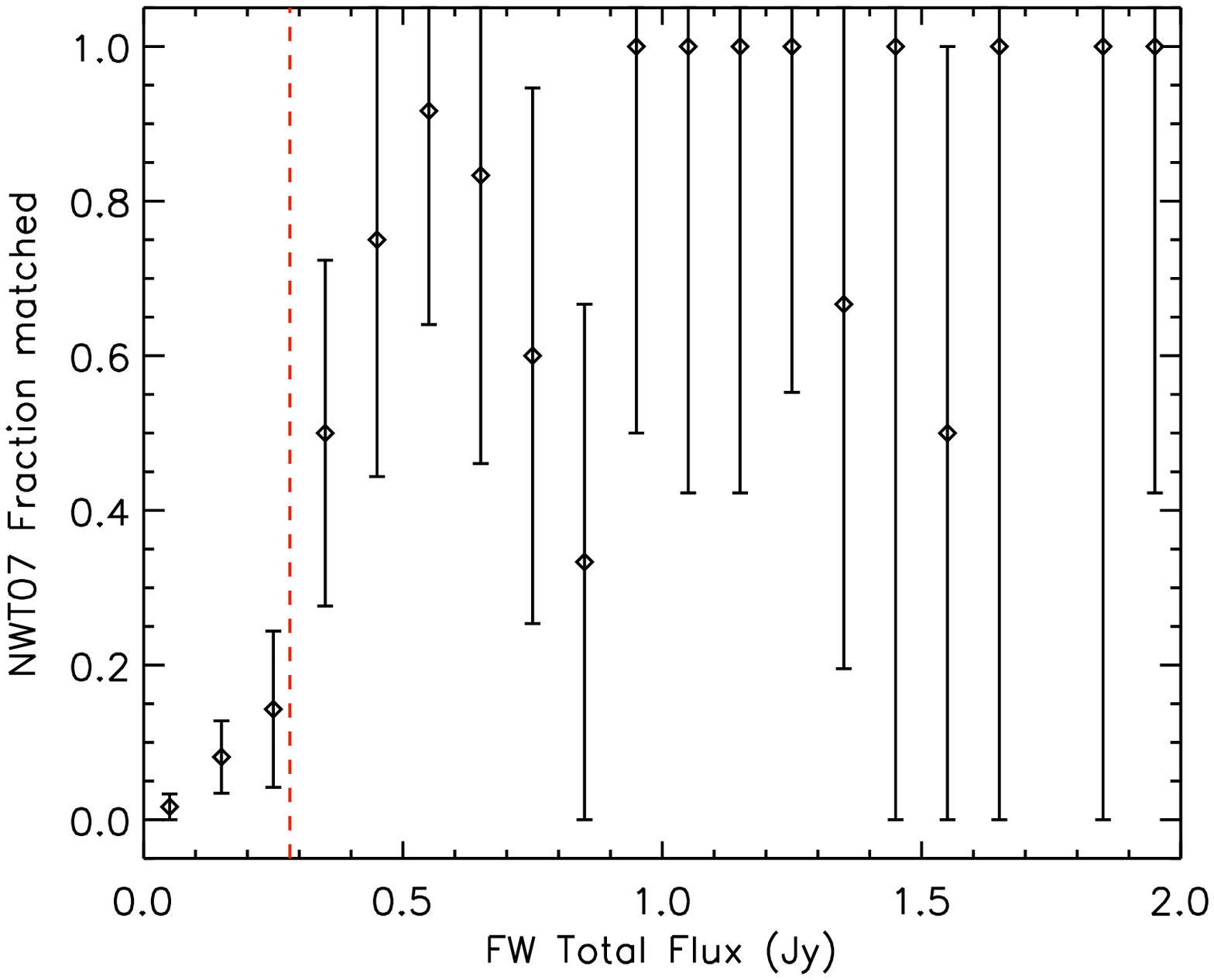} \\
\end{tabular}
\caption{The fraction of our SCUBA-2 cores which have a match in the SCUBA
	catalogue of NWT07 for \None\ (top row) and \Ntwo\ (bottom row).  
	The left panels show the fraction as a function of 
	peak flux, binned to 0.025~Jy~bm$^{-1}$, while the right panels show the
	fraction as a function of total flux, binned to 0.1~Jy.  The vertical bars denote
	$\sqrt(N)$ counting statistics within the NWT07 catalogue, while
	the vertical (red) dashed lines give the approximate completeness levels
	in NWT07 (see text for details).
	}
\label{fig_compl_frac}
\end{figure*}

Finally, we show the effective completeness levels in the SCUBA catalogues by
determining the fraction of SCUBA-2 cores found.  Figure~\ref{fig_compl_frac}
shows the fraction of cores NWT07 identify in our SCUBA-2 catalogue
as a function of peak flux (left panel) and total flux (right panel).  
The bin sizes adopted were chosen to ensure most bins were reasonably populated, while still
showing sufficient detail at low fluxes.  The vertical lines in both panels show the 
statistical / counting uncertainty based on the number of NWT07 cores in each 
bin, illustrating that most bins are not strongly populated.  Still, it is clear that at
higher peak and total flux values, the fraction of matches is generally good.  We visually
inspected the instances of fewer correspondences at higher peak and total fluxes and
found that these were attributable to differences in how the FellWalker and ClumpFind
algorithms divided complex emission structures into individual cores.  

In the lower
peak and total flux regimes, we find that the NWT07 catalogue is complete to
roughly 40\% to 60\% of the cores in our catalogue at their nominal completeness level.  For the
peak flux, the completeness level shown (vertical dashed line) is five times the SCUBA
noise level as listed in NWT07, which they state was used as the minimum 
ClumpFind threshold.  The total flux completeness level is more difficult to determine,
since it varies with core size.  We roughly estimated the completeness level by 
taking a flux of three times the noise level across the average area of their 
cores\footnote{Normally, ClumpFind includes pixels within cores that have
fluxes down to 2~$\sigma$ below the specified minimum peak flux value.}.
Although there is a significant amount of uncertainty introduced by the complex emission
structure, these comparisons suggest that the NWT07 catalogue would be 
roughly 90\% complete at a completeness limit of approximately 50\% higher 
than quoted in their paper.
 
These comparisons show that the SCUBA-2 data of Orion~B analyzed here provide significantly 
more sensitive coverage
than the SCUBA data (as well as also covering a larger area, as is evident comparing, e.g.,
Figures~\ref{fig_N2023} and \ref{fig_compl_im_2023}).  With a typical noise level of 
$\sim$3.7~mJy~bm$^{-1}$, our nominal 3~$\sigma$ completeness is roughly 11~mJy~bm$^{-1}$ in
peak flux, although we expect the actual value to be slightly larger, since faint cores
lying outside of our external mask will have somewhat diminshed flux levels.

\acknowledgements{
The authors thank the referee for a constructive report.
The authors wish to recognize and acknowledge the very significant cultural role 
and reverence that the summit of Maunakea has always had within the indigenous 
Hawaiian community.  We are most fortunate to have the opportunity to conduct 
observations from this mountain.
The JCMT has historically been operated by the Joint Astronomy Centre on behalf of the 
Science and Technology Facilities Council of the United Kingdom, the National Research 
Council of Canada and the Netherlands Organisation for Scientific Research. Additional 
funds for the construction of SCUBA-2 were provided by the Canada Foundation for 
Innovation. The identification number for the programme under which the SCUBA-2 data 
used in this paper is MJLSG41\footnote{One scan in \None\ (target OrionB\_450\_S) 
is presently mis-labelled
in CADC with the project code MJLSG31 (the designation for Orion A).  Observations
taken during science verification across all GBS regions falls under the project code MJLSG22}.  
The authors thank the JCMT staff for their support of
the GBS team in data collection and reduction efforts.
The Starlink software \citep{Currie14} is supported by 
the East Asian Observatory.  These data were reduced using a development version from 
December 2014 (version 516b455a).
This research used the services of the Canadian Advanced Network for
Astronomy Research (CANFAR) which in turn is supported by CANARIE,
Compute Canada, University of Victoria, the National Research Council of
Canada, and the Canadian Space Agency.
This research used the facilities of the Canadian Astronomy Data Centre operated by the 
National Research Council of Canada with the support of the Canadian Space Agency.
Figures in this paper were creating using the NASA IDL astronomy library
\citep{idlastro} and the Coyote IDL library ({\tt http://www.idlcoyote.com/index.html}).
}

\bibliographystyle{apj}
\bibliography{orionbib}{}

\input{tab4}

\end{document}

%% file: tab1.tex
\begin{deluxetable*}{ccccccccc}
\tablecolumns{9}
\tablewidth{0pc}
\tabletypesize{\scriptsize}
\tablecaption{Noise per area observed\label{tab_pong_noise}}
\tablehead{
\colhead{Region} &
\colhead{Name\tablenotemark{a}} &
\colhead{R.A.\tablenotemark{b}} &
\colhead{decl.\tablenotemark{b}} &
\colhead{$\sigma_{850}$\tablenotemark{c}} &
\colhead{$\sigma_{450}$\tablenotemark{c}} &
\colhead{$\sigma_{850}$\tablenotemark{d}} &
\colhead{$\sigma_{450}$\tablenotemark{d}} &
\colhead{N$_{obs}$\tablenotemark{e}} \\
\colhead{} & \colhead{} &
\colhead{(J2000.0)} & \colhead{(J2000.0)} &
\multicolumn{2}{c}{(mJy~arcsec$^{-2}$)} & 
\multicolumn{2}{c}{(mJy~bm$^{-1}$)} & 
\colhead{}
}
\startdata
 LDN~1622 &  ORIONBN\_850\_solo & 5:54:33 & 1:49:34  & 0.053 &  2.0 & 3.9 & 98 & 6 \\
 \Ntwo &  ORIONBN\_450\_E    & 5:47:55  &   0:13:60  & 0.050 &  1.0 & 3.7 & 49 & 6 \\
 \Ntwo &  ORIONBN\_450\_S    & 5:46:17  &   0:06:30  & 0.050 &  1.2 & 3.7 & 59 & 6 \\
 \Ntwo &  ORIONBN\_450\_W    & 5:45:55  &   0:24:42  & 0.055 &  1.7 & 4.0 & 84 & 6 \\
 \Ntwo &  ORIONBN\_850\_N    & 5:47:33  &   0:45:26  & 0.047 &  0.9 & 3.4 & 44 & 6 \\
 \None &  ORIONBS\_450\_E    & 5:42:38  &  -1:54:19  & 0.049 &  1.1 & 3.6 & 54 & 6 \\
 \None &  ORIONBS\_450\_S    & 5:41:16  &  -2:18:26  & 0.051 &  0.8 & 3.7 & 39 & 4 \\
 \None &  ORIONBS\_450\_W    & 5:40:34  &  -1:48:26  & 0.052 &  0.9 & 3.8 & 44 & 4 \\
 \None &  ORIONBS\_850\_N    & 5:43:39  &  -1:09:11  & 0.047 &  1.0 & 3.4 & 49 & 6 \\
 \None &  ORIONBS\_850\_S    & 5:41:53  &  -1:24:41  & 0.043 &  1.2 & 3.1 & 59 & 7 \\
\enddata
\tablenotetext{a}{Observation designation chosen by GBS team, denoted as Target Name in
	the CADC database at {\tt http://www3.cadc-ccda.hia-iha.nrc-cnrc.gc.ca/en/jcmt/} }
\tablenotetext{b}{Central position of each observation}
\tablenotetext{c}{Pixel-to-pixel (rms) noise for the final mosaic of all of the observed
	PONG 1800s for the given area at 850~\mum\ and 450~\mum\ respectively.}
\tablenotetext{d}{Effective noise per beam (i.e., point source sensitivity) for the final 
	mosaic of all of the observed
	PONG 1800s for the given area at 850~\mum\ and 450~\mum\ respectively.}
\tablenotetext{e}{Total number of PONG 1800 observations taken at each wavelength.  Note that
	this count may include partially completed scans.}
\end{deluxetable*}

%% file: tab2.tex
\begin{deluxetable}{ccccccc}
\tablecolumns{7}
\tablewidth{0pc}
\tabletypesize{\footnotesize}
\tablecaption{Ratio of Starless to Protostellar Cores \label{tab_proto_ratio}}
\tablehead{
\colhead{Density Range\tablenotemark{a}} &
\multicolumn{3}{c}{All Cores\tablenotemark{b}} &
\multicolumn{3}{c}{Cores $> 0.1$~\Msol \tablenotemark{c}} \\
\colhead{(cm$^{-3}$)} & 
\colhead{N$_{sl}$} &
\colhead{N$_{p}$} &
\colhead{Ratio} &
\colhead{N$_{sl}$} &
\colhead{N$_{p}$} &
\colhead{Ratio} 
}
\startdata
$>10^5$             & 34  & 31 & 1:1  & 34 & 31 & 1:1 \\
$10^{4.5} - 10^{5}$ & 270 & 29 & 9:1  & 205 & 29 & 7:1\\
$10^{4} - 10^{4.5}$ & 546 & 5  & 109:1 & 392 & 5 & 78:1\\
\enddata
\tablenotetext{a}{Mean core densities calculated using the total mass and effective radius.}
\tablenotetext{b}{Number of starless cores, protostellar cores, and their ratio for the full
	dense core sample.}
\tablenotetext{c}{Number of starless cores, protostellar cores, and their ratio for dense cores
	above 0.1~\Msol.}
\end{deluxetable}

%% file: tab3.tex
\begin{deluxetable}{ccccccccc}
\tablecolumns{9}
\tablewidth{0pc}
\tabletypesize{\footnotesize}
\tablecaption{Number of cores matched between SCUBA and SCUBA-2 maps\label{tab_core_match}}
\tablehead{
\colhead{Region} &
\multicolumn{2}{c}{NWT07\tablenotemark{a}} &
\multicolumn{2}{c}{SLC\tablenotemark{b}} &
\multicolumn{2}{c}{FW-NWT07\tablenotemark{c}} &
\multicolumn{2}{c}{FW-SLC\tablenotemark{d}} \\
\colhead{ } &
\colhead{match} &
\colhead{no mat.} &
\colhead{match} &
\colhead{no mat.} &
\colhead{match} &
\colhead{no mat.} &
\colhead{match} &
\colhead{no mat.} 
}
\startdata
 \None & 59 & 0 & 90 & 9 & 57 & 312 & 80 & 291 \\ 
 \Ntwo & 100 & 0 & 87 & 2 & 90 & 147 & 80 & 157 \\
\enddata
\tablenotetext{a}{Number of cores in \citet{Nutter07} which were and were not matched to
	a SCUBA-2 FellWalker core.}
\tablenotetext{b}{Number of cores in the SCUBA Legacy Catalog \citep{DiFrancesco08} 
	which were and were not matched to a SCUBA-2 FellWalker core.}
\tablenotetext{c}{Number of SCUBA-2 FellWalker cores associated with one or more cores
	in \citet{Nutter07}.}
\tablenotetext{d}{Number of SCUBA-2 FellWalker cores associated with one or more cores
	in the SCUBA Legacy Catalog.}
\end{deluxetable}

%% file: tab4.tex
\begin{deluxetable}{lclcclllllccccc}
\tablecolumns{15}
\tablewidth{0pc}
\tabletypesize{\footnotesize}
\tablecaption{Dense Cores Identified with Fellwalker\label{tab_cores}}
\tablehead{
\colhead{Source} &
\colhead{ } &
\colhead{ } &
\colhead{ } &
\colhead{ } &
\multicolumn{3}{c}{850~\mum} &
\multicolumn{2}{c}{450~\mum} &
\colhead{ } &
\multicolumn{3}{c}{CO Contam.} &
\colhead{Bkgrd} \\
\colhead{Name} &
\colhead{Region} & 
\colhead{Index} &
\colhead{R.A.\tablenotemark{a}} &
\colhead{Dec.\tablenotemark{a}} &
\colhead{Peak \tablenotemark{b}} &
\colhead{Total \tablenotemark{b}} &
\colhead{$R_{eff}$\tablenotemark{b}} &
\colhead{Peak \tablenotemark{c}} &
\colhead{Total \tablenotemark{c}} &
\colhead{YSO\tablenotemark{d}} &
\colhead{Cov.\tablenotemark{e}} &
\colhead{Peak\tablenotemark{e}} &
\colhead{Total\tablenotemark{e}} &
\colhead{Total\tablenotemark{f}} \\
\colhead{MJLSG} &
\colhead{ } &
\colhead{ } &
\colhead{(J2000.0)} & 
\colhead{(J2000.0)} &
\colhead{(Jy~bm$^{-1}$)} &
\colhead{(Jy)} &
\colhead{(pc)} &
\colhead{(Jy~bm$^{-1}$)} &
\colhead{(Jy)} &
\colhead{ } & 
\colhead{ } &
\colhead{(Jy~bm$^{-1}$)} &
\colhead{(Jy)} &
\colhead{(Jy)} 
}
\startdata
    J055424.5+014419 &      L1622 &    1 &    5:54:24.60 &   +01:44:19.7 &    1.200 &    4.889 &    0.121 &    4.656 &   27.708 & Y   & 0.00 &      -1 &      -1 &   4.245 \\
    J055412.3+014234 &      L1622 &    2 &    5:54:12.39 &   +01:42:34.7 &    0.202 &    2.013 &    0.108 &    0.985 &   10.506 & Y   & 0.00 &      -1 &      -1 &   1.528 \\
    J055440.2+015404 &      L1622 &    3 &    5:54:40.21 &   +01:54:04.7 &    0.138 &    1.714 &    0.135 &    0.772 &    6.288 & N   & 0.00 &      -1 &      -1 &   0.952 \\
    J055403.3+014025 &      L1622 &    4 &    5:54:03.39 &   +01:40:25.7 &    0.115 &    0.174 &    0.043 &    0.647 &    0.775 & N   & 0.00 &      -1 &      -1 &   0.084 \\
    J055406.1+014255 &      L1622 &    5 &    5:54:06.19 &   +01:42:55.7 &    0.110 &    0.665 &    0.073 &    0.730 &    3.043 & N   & 0.00 &      -1 &      -1 &   0.387 \\
    J055429.7+014737 &      L1622 &    6 &    5:54:29.80 &   +01:47:37.7 &    0.101 &    0.782 &    0.081 &    0.746 &    3.132 & Y   & 0.00 &      -1 &      -1 &   0.432 \\
    J055453.4+015837 &      L1622 &    7 &    5:54:53.41 &   +01:58:37.7 &    0.088 &    0.273 &    0.057 &    0.608 &    0.978 & N   & 0.00 &      -1 &      -1 &   0.121 \\
    J055440.0+014340 &      L1622 &    8 &    5:54:40.00 &   +01:43:40.7 &    0.088 &    0.241 &    0.060 &    0.604 &    0.369 & N   & 0.00 &      -1 &      -1 &   0.090 \\
\enddata
\tablenotetext{a}{Position of peak emission in the dense core}
\tablenotetext{b}{Peak flux, total flux, and effective radius of the dense cores, as measured
	at 850~\micron.  The effective beamsize at 850~\mum is 14.6\arcsec.}
\tablenotetext{c}{Peak flux and total flux at 450~\mum within the dense core boundaries defined
	on the 850~\mum image (note that the total flux may be negative in regions without
	a detected 450~\mum signal).  The effective beamsize at 450~\mum is 9.8\arcsec.}
\tablenotetext{d}{Is the dense core associated with a protostar 
	in {\it Spitzer} \citep{Megeath12} or {\it Herschel} \citep{Stutz13}?}
\tablenotetext{e}{CO contamination: fraction of dense core footprint where CO observations exist,
	peak and total flux at 850~\mum in CO subtracted mosaic (-1 indicates null values).}
\tablenotetext{f}{Background subtracted total flux at 850~\micron.}

\end{deluxetable}